# *In situ* Gas-Cell Electron Microscopy Reveals Pressure-Selected Restructuring Pathways in AuRu Ammonia Catalysts

Amy S. McKeown-Green[#†*], Parivash Moradifar[#‡*], Zisheng Zhang[¶,§], Cedric Lim[‡], Andrew Barnum[‖], Lin Yuan[‡], Robert Sinclair[‡], Frank Abild-Pedersen[§], Colin Ophus[‡], Jennifer A. Dionne[‡*].

†*Department of Chemistry, Stanford University, Stanford, California 94305, United States.*

‡*Department of Materials Science and Engineering, Stanford University, Stanford, California 94305, United States.*

¶*Department of Chemical Engineering, Stanford University, Stanford, California 94305, United States.*

§*SUNCAT Center for Interface Science and Catalysis, SLAC National Accelerator Laboratory, Menlo Park, California 94025, United States.*

‖*Stanford Nano Shared Facilities, Stanford University, Stanford, California 94305, United States.*

[#]These authors contributed equally to this work

*Corresponding authors. Emails: amygreen@stanford.edu; pmoradi@stanford.edu; jdionne@stanford.edu

## Abstract

Bimetallic catalysts provide new routes toward sustainable ammonia synthesis, but the nanoscale structural dynamics under reaction-relevant conditions remain poorly understood. Here, we combine *in situ* gas-cell and multimodal electron microscopy to determine how temperature, gas pressure, and chemistry select among distinct restructuring pathways in AuRu nanocrystal catalysts. Initially, the AuRu nanocrystals form polycrystalline face-centered cubic (FCC) alloys with Au/Ru intermixing. Elevated temperature (≥350 °C) induces intraparticle phase segregation into distinct Au-rich (FCC) and Ru-rich hexagonal close-packed (HCP) domains that exhibit localized plasmonic modes. Atmospheric-pressure 3:1 $H_2$:$N_2$ gas unlocks a distinct restructuring regime absent at lower pressures, characterized by pronounced faceting and nanovoid formation. Systematic gas variation identifies hydrogen as the dominant driver. Density functional theory-trained machine-learning interatomic potentials and grand-canonical Monte Carlo simulations reveal that H–Ru interactions enhance the Au/Ru diffusivity mismatch, promoting vacancy accumulation and nanovoid formation. Together, these results show that, rather than simply accelerating the thermally driven phase segregation observed at lower pressures, atmospheric-pressure $H_2$:$N_2$ gas redirects restructuring toward faceting and nanovoid formation through a gas-mediated Kirkendall-type mechanism.

## Introduction

Bimetallic nanocrystals unlock emergent structures and properties unavailable to single-component materials by coupling metals with distinct chemical, electronic, and optical characteristics[1–3]. This is particularly valuable in catalysis, where complementary functions within a single nanocrystal can provide precise control over reactant activation, charge transfer, and selectivity[3–6]. Yet this synergy comes at a cost. Bimetallic nanocrystals are structurally complex, exhibiting strain, chemical potential variations, and compositional gradients that promote dynamic restructuring under reactive conditions[2,7,8].

The tension between catalytic multifunctionality and structural stability is especially relevant to ammonia synthesis, where reducing the energetic demands of the Haber–Bosch process has motivated the development of multicomponent catalysts that can activate $N_2$ and mediate hydrogenation under mild conditions[9,10]. Recently, bimetallic catalysts such as AuRu, AuOs, and CuFe have coupled strong N-binding metals with optically active hosts, enabling $NH_3$ synthesis at atmospheric pressure.[6,11–13] While the benchtop reactor performance of these platforms is increasingly well documented, our understanding of their nanoscale structure, stability, and evolution under reaction-relevant conditions remains limited. Probing this behavior is critical as previous studies of related Au- and Pt-based bimetallic systems have shown them to be highly dynamic, with particle size and operating temperature strongly influencing whether the constituent metals remain intermixed or rearrange into core–shell or Janus-type morphologies[14–16]. Determining the processes governing the nanoscale structure of ammonia-relevant bimetallic catalysts therefore requires direct, *in situ* observation under operating conditions.

Environmental transmission electron microscopy (ETEM) is a powerful tool for capturing nanoscale structural, optical, and chemical changes in reactive gas environments[17–23]. Its application to $NH_3$ synthesis for Ru-based catalysts has resolved how insulating supports stabilize catalytically relevant sites and how these catalysts can sinter and grow in reactive $H_2$:$N_2$ gas environments[24,25]. However, traditional differential pumping of the TEM column has confined such studies to pressures of $10^{-5}$-$10^{-3}$ atm[24–27] (Figure S1), which are well below those used in emerging benchtop reactors (1 atm)[6,13,28] and far below those employed in industrial Haber–Bosch reactors (50-200 atm)[29,30]. As gas pressure can alter adsorbate coverage, adsorption kinetics, and surface diffusion,[31] higher-pressure environments can enable distinct restructuring regimes, rather than simply accelerate structural changes observed at reduced pressures.[7,32,33] Studying catalyst evolution across a broad range of pressures is therefore essential for identifying how pressure selects among competing restructuring pathways and which predominate under reactor-relevant conditions.

To directly compare catalyst structure across these pressure regimes, we employ an *in situ* gas-cell scanning transmission electron microscopy (STEM) platform that supports reactive gas mixtures at pressures up to 1 atm. We select AuRu nanocrystals as a representative ammonia-relevant bimetallic catalyst with demonstrated benchtop catalytic activity,[6] and visualize their evolution in real time under vacuum, sub-atmospheric, and atmospheric pressures in $H_2$:$N_2$ (3:1) gas mixtures. As *in situ* STEM imaging is primarily sensitive to atomic number (Z) contrast, resolving these transformations requires correlative measurements that connect chemical, structural, optical, and morphological information. We combine atomic-resolution imaging with electron spectroscopy, diffraction, and tomography in a multimodal framework that maps changes across individual particles. The resulting multidimensional datasets connect real-time morphological changes with the underlying structural and chemical transformations that define each restructuring pathway.

Using this multimodal electron microscopy framework, we systematically decouple the effects of temperature, pressure, and gas chemistry on AuRu nanocrystals.

As-synthesized AuRu (1:0.2) nanocrystals exhibit a polycrystalline face-centered cubic (FCC) crystal structure with Au–Ru intermixing. Vacuum annealing drives phase segregation into Au-rich FCC and Ru-rich hexagonal-close-packed (HCP) domains. We then introduce a $H_2$:$N_2$ (3:1) gas environment at varying pressures to distinguish pathways that persist under reduced-pressure conditions from those that emerge only at higher pressures. Within this gas environment, phase segregation occurs across all pressures, whereas increased nanocrystal faceting and internal nanovoid formation emerge only at atmospheric pressure. We subsequently vary gas chemistry at atmospheric pressure and find that Ar and $N_2$ reproduce vacuum-like restructuring, whereas $H_2$ promotes nanovoid formation, implicating hydrogen as the key chemical driver. Calculations using Density Functional Theory (DFT)-trained machine-learning interatomic potentials reveal relative diffusion barriers for Au and Ru that support a hydrogen-mediated Kirkendall-type mechanism in which absorbed H atoms alter Au–Ru interdiffusion and promote void formation. Collectively, these correlative *in situ* and multimodal measurements show that pressure can alter the nature, not only the extent, of bimetallic nanocrystal evolution by enabling hydrogen-mediated restructuring absent under vacuum, inert, and sub-atmospheric conditions.

## Results and Discussion

### Thermally driven phase segregation restructures AuRu nanocrystal catalysts

In their pristine state, the AuRu nanocrystals are polycrystalline and frequently pentatwinned with an FCC lattice (Figure 1b). Atomic high-angle annular dark-field (HAADF)-STEM reveals a lattice constant of 0.405 nm, calculated from the measured {111} lattice-plane spacing of 0.234 nm (Figure 1c). This calculated lattice constant is slightly lower than that of pure Au (a = 0.408 nm) and is consistent with Ru-alloying. X-EDS elemental mapping shows nanoscale Au–Ru intermixing within individual pristine nanocrystals despite the bulk immiscibility of Au and Ru (Figure 1d)[1,34]. While Au and Ru do not form a solid solution, the characteristic Au Lα/Mα and Ru Kα/Lα signals show that Au and Ru intermix and are distributed throughout the nanocrystals, with modest local variation in Ru content (Figure 1d and Figures S2–S4). Representative X-EDS quantification of the nanocrystal in Figure 1d gives an Au:Ru ratio of 1:0.15 (Table S1), comparable to the nominally prepared ratio of 1:0.2.

Monochromated low-loss electron energy loss spectroscopy (EELS) reveals a persistent Au plasmon appearing near 2.5 eV (Figure 1e)[6]. 4D-STEM dark-field imaging shows structural heterogeneity across the AuRu nanocrystal ensemble, including variations in internal grain structure and faceting (Figure 1f). Single-point nanobeam electron diffraction patterns collected from selected probe positions confirm an FCC crystal structure (Fm-3m) with varied crystallographic orientations (Figure 1g and Figures S5–S6). The in-plane and out-of-plane orientations were determined using automated crystal orientation mapping (ACOM) with the py4DSTEM Python package and are shown in Figure S5[35,36]. The structural diversity likely reflects alloying-induced strain, which facilitates the formation of stacking faults and twin boundaries[37]. Additional HAADF-STEM characterization shows a mean nanocrystal diameter of 25 ± 6 nm, indicating a relatively narrow size distribution despite particle-to-particle structural heterogeneity (Figure S7).

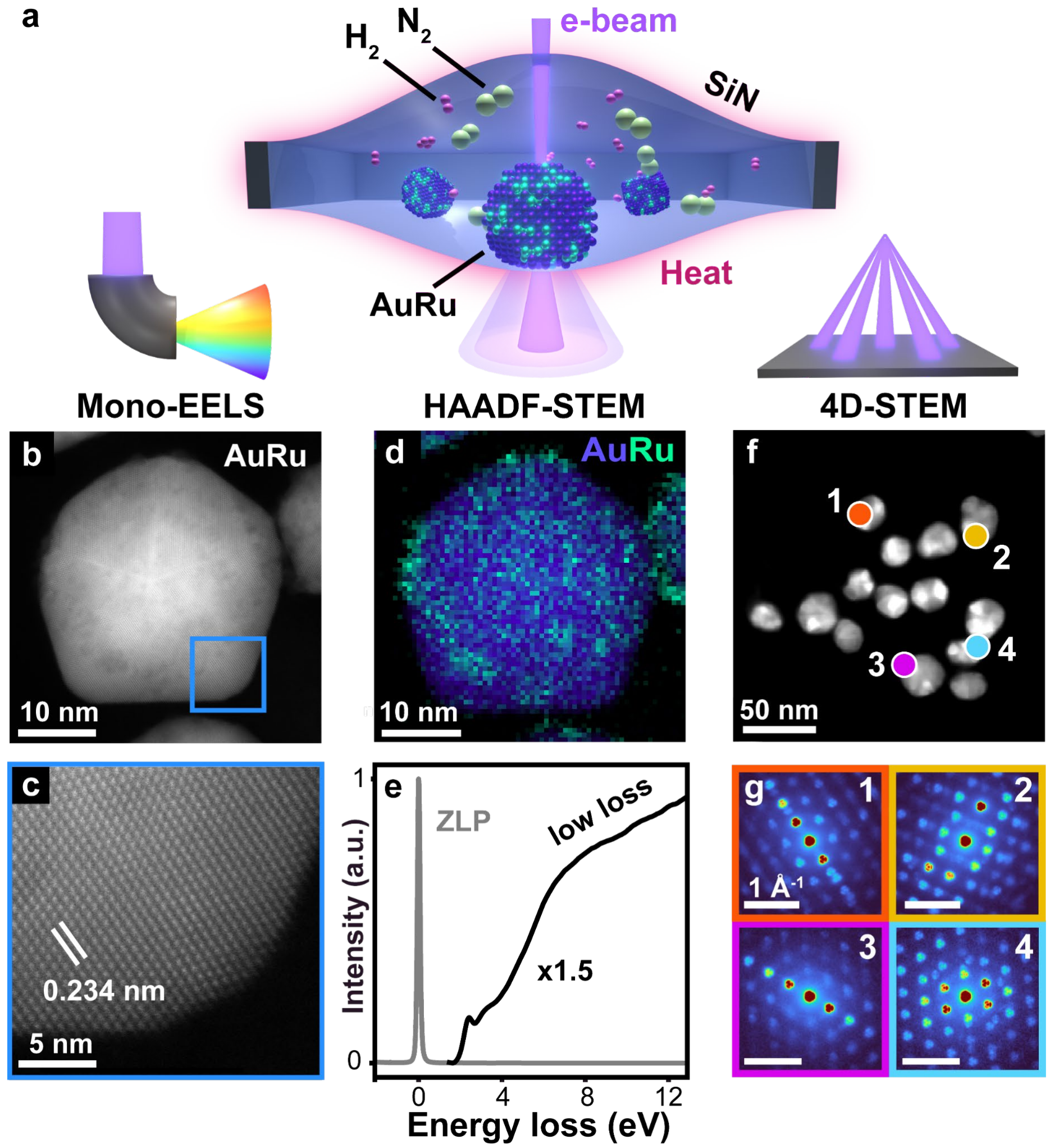

Figure 1. Multimodal characterization of pristine AuRu nanocrystals. (a) Schematic of the experimental setup showing the high-pressure gas-cell integrated with HAADF-STEM, along with monochromated EELS and 4D-STEM modalities used for correlative structural and optical characterization. (b) Atomic-resolution HAADF-STEM image of a representative polycrystalline AuRu (1:0.2) bimetallic nanocrystal, with the boxed region enlarged in (c) showing atomic columns and measured {111} lattice spacing of 0.234 nm. (d) X-EDS elemental map of the nanocrystal in (b), showing intermixing of Au and Ru based on the characteristic Au Lα/Mα and Ru Kα/Lα x-ray lines. For display, each X-ray line, Au Lα/Mα and Ru Kα/Lα, was independently scaled to its 99.9th percentile intensity. Figure S3 shows individual maps. (e) Monochromated dual-EELS spectrum showing the zero-loss peak (ZLP) and low-loss region with the AuRu plasmon resonance near 2.5 eV. (f) 4D-STEM dark-field image of pristine AuRu nanocrystals showing

internal grain structure. (g) Representative diffraction patterns collected from the numbered probe positions marked in (f), with colored borders corresponding to the marker colors in (f).

Our multimodal characterization has shown that, in their pristine state, the AuRu nanocrystals have an FCC crystal structure, exhibit nanoscale Au–Ru intermixing, support a 2.5 eV plasmon resonance, and contain diverse grain structures. We next examine how AuRu catalytic nanocrystals evolve under increasing temperature in vacuum, allowing temperature-driven restructuring to be isolated from gas-induced changes. We find that under vacuum annealing, Au and Ru phase-segregate into Ru-rich and Au-rich regions with different crystal structures.

To track the onset and progression of this temperature-driven restructuring, we performed stepwise vacuum annealing with the electron beam blanked during heating and briefly unblanked to image the same nanocrystal at 25, 350, 450, 550, 650, and 750 °C. The sample was ramped to each target temperature at 20 °C/min, equilibrated at that temperature for 5 min with the beam blanked, and then imaged for 5 min, giving a total hold time of 10 min. Pre-annealing X-EDS maps of the target nanocrystals show nanoscale Au–Ru intermixing with local variation in Ru content rather than complete phase separation (Figure 2a). This variation is evident in particle #1, which contains a modest Ru-enriched region in the lower half of the nanocrystal, whereas particle #2 shows a more even Ru distribution (Figure 2a). In the HAADF-STEM image series, Au–Ru phase segregation is first evident between 350 and 450 °C as lower-Z, dark-contrast domains that progressively increase in size with temperature (Figure 2b, blue arrows; Figure S8). HRTEM imaging provides complementary evidence of phase segregation, showing the emergence of a lighter-contrast Ru-rich domain within the darker-contrast Au-rich nanocrystal (Figure 2c, orange arrows). Post-annealing X-EDS elemental mapping confirms near-complete phase segregation across the analyzed particles, resolving spatially distinct Au- and Ru-rich regions (Figure 2d). Pre- and post-annealing spectra and local quantification further corroborate this near-complete compositional segregation (Figures S9–S11 and Tables S2–S3).

To test whether repeated electron-beam exposure contributes to the observed phase segregation, we compared AuRu nanocrystals imaged throughout the full annealing series with nanocrystals first exposed to the beam only at later temperature steps in both STEM and HRTEM modes (Figures S12–S13). This comparison is important because electron-beam exposure can influence *in situ* TEM measurements through atomic displacement, ionization/radiolysis, beam heating, and contamination[38]. In both imaging modes, nanocrystals with and without prior beam exposure show the same temperature-dependent evolution, including Au–Ru phase segregation and comparable final segregated morphologies as confirmed by X-EDS elemental maps (Figures S12–S13). These controls indicate that repeated electron-beam exposure does not drive phase segregation. Additionally, catalyst restructuring is known to be strongly influenced by catalyst–support interactions. To investigate whether the presence of support was sufficient to stabilize the catalyst, we performed identical STEM-mode vacuum-annealing experiments on AuRu nanocrystals supported on MgO (Figure S14). Phase segregation is also observed in this AuRu/MgO catalyst-support system using X-EDS elemental mapping (Figure S14), indicating that the presence of MgO alone is insufficient to prevent this thermally enabled restructuring.

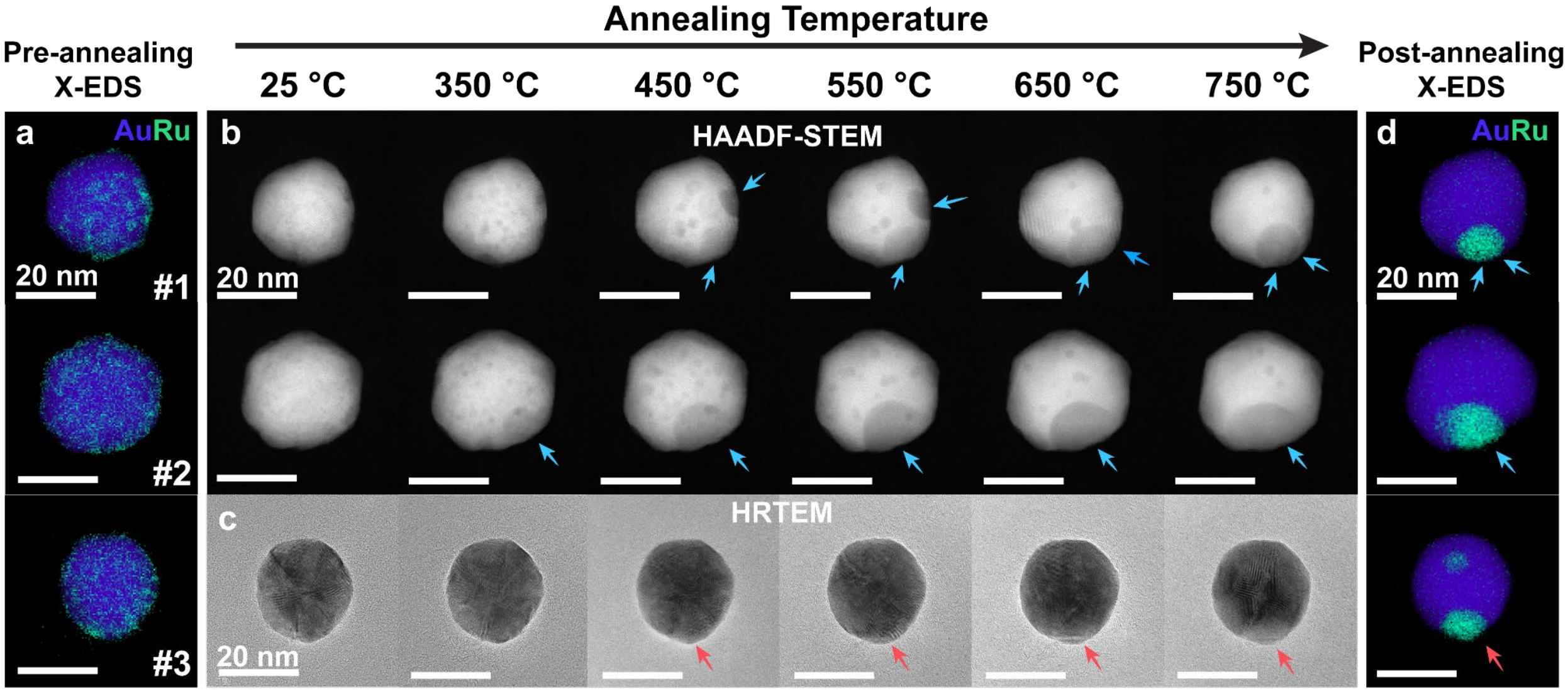

Figure 2. Vacuum annealing induces Au–Ru phase segregation. (a) Pre-annealing X-EDS elemental maps of AuRu nanocrystals #1, #2, and #3, showing variable Au and Ru intermixing in the pristine state. (b) HAADF-STEM image series of nanocrystals #1 and #2 during stepwise vacuum annealing from room temperature to 750 °C, showing progressive Au–Ru phase segregation into regions of distinct Z-contrast, with Au-rich regions appearing bright and Ru-rich regions appearing dark. Blue arrows indicate the emergence of Ru-rich domains. (c) HRTEM image series of nanocrystal #3 under the same heating protocol, showing analogous phase segregation with contrast reversed relative to HAADF-STEM. Orange arrows indicate emerging Ru-rich domains. (d) Post-annealing X-EDS elemental maps of nanocrystals #1, #2, and #3, confirming the near-complete spatial segregation of Au and Ru. For all composite X-EDS maps, each X-ray line, Au Lα/Mα and Ru Kα/Lα, was independently scaled to its 99.9th percentile intensity. Figure S9 shows individual maps for particle #1 in Figure 2a.

Having established that temperature-driven Au–Ru phase segregation occurs across different imaging conditions and support environments, we next examine how this restructuring changes the atomic structure, crystallography, and optical response of AuRu catalytic nanocrystals. Atomic-resolution HAADF-STEM directly resolves the segregated Au–Ru interface, where we observe a close-packed plane orientation between the close-packed Au FCC {111} planes (0.235 nm measured) and Ru HCP {0002} planes (0.216 nm)[39,40]. This is consistent with an FCC-to-HCP phase transition of Ru during phase segregation (Figure 3a), which has been observed in related AuRu systems and reflects Ru's thermodynamic preference for the HCP structure[41].

To obtain statistically representative insight into phase segregation across the nanocrystal ensemble, we employ 4D-STEM with a patterned bullseye aperture across a sample region containing more than 20 nanocrystals[42]. The patterned aperture enhances Bragg peak discrimination against the amorphous background. Although the summed diffraction pattern is dominated by FCC reflections (Figure 3b), analysis of the Bragg vector histogram resolves contributions from both Au FCC and Ru HCP phases (Figure 3c). Automated crystal orientation mapping (ACOM) with FCC/HCP phase assignment yields spatial maps of distinct Au-rich and Ru-rich regions (Figure 3d)[36]. Representative single-point diffraction patterns are shown in Figure 3e–f, with assignments provided in Figures S15–S16. Quantification yields Ru domains with an average size of 7 ± 4 nm, with 82% and 18% of assigned probe positions indexed as Au FCC and Ru HCP, respectively. This gives a Au:Ru ratio of 1:0.22, which is consistent with the nominal composition of Au:Ru = 1:0.2 and supports near-complete phase separation without measurable Ru loss. This Ru redistribution and FCC-to-HCP transition transforms the coordination environment of Ru atoms, with likely consequences to catalytic activity.

Beyond altering the elemental distribution, this phase segregation also reshapes the electronic and optical response of the AuRu nanocrystals, which has previously been shown to influence catalytic activity[6,13]. Using monochromated EELS, we probe the nanocrystals' plasmonic response using the electron beam to assess how phase segregation influences their optical behavior. Before phase segregation, monochromated low-loss EELS resolves a ~2.5 eV plasmon at the single-particle level (Figure 1e), consistent with ensemble UV–Vis measurements (Figure S17). Post-segregation, in a particle with segregated Ru-rich regions at both ends (Figure 3g), we observe that the Au surface plasmon near 2.5 eV persists across the phase-segregated nanocrystal. However, the plasmon is attenuated within Ru-rich regions, with slight spectral broadening at the Au–Ru interface (Figure 3h). We also observe an additional electron energy loss feature between 10 and 11 eV that is strongly localized to the Ru-rich regions (Figure 3i). Electronic loss-function calculations based on bulk Ru refractive index data identify this feature as a Ru bulk plasmon (Figure S18)[43], consistent with experimental observations[44]. In a second particle with less symmetric Ru regions, spatial maps more clearly reveal an inverse relationship between the Au and Ru plasmonic responses (Figure 3j). The Au surface plasmon is strongest in Au-rich regions, while the Ru plasmon predominantly localizes to Ru-rich domains (Figure 3k–l). Together, these spatial distributions yield a clear anticorrelation between the Au and Ru plasmonic responses (Figure S19). The faint edge-adjacent Ru-plasmon intensity is consistent with plasmon delocalization and edge-enhanced inelastic scattering in low-loss EELS, compounded by convolution with the SiN support. Collectively, these measurements indicate that Au–Ru phase segregation reshapes the plasmonic landscape, introducing a localized deep-UV Ru plasmon and attenuating the Au plasmon in Ru-rich regions, with implications for plasmon-mediated catalysis.

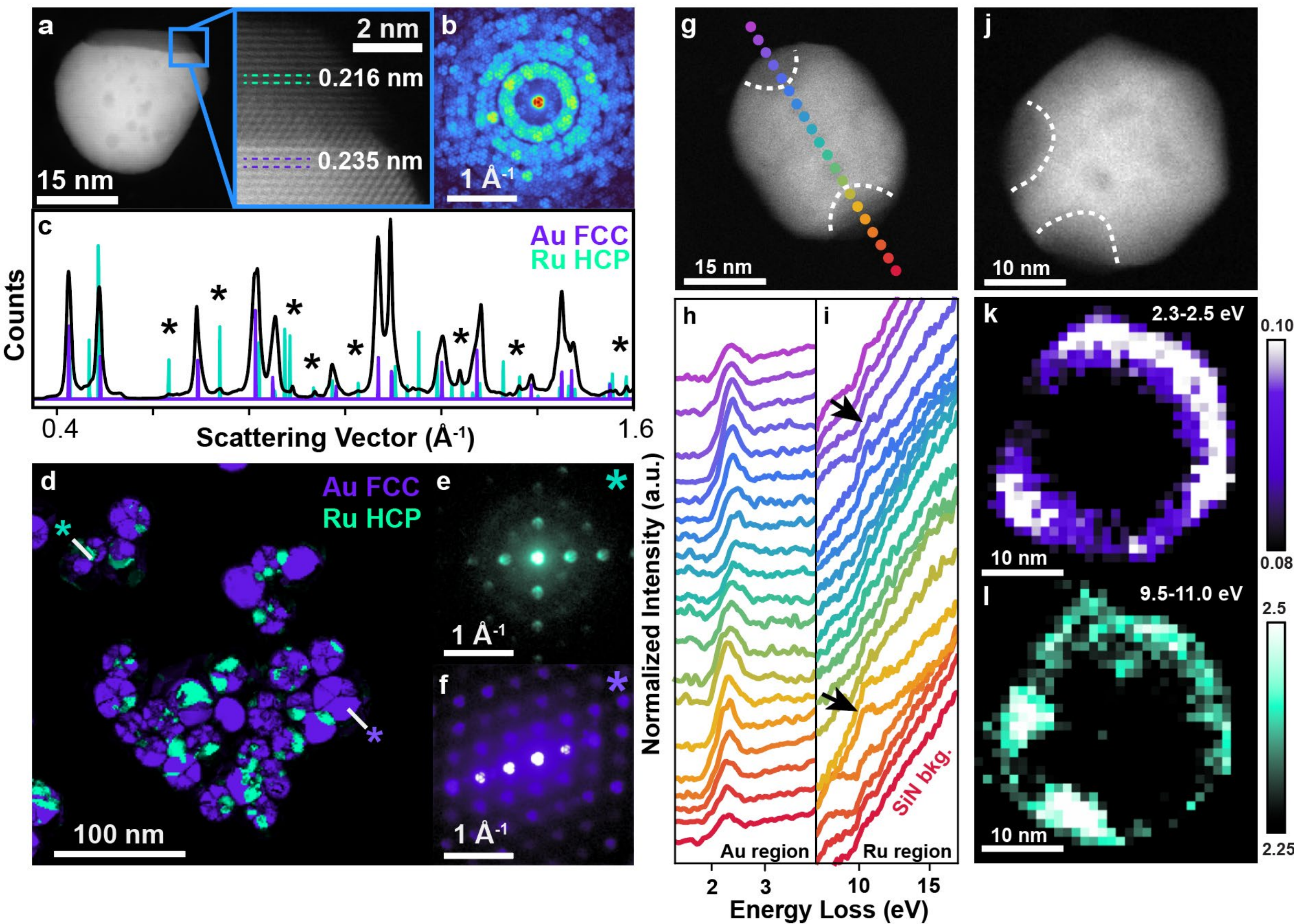

Figure 3. Crystallographic phase segregation and plasmonic reconfiguration in Au–Ru nanocrystal catalysts following vacuum annealing. (a) Atomic-resolution HAADF-STEM image of a representative phase-segregated Au–Ru nanocrystal after vacuum annealing. Inset highlights lattice spacings corresponding to Ru HCP {0002} (0.216 nm) and Au FCC {111} (0.235 nm), consistent with intraparticle phase segregation. (b) Accumulated maximum diffraction pattern from the 4D-STEM scan in (d) showing the patterned aperture and presence of mixed FCC and HCP phases. (c) Integrated powder diffraction histogram compiled from 4D-STEM data across all particles, resolving reflections from Au FCC (purple) and Ru HCP (green) phases (asterisks denote prominent Ru HCP peaks). (d) Phase map generated by ACOM for 20+ particles, showing spatial distribution of Ru HCP (green) and Au FCC (purple) domains. (e-f) Representative single-point diffraction patterns indexed to Ru HCP (e) and Au FCC (f) regions marked in (d). Phase-matching is shown in Figures S15–S16. (g) HAADF-STEM image with an overlaid monochromated EELS linescan acquired across a phase-segregated particle. Dashed white outlines indicate Ru-rich regions identified by Z-contrast. (h) Corresponding low-loss EELS spectra extracted along the linescan in (g), showing the Au surface plasmon near 2.5 eV, with enhanced intensity at particle edges, which attenuates slightly in Ru regions. (i) Similarly processed EELS spectra identify a Ru plasmon near 11 eV, which lies atop the SiN background yet remains predominantly localized to Ru-rich regions at the particle edges. (j) HAADF-STEM image of a second particle containing two asymmetric Ru-rich domains (dashed outlines), used for spatial plasmon mapping. (k, l) Integrated plasmon intensity maps for two energy windows (2.3–2.5 eV for Au and 9.5–11.0 eV for Ru), revealing the spatial distribution of the Au and Ru plasmonic response.

**Atmospheric-pressure $H_2$:$N_2$ unlocks a distinct restructuring regime**

To determine the onset and pressure dependence of gas-mediated restructuring of the AuRu nanocrystal catalysts, we employ *in situ* gas-cell HAADF-STEM under $H_2$:$N_2$ (3:1) across pressures from 50 to 780 Torr. Similar to our vacuum experiments, the gas-phase experiments used stepwise annealing with a 20 °C/min ramp rate and 10-min isothermal holds at each target temperature. For the gas experiments, the catalysts were imaged at 25, 150, 350, 450, 550, 650, and 750 °C under the selected gas environment and pressure. We operate the gas cell under continuous flow (~0.1 sccm) to maintain a steady-state chemical environment and better reflect reactor-relevant conditions. HAADF-STEM imaging relies on incoherent high-angle scattering, which provides high-contrast imaging through both gas-cell SiN membranes and the gas environment, enabling particle visualization across pressures ranging from 50 to 780 Torr (Figure 4a–c).

We find that sub-atmospheric pressure $H_2$:$N_2$ (3:1) preserves the thermally driven restructuring pathway observed during vacuum annealing. At 50 Torr (0.07 atm), particle facets begin to round at 350 °C, and Au–Ru phase segregation emerges at 450 °C (Figure 4a). Increasing the pressure to 350 Torr (0.46 atm) does not substantially change catalyst evolution (Figure 4b). Facet rounding and segregation proceed similarly to the low-pressure case, with only a slight increase in facet definition within Ru-segregated regions. The final catalyst morphology remains comparable to that observed after vacuum annealing. We quantify changes in nanocrystal size at each temperature using automated radial edge detection, with error bars reflecting uncertainty in particle-boundary identification from the radial intensity profiles (Figure 4d–f; see Figures S20–S21 and Methods for details). Particle-size changes are negligible at both 50 and 350 Torr, with measured changes of 2 ± 1% and 0.6 ± 3%, respectively (Figure 4d–e). These slight changes are attributed to morphological evolution altering the particle's projected 2D area.

In contrast to these sub-atmospheric pressures, the introduction of an atmospheric-pressure (780 Torr; ~1.0 atm) $H_2$:$N_2$ gas environment unlocks a distinct restructuring pathway (Figure 4c). Qualitatively, the particle facets remain sharply defined rather than rounding with increased temperature. Quantitatively, the particle size begins to increase at 550 °C and expands by 18 ± 1% by 750 °C relative to its room-temperature size (Figure 4f). This size increase coincides with the appearance of a central dark-contrast feature that emerges at 650 °C and continues to grow to 750 °C. Based on HAADF-STEM alone, this feature is consistent with an internal nanovoid and is confirmed by X-EDS elemental mapping, which shows an absence of both Au Lα/Mα and Ru Kα/Lα X-ray signal in the center of the nanocrystal (Figure 4g). Au–Ru phase segregation still occurs, as in the lower-pressure series, initiating between 350 and 450 °C and preceding nanovoid nucleation. During *in situ* imaging, 1 of the 14 particles we observe exhibits this behavior, but post-characterization of particles not previously exposed to the electron beam during heating indicates that nanovoid formation occurs across the sample and not driven by beam effects (Figure S22). Additional *in situ* pressure series are shown in Figures S23–S26.

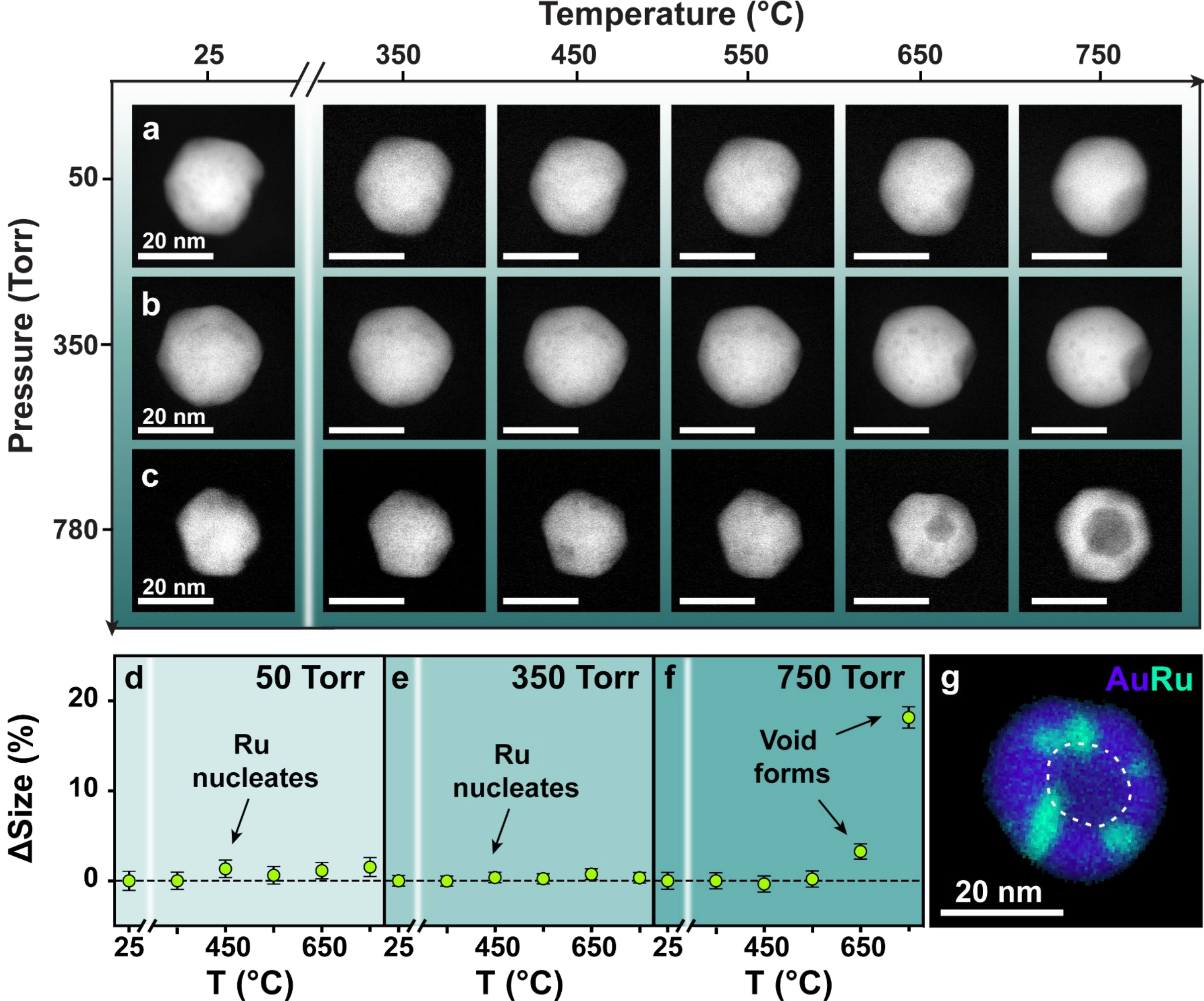

Figure 4. *In situ* gas-cell imaging of pressure-dependent restructuring in AuRu nanocrystals in a $H_2$:$N_2$ (3:1) gas environment. (a-c) HAADF-STEM images of three representative AuRu nanocrystals acquired during stepwise heating to 750 °C in 100 °C increments in a $H_2$:$N_2$ (3:1) gas mixture. (a) At 50 Torr (0.07 atm), Au–Ru phase segregation resembles that of vacuum annealing; see Figure 2b. (b) At 350 Torr (0.46 atm), progressive Au–Ru phase segregation persists with slightly more pronounced faceting on the Ru-rich region. (c) At 780 Torr (~1.0 atm), a central dark-contrast feature emerges around 650 °C and continues to grow as the nanocrystal is heated to 750 °C, accompanied by concurrent particle expansion. (d–f) Percent change in particle size as a function of temperature at 50 Torr (d), 350 Torr (e), and 780 Torr (f), relative to the initial particle size. Particle size was determined from the projected nanocrystal area using automated radial edge detection, and error bars reflect uncertainty in identifying the particle boundary from the radial intensity profiles. See Methods for details. (g) X-EDS elemental map of the particle in (c) showing Au (purple) and Ru (green) phase segregation and the absence

of Au Lα/Mα and Ru Kα/Lα peaks from the dashed central region, consistent with a central nanovoid. For all composite X-EDS maps, each X-ray line, Au Lα/Mα and Ru Kα/Lα, was independently scaled to its 99.9th percentile intensity.

**Hydrogen activates nanovoid formation via gas-mediated Kirkendall effect**

To isolate the chemical drivers behind nanovoid formation, we perform a series of controls by annealing AuRu nanocrystals independently in three separate pure-gas environments: Ar, $N_2$, and $H_2$, each at 1 atm (Figure 5a–c). Under the same temperature-ramping protocol, Ar annealing produces vacuum-like restructuring, with Au–Ru phase segregation beginning around 450–550 °C and no observable nanovoid formation (Figure 5a). $N_2$ annealing reveals comparable behavior to Ar, with phase segregation and no observable nanovoid formation (Figure 5b). By contrast, annealing in atmospheric-pressure $H_2$ induces nanovoid formation in AuRu nanocrystals, consistent with the restructuring pathway observed under atmospheric-pressure $H_2$:$N_2$ (3:1) (Figure 5c; Figures S27–S28). Post-reaction imaging of an entire SiN window that was not exposed to the electron beam during heating identified voids in 60 of 528 AuRu nanocrystals, corresponding to approximately 11% of the analyzed population (Figure S29). Across both pure $H_2$ and $H_2$:$N_2$ (3:1) environments, nanovoids consistently adopt near-faceted geometries that exhibit a recurring intersection with Ru-rich regions (Figures S22 and S27–S28). Together, the absence of nanovoids under 1 atm Ar and $N_2$ and their reproducible formation outside the electron-beam-exposed region after 1 atm $H_2$ heating indicate that void formation requires a hydrogen environment, rather than elevated pressure or continuous beam exposure alone (Figures S22 and S27–S29).

To resolve the three-dimensional (3D) relationship between the Ru-rich regions and nanovoids, we reconstructed the 1 atm $H_2$:$N_2$-annealed nanocrystal shown in Figure 4c from a HAADF-STEM tilt series (Figure S30) using a machine-learning (ML)-accelerated tomographic reconstruction workflow[45]. The ML-accelerated workflow reduces artifacts associated with the missing wedge due to the ±40° tilt limitation[45], while the 2° tilt increment provides dense angular sampling that helps limit sparse-tilt artifacts (Figure 5e–f)[46]. Three-dimensional isosurface rendering reveals the particle exterior (magenta) and interior (yellow), showing that the nanovoid extends through the particle and intersects two free surfaces in the axial view (Figure 5e–f). The reconstruction also reveals a branched internal channel extending along the y-axis (Figure 5f; side view). As vacuum and Ru-rich regions both exhibit reduced HAADF-STEM contrast, we assign this channel by co-registering the tomographic reconstruction with the HAADF-STEM tilt series and X-EDS elemental map (Figure 4g), which identifies the channel as Ru-rich (Figure S31). Orthogonal slices through the reconstructed volume further show the direct intersection between this Ru-rich channel and the nanovoid (Figure 5g).

Although the reconstructed nanovoid in Figure 5f extends through the nanocrystal, this geometry is not universal. Applying the same tomographic workflow to a second void-containing nanocrystal annealed under pure $H_2$ reveals an entirely internal, off-center cavity (Figures S32–S33). Across all voided particles, tomography, HAADF-STEM, and X-EDS analyses indicate the presence of both fully internal nanovoids and nanovoids that extend to free surfaces.

Our combined gas-specific annealing experiments and 3D reconstructions establish two key constraints for any mechanism of nanovoid formation. First, nanovoid formation is specific to high-pressure $H_2$-containing environments. Under standard preparation conditions, nanovoids form only in $H_2$-containing environments at partial pressures of 0.75-1 atm. They are not observed under vacuum or inert gases, despite comparable Au–Ru phase segregation across all conditions. Second, nanovoids repeatedly intersect Ru-rich domains and are often located near the particle center rather than at the external surface (Figures S28, S31, and S33–S34), consistent with void formation at internal Au/Ru interfaces generated during phase segregation, rather than through a mechanism dominated by surface etching or volatilization.

Population-level size, shape, and composition analyses further support this interpretation. Void-containing particles have an average diameter of 26 ± 5 nm (n = 60), comparable to an average diameter of 25 ± 5 nm (n = 528) for the full post-heating particle population, indicating that void formation is not strongly restricted to a distinct nanocrystal-size population (Figure S29d). Quantitative shape analysis additionally identifies a moderate population-level association between nanovoid formation and hexagonal faceting (Figures S35–S36). Further X-EDS quantification shows that void-containing nanocrystals retain Ru:Au ratios comparable to the pristine ensemble, indicating that void formation is not accompanied by measurable net Ru loss within the uncertainty and sampling of these measurements (Figure S37 and Table S4). Notably, we repeat identical temperature ramping with pure Au nanocrystals under 1 atm $H_2$ and observe no nanovoid formation (Figure S38), underscoring the necessity of Au–Ru interfaces for nanovoid nucleation.

These constraints narrow the possible mechanism and distinguish the observed voiding from several known routes to hollow or voided nanoparticles. These features differ from the classical nanoscale Kirkendall effect, where oxidation in metallic systems or galvanic replacement produces hollow particles[47–49]. As the nanovoids observed here form under strongly reducing conditions, an oxidation-driven pathway would most likely require trace oxidative impurities. Such a mechanism is unlikely because all experiments used ultrahigh-purity gases, including $H_2$ (99.9999% purity) and Ar and $N_2$ (both 99.999% purity), and the gas-delivery system was baked under turbomolecular pumping before each experiment to minimize volatile contaminants (see the Methods section). Moreover, 1 atm Ar and $N_2$ controls produced no nanovoids despite using the same gas-cell platform and gas-delivery system, arguing against gas-system-derived impurities as the cause of void formation. Oxidation and loss of Ru via volatile $RuO_x$ is also unlikely as $RuO_x$ is reduced to metallic Ru at $H_2$ pressures as low as 4.5 Torr at 250 °C, well below the conditions of our 50 Torr experiment, where nanovoids are not observed[50,51]. The observed nanovoid formation also differs from some bimetallic cases in which thermal annealing alone drives interface-trapped nanovoids through asymmetric vacancy formation energies[52–54]. Here, identical thermal conditions are insufficient. Nanovoids emerge *only* under high-pressure $H_2$.

One possible explanation is that initial Au–Ru phase nucleation establishes internal metal/metal interfaces that enable asymmetric interdiffusion, with continued Au/Ru interdiffusion driving vacancy accumulation. Transition state calculations on a large-scale Au/Ru interface model using a DFT-trained machine-learning interatomic potential (MLIP) quantify this asymmetry, yielding diffusion energy barriers of 0.41 eV for Au in Ru and 0.23 eV for Ru in Au. Under inert or vacuum conditions, we propose that this weaker asymmetry maintains vacancies at low equilibrium concentrations, where they are efficiently annihilated at free surfaces, suppressing nanovoid nucleation. In the presence of adsorbed hydrogen, the diffusion barriers change to 0.87 eV and 0.18 eV, respectively. Thus, hydrogen amplifies Au–Ru diffusion asymmetry, a hallmark of bimetallic Kirkendall processes[47,52,55,56], promoting vacancy clustering and nanovoid nucleation at internal interfaces. Atomic hydrogen is also expected to stabilize vacancies and increase their mobility, particularly in Ru-rich regions where hydrogen solubility is higher[57,58]. The consistent void–Ru co-localization supports such an interface-mediated mechanism.

To further elucidate the impact of H on the structural evolution, we perform grand canonical Monte Carlo (GCMC) simulations on a large-scale slab model of the FCC AuRu alloy (1:0.2), allowing for both metal rearrangement and H exchange with a reservoir at fixed chemical potential. Under $H_2$-free conditions, GCMC-simulated H-coverage quickly converges to zero on both the surface and sub-surface (Figure S39). Under 1 atm $H_2$ conditions, Ru aggregates at the surface are stabilized by high local H coverage, with some H permeating the Ru region's sub-surface. Notably, bulk Ru hydrides do not form under these pressures[59], indicating local hydrogen stabilization rather than hydride-phase formation. These simulations show that hydrogen preferentially populates Ru-rich surface and subsurface environments under 1 atm $H_2$, establishing the local chemical conditions required for hydrogen to modify metal diffusion and vacancy behavior near evolving Au/Ru interfaces.

Given that volatile Ru oxides are not stable under the reducing conditions used here[51], and that nanovoids are observed only in the presence of hydrogen, our results reveal a distinct variant of the Kirkendall effect in which gas-phase hydrogen controls vacancy accumulation and nanovoid nucleation. To our knowledge, this represents the first *in situ* demonstration that hydrogen pressure can regulate vacancy-mediated nanovoid formation in multimetallic nanocrystals. These findings extend the Kirkendall framework to regimes in which reactive gases influence vacancy kinetics rather than chemical conversion and highlight a new pathway for nanovoid generation in alloy nanocrystals under reaction-relevant conditions.

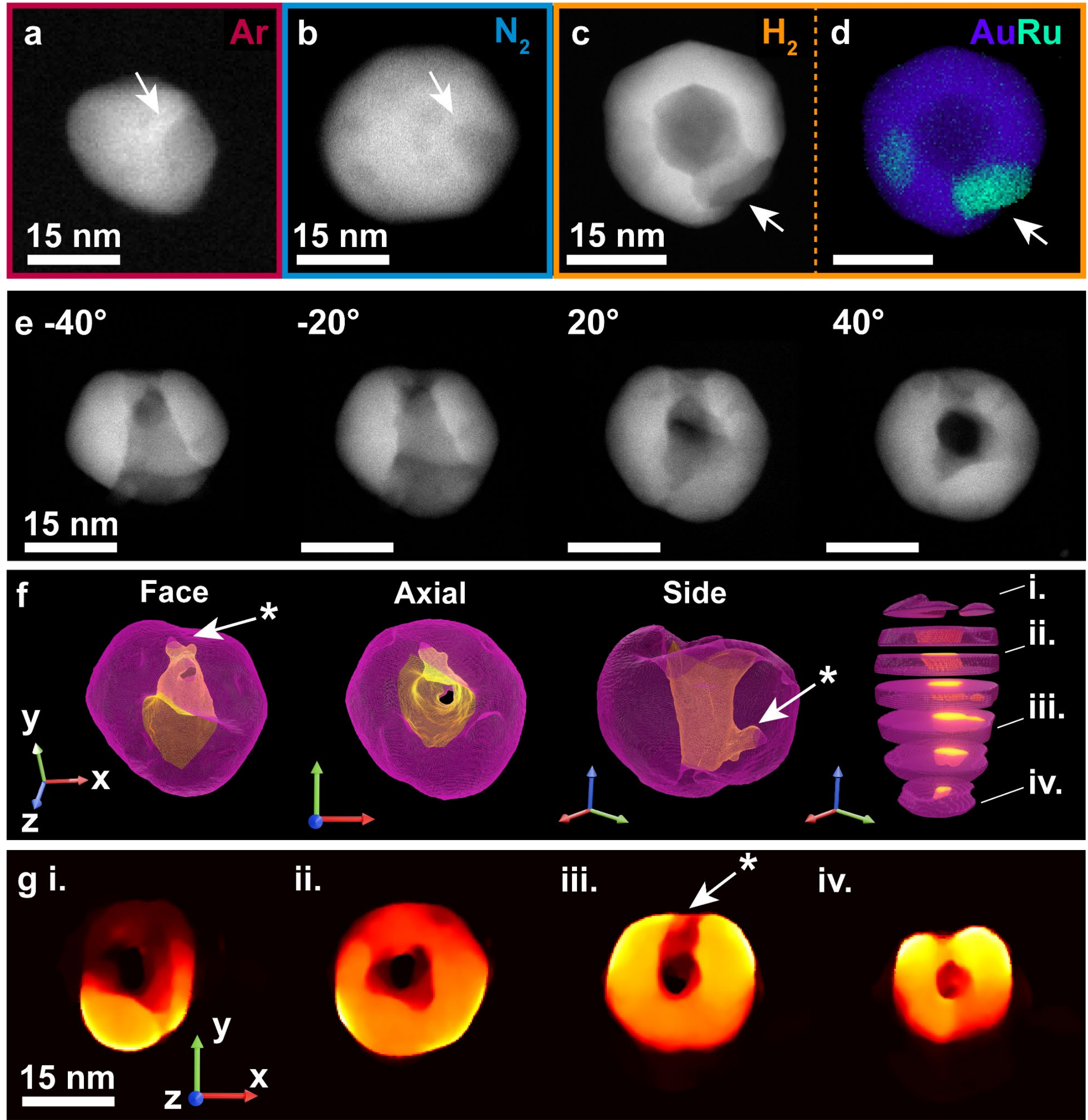

**Figure 5. Hydrogen drives nanovoid nucleation in Au–Ru nanocrystals.** (**a–c**) HAADF-STEM images of representative Au–Ru nanocrystals heated to 750 °C in Ar (**a**), $N_2$ (**b**), and $H_2$ (**c**) at 1 atm, and white arrows point to identified Ru-rich domains. Pronounced nanovoid formation emerges only under $H_2$. (**d**) X-EDS maps of the particle in (c) show Au–Ru segregation and a complete absence of Au (Lα) and Ru (Kα) counts in the central region, consistent with an internal nanovoid. (**e**) HAADF-STEM tilt series of the particle from Figure 4c after reaction, showing that the void extends through the whole particle. (**f**) Machine-learning-assisted tomographic reconstruction from multiple viewing directions reveals the catalyst's internal cavity (yellow) and exterior (magenta) 3D geometry. (**g)** Orthogonal tomographic slices reveal that the cavity tapers asymmetrically along the length of the particle. Intensity reflects reconstructed voxel values. (**f-g**) White arrows indicate a low-contrast side channel, attributed to Ru.

## Conclusion and Outlook

In summary, correlated *in situ* and multimodal electron microscopy reveals two distinct restructuring pathways in AuRu nanocrystals under reaction-relevant conditions. As synthesized, the polycrystalline AuRu nanocrystals exhibit Au/Ru intermixing with an FCC crystal structure and upon heating, undergo a thermally driven transformation into distinct Au-rich FCC and Ru-rich HCP domains. At 1 atm $H_2$:$N_2$ (3:1), a second restructuring regime emerges, characterized by persistent faceting and selective nanovoid formation that is not observed under vacuum or sub-atmospheric pressures. By varying the chemical environment from inert Ar/$N_2$ to $H_2$ gas, we identify $H_2$ as the key chemical driver of this restructuring pathway. Mechanistically, diffusion barriers calculated using DFT-trained machine-learning potentials suggest that hydrogen amplifies Au/Ru diffusion asymmetry, promoting vacancy accumulation at internal Au/Ru interfaces and supporting a hydrogen-activated variant of the nanoscale Kirkendall effect as the mechanism underlying nanovoid formation.

Our results establish that pressure and gas chemistry can select among distinct restructuring pathways in bimetallic nanocrystal catalysts. Thermally driven Au–Ru phase segregation persists from vacuum through sub-atmospheric $H_2$:$N_2$ (3:1) environments. However, atmospheric-pressure $H_2$:$N_2$ (3:1) does not simply accelerate this transformation. Instead, it selects a distinct pathway characterized by persistent faceting and nanovoid formation. Thus, pressure changes not only the extent of restructuring but also the pathway through which it proceeds. This demonstrates that reduced-pressure observations may accurately capture one restructuring regime while failing to reveal pathways that emerge only under atmospheric-pressure reactive environments. More broadly, the ability of hydrogen to modify Au/Ru diffusion asymmetry suggests that reactive gases can govern internal nanocrystal restructuring, rather than acting solely through surface adsorption or reaction. The multimodal electron microscopy framework developed here provides a systematic strategy for resolving how temperature, pressure, and gas chemistry select restructuring pathways in structurally complex catalysts. This framework is especially well suited to pressure-sensitive catalytic systems such as Fischer–Tropsch synthesis, $CO_2$ hydrogenation, and selective acetylene hydrogenation, where changes in reactant pressure and composition can alter adsorbate coverage, catalyst phase, interfacial structure, and subsurface hydrogen or carbon content, thereby selecting distinct structural states and restructuring pathways.

## Methods

### Nanocrystal Synthesis

Bimetallic AuRu alloy nanocrystals were synthesized using a polyol reduction method with native polyvinylpyrrolidone (PVP) ligands, as described by Yuan and colleagues[6]. The average particle size was found to be 25 ± 6 nm.

### Electron Microscopy Supports, Holders, Handling, and Sample Preparation

*TEM grid, MEMS chip, and holder configurations*

Characterization of the pristine AuRu nanocrystals shown in Figure 1b–d was performed using carbon-coated Formvar TEM grids, whereas monochromated EELS measurements shown in Figure 1e were performed using silicon nitride (SiN) TEM chips. The carbon-coated Formvar TEM grids consisted of a 15–20 nm-thick carbon film supported on 400-mesh Cu grids with a 3.05 mm outer diameter (Ted Pella, Inc.). The nonporous SiN chips had a 2.9 mm-wide hexagonal frame with a thickness of 0.1 mm and contained a 10 nm-thick SiN membrane with eight 100 × 100 µm windows and one 100 × 350 µm window (SimPore Inc.).

Vacuum annealing experiments were performed using a Protochips Aduro heating holder with both holey carbon and silicon nitride (SiN) support films. The holey carbon and SiN MEMS chips were both patterned with heating contacts and measured 4.0 mm × 4.65 mm with a thickness of 300 µm. Each chip contained nine 8 µm-diameter viewing holes. For the holey carbon chips, the viewing holes were covered by an approximately 18 nm-thick amorphous carbon support film, whereas for the SiN chips, the viewing holes were covered by 40 nm-thick amorphous SiN. Holey carbon chips were used for the direct imaging experiments shown in Figure 2, and SiN chips were used to prepare samples for post-heating monochromated EELS measurements shown in Figure 3.

Gas-phase experiments were conducted using a Protochips Atmosphere holder with $H_2$:$N_2$ (3:1), $H_2$, $N_2$, or Ar gas environments at pressures ranging from 50 to 780 Torr. The gas cell consisted of two MEMS chips, referred to as the top and bottom chips, separated by a 5 µm spacer. The top chip with heating contacts measured 6.0 mm × 4.5 mm with a thickness of 300 µm and contained a thinned, square 300 µm × 300 µm silicon carbide (SiC) film supporting six circular 8 µm-diameter viewing windows composed of 50 nm-thick amorphous SiN. The bottom chip measured 2 mm × 2 mm with a thickness of 300 µm and contained a 5 µm spacer for gas flow, with a 300 µm × 300 µm thinned SiN window region that was 50 nm thick. The gas cell holder and assembly order are shown in Figure S40.

*Grid and chip handling*

Self-closing polyvinylidene fluoride (PVDF)-tip tweezers (Style 5X, Oxford Instruments) were used to handle the MEMS chips during ligand removal and rinsing to reduce the risk of damaging or shattering the chips during sequential immersion in $NaBH_4$, water, and isopropyl alcohol solutions.

Self-closing stainless-steel tweezers (Dumont Style 4ACN, Ted Pella, Inc.) were used to handle the carbon-coated Formvar TEM grids during ligand removal and rinsing because their finer tips were better suited for handling the thinner carbon-based grids.

*Nanocrystal deposition and ligand removal*

S/TEM samples were prepared by drop-casting 5 µL of a dilute aqueous dispersion of AuRu nanocrystals onto a TEM grid/chip and air-drying for ≥12 hrs. On-grid/chip ligand removal was performed by dipping the particle-coated grids/chips in an aqueous 20 mM sodium borohydride ($NaBH_4$, Fisher) solution (10 s for carbon-coated grids/chips, 30 s for SiN grids/chips), Milli-Q deionized water (Millipore) cooled to 5–10 °C in an ice bath (10 s), and IPA (10 s). This method significantly reduced observed carbon contamination by removing the nanocrystals' native PVP ligands. In the case of the top and bottom chips for the gas cell, the sample was always drop-cast on the top chip, which was patterned with contacts to allow for *in situ* heating.

Although a minor interaction between $NaBH_4$ and the Cu support, such as reduction of the native copper oxide layer, cannot be excluded, X-EDS maps and spectra acquired from AuRu nanocrystals prepared on carbon-coated Cu grids and holey carbon MEMS chips show comparable Au and Ru signals and spatial distributions (Figures S4 and S9a–d). The only additional spectral features in the Cu-supported samples are the expected Cu Kα and Kβ peaks from the support grid, indicating that the choice of support does not measurably affect the conclusions regarding the Au/Ru distribution.

*Plasma cleaning of TEM grids and MEMS chips*

The Formvar TEM grids were subjected to $O_2$ plasma cleaning for 10 s prior to sample drop-casting. The SimPore SiN chips used for pristine-state characterization, which contained 10 nm-thick SiN windows, were plasma cleaned for 30 s prior to sample deposition. The Protochips Aduro MEMS heating chips with holey-carbon windows were also subjected to 30 s of $O_2$ plasma cleaning prior to sample drop-casting.

The SiN MEMS chips used for vacuum annealing and gas-cell experiments were subjected to 3 min of $O_2$ plasma cleaning prior to sample deposition. For gas-cell experiments, the particle-coated top chip was additionally plasma cleaned for 60 s following $NaBH_4$ ligand removal to further minimize contamination before gas-cell assembly.

## Electron microscopy

All STEM and TEM characterization was performed using an image- and probe-corrected Thermo Fisher Spectra 300 operated at 300 kV. Electron illumination and acquisition conditions for each dataset are summarized in Table 1.

## ***In situ* imaging**

*STEM*

Probe-corrected HAADF-STEM images were acquired using a spot-size setting of 9, a probe convergence semi-angle of 30 mrad, and an incident beam current of 100 pA. 2048 × 2048-pixel images were acquired with a per-pixel dwell time of 110 ns using a high-angle annular dark field (HAADF) detector with detection half-angles spanning from 63 to 200 mrad. Under atmospheric-pressure gas-cell conditions, lattice resolution was achieved in selected cases, as demonstrated by resolved AuRu {111} lattice fringes (d = 0.24 nm; Figure S41).

*HRTEM*

Aberration-corrected high-resolution transmission electron microscopy (HRTEM) imaging was performed with a maximum electron dose rate of 800 e- $Å^{-2}$ $s^{-1}$. 4096 × 4096-pixel images were acquired with an exposure time of 1 s using a Gatan GIF CMOS camera. Gatan-drift correction was employed during imaging to enable lattice resolution at elevated temperatures.

## ***In situ* experiments**

*Vacuum annealing*

Both S/TEM and HRTEM annealing experiments were carried out using a Protochips Aduro 500 Heating Holder with MEMS-based, carbon-coated grids. For the data shown in Figure 2b–c, samples were heated at 20 °C/min from room temperature to 350 °C and then from 350 to 750 °C in 100 °C increments. At each target temperature, heating was paused for a 5 min equilibration period followed by approximately 5 min of imaging, giving a total hold time of 10 min. Post-annealing, heating was stopped, and the sample was allowed to return to room temperature.

MEMS-based silicon nitride (SiN) grids were used for monochromated STEM-EELS to allow observation of the gold (Au) plasmon. To further remove volatile surface contaminants, samples first underwent 1 hr of beam showering (5.6 nA screen current, ~50 µm × 50 µm illuminated area). Annealing was performed from 350 °C to 750 °C in 100 °C increments using the same 20 °C/min ramp rate. Following each 100 °C step, the sample was allowed to equilibrate for 10 min before being ramped to the next temperature. Low-loss STEM-EELS spectra were collected at room temperature post-annealing.

*Gas-phase reaction*

Prior to each gas-cell experiment, the Protochips gas cart was baked at 50 °C for 15 h under turbomolecular pumping to pressures <1 Torr to minimize the presence of volatile contaminants. Gas experiments were performed using ultra-high-purity $H_2$ gas (6.0 purity, 99.9999% Praxair), ultra-high-purity $N_2$ gas (99.999% UHP, Linde #: NI 5.0UH-G), and ultra-high-purity Ar gas (99.999% UHP, Linde #: AR 5.0UH-K) to further minimize the presence of any contaminant species.

At the start of each experiment, the sample was drop-cast onto the top chip, ligand stripped using the $NaBH_4$ procedure, and $O_2$ plasma cleaned for 60 s. After gas cell assembly and holder insertion into the microscope, the sample underwent 10 evacuation–backfill cycles with inert Ar gas, achieving a minimum evacuation pressure of 1 Torr and a maximum backfill pressure of 760 Torr. This process serves to further remove volatile organic and oxidative species from the gas cell. At each gas pressure, samples were heated using the same 20 °C/min ramp rate and 10 min hold protocol, consisting of 5 min of equilibration followed by approximately 5 min of imaging. Images were acquired at 25, 150, 350, 450, 550, 650, and 750 °C. This imaging procedure was identical to the vacuum-annealing procedure, except that an additional imaging point at 150 °C was included to determine whether the gas environment caused structural changes at lower temperatures. No such changes were observed. Post-annealing, heating was stopped, and the sample was allowed to return to room temperature in the same gas environment for further characterization.

### Characterization Before and After *In Situ* Annealing and Gas-cell reactions

#### *Atomic-resolution STEM imaging*

For pre- and post-annealing characterization, including measurements performed after gas-cell experiments, the imaging conditions were optimized for atomic-resolution characterization through a single membrane. Atomic-resolution HAADF-STEM images were acquired using a spot-size setting of 9, a probe convergence semi-angle of 24 mrad, and an incident beam current of 50 pA. 4096 × 4096-pixel images were acquired with per-pixel dwell times of 100–500 ns using a high-angle annular dark field (HAADF) detector with detection half-angles spanning from 63 to 200 mrad.

#### *X-EDS spectral image acquisition*

X-EDS spectral images were acquired using a spot-size setting of 6, a probe convergence semi-angle of 30 mrad, and an incident beam current of 130–150 pA. 256 × 256-pixel spectral images were acquired with a per-pixel dwell time of 50 μs using a SuperX energy dispersive spectroscopy detector with a 0.7 sr X-EDS solid angle. 40–60 scans were performed for each map to obtain enough counts for X-EDS quantification. Chemical distributions were obtained using both the HyperSpy and eXSpy open-source Python packages.

For AuRu nanocrystal pristine-state characterization, X-EDS data sets were collected using a Thermo Fisher single-tilt holder. For pre- and post-annealing elemental mapping of vacuum-annealed AuRu nanocrystals, the Protochips Aduro holder was used. For post-reaction characterization, the gas cell was disassembled, and the sample-containing top chip was mounted in the Protochips Inspection holder, resulting in brief exposure to ambient air. An air-exposure control was performed using a vacuum-annealed AuRu nanocrystal. X-EDS maps were acquired immediately after heating, before air exposure, and again after the sample was removed from the microscope, exposed to ambient air on the benchtop for 5 min, and reinserted. The corresponding elemental maps are shown in Figure S42.

*4D-STEM nanobeam electron diffraction*

Four-dimensional scanning transmission electron microscopy (4D-STEM) was performed on the AuRu nanocrystals using a spot-size setting of 6, a probe convergence semi-angle of 1.25 mrad, and an incident beam current of 50 pA. Diffraction patterns were acquired over a 256 × 256 grid of real-space probe positions with a dwell time of 500 μs per probe position using a DECTRIS ARINA (Si) direct-electron detector. For pre- and post-annealing AuRu characterization, we employed a 10 μm patterned bullseye aperture[36]. The 185 mm camera length allowed for full resolution of the bullseye pattern on the DECTRIS 192 × 192 pixel detector, yielding a calibrated k-space pixel size of 0.0151 $Å^{-1}$. In the post-annealing case, the patterned aperture enabled the precise measurement of the Bragg diffraction spots used to distinguish Au (FCC) and Ru (HCP) phases.

To further observe their internal twin-boundary structure with higher spatial resolution, the pristine state AuRu nanocrystals were examined using a 20 μm aperture and camera length of 73 mm. This enabled a nominal convergence semi-angle of 2.5 mrad and k-space pixel size of 0.0384 $Å^{-1}$. The 256 × 256 image in Figure S6 was acquired with a per-pixel dwell time of 500 μs, a spot size of 6, and an incident beam current of 70 pA.

*Monochromated STEM EELS spectra and map acquisition*

Electron energy loss spectroscopy (EELS) was performed using a monochromated EELS system equipped with a dual EELS Gatan Imaging Filter (GIF) Continuum spectrometer. Mono-EELS measurements were acquired using a spot-size setting of 14, a probe convergence semi-angle of 21.4 mrad, and an incident beam current of 130–135 pA. 32 × 32-pixel spectral maps were acquired using exposure times of 2.0 ms for the zero-loss peak and 0.8 s for the low-loss spectrum. Additionally, 40 × 1-pixel line scans were acquired using exposure times of 2.0 ms for the zero-loss peak and 0.5 s for the low-loss spectrum. All spectra were acquired with an energy dispersion of 10 meV per channel, an EELS entrance aperture of 1 mm, and a collection semi-angle of 28.0 mrad.

*Tomography*

To perform a tomographic tilt series of the AuRu nanocrystals post-gas-phase reaction, we first beam-showered the sample in TEM mode (30 min, 5.56 nA screen current, ~50 µm × 50 µm illuminated area). HAADF-STEM tomography images were acquired using a spot-size setting of 9, a probe convergence semi-angle of 30 mrad, and an incident beam current of 50 pA. 4096 × 4096-pixel images were acquired with a per-pixel dwell time of 500 ns using a high-angle annular dark field (HAADF) detector with detection half-angles spanning from 63 to 200 mrad. The tilt series was conducted on *in situ* gas chips mounted in the Protochips Inspection holder, which constrained the tilt range to ±40°. The limited ±40° tilt range produces a substantial missing wedge, which can cause anisotropic artifacts, including elongation along the missing-wedge direction (the blue *z*-direction in Figure 5f). The implicit-neural-representation reconstruction of Lim et al. used in this work reduces streaking and elongation but cannot fully recover the missing information[45]. At each tilt angle, orthogonal 0° and 90° scan pairs were acquired to correct nonlinear drift and scan-direction-dependent distortions[60]. A 2° tilt increment provided dense sampling across the accessible tilt range, limiting sparse-tilt artifacts such as multidirectional streaking and blurred feature boundaries[46].

Table 1. Electron illumination conditions applied to each dataset.

| Imaging Mode | Incident beam current or dose rate | Dwell or exposure time | Image dimensions |
|---|---|---|---|
| *In situ* STEM imaging | 100 pA | 110 ns pixel$^{-1}$ | 2048 × 2048 |
| *In situ* HRTEM imaging | $\leq$800 e$^{-}$ Å$^{-2}$ s$^{-1}$ | 1 s | 4096 × 4096 |
| Atomic resolution STEM | 50 pA | 100–500 ns pixel$^{-1}$ | 4096 × 4096 |
| X-EDS | 130–150 pA | 50 µs pixel$^{-1}$ | 256 × 256 |
| 4D-STEM | 50–70 pA | 500 µs probe position$^{-1}$ | 256 × 256 |
| Mono-EELS | 130–135 pA | 2.0 ms for ZLP; 0.8 s for low loss (map) | 32 × 32 |
| | | 2.0 ms for ZLP; 0.5 s for low loss (line scan) | 40 × 1 |
| Tomography | 50 pA | 500 ns pixel$^{-1}$ | 4096 × 4096 |

*Computational details*

The training dataset for the machine learning interatomic potential (MLIP) was generated using the GOCIA package[61], which performed Grand Canonical Global Optimization (GCGO) sampling for the (111) facets of both the pure metals (Au and Ru) and alloys (Au:Ru=1:0.2) with randomly generated numbers of adsorbates or defects. The sampled configurations were then locally optimized using density functional theory calculations with the VASP program (version 6.4.1)[62,63]. The PBEsol functional[64] was adopted with PBE_PAW pseudopotentials[63] and D3 correction[65] to balance its performance on both solid-state and surface adsorption energetics. The convergence criteria for electronic and force minimization were $10^{-5}$ eV and $2.5\times10^{-2}$ eV/Å, respectively. The Brillouin zone was sampled using Γ-centered 3×3×1 *k*-points, and the kinetic energy cutoff for the planewaves was 400 eV. The local optimization trajectories from GCGO sampling were filtered based on force magnitude and structural similarity. The curated dataset was then used to fine-tune the MACE-MP-0-medium foundation model[66] for 100 epochs. Dataset distribution and model performance are shown in Figures S43–S44.

Using the trained MLIP, transition state searches for large-scale interfacial vacancy diffusion were performed via the nudged elastic band method[67]. Grand Canonical Monte Carlo (GCMC)[68] simulations, as implemented in GOCIA, were also performed using the MLIP on a 5×5×10 supercell of the (111) facet of the alloy (Au:Ru=1:0.2). Each simulation started from a randomized alloy configuration with 1 monolayer (ML) of hydrogen atoms on the top surface. The system was relaxed to a local minimum after every GCMC move (hydrogen atom insertion/deletion, site-swapping, rattling). The chemical potential of hydrogen was calculated using $\mu_H = 1/2\ [G(H_2, gas) + k_B T\ \ ln\ (p/p^\circ)]$.

**Data analysis**

*EELS*

Dual-EELS spectra containing simultaneously acquired low-loss and high-loss signals were processed using the HyperSpy library. The zero-loss peak (ZLP) position was determined on a per-pixel basis, and energy drift across the scan was corrected by aligning all spectra to a common ZLP reference. Following alignment, spectra were normalized by the integrated ZLP intensity to account for sample thickness and probe-current variations. The ZLP tail was removed from the low-loss spectra by fitting a power-law background over the 1.2–1.7 eV energy range and subtracting the fitted background.

Energy-filtered maps were generated by integrating ZLP-normalized spectra over selected energy-loss windows using channel-averaged intensities. Au-associated plasmon maps were constructed by integrating over 2.3–2.5 eV, corresponding to the Au surface plasmon resonance, while Ru-associated maps were generated over 9.5–11 eV, corresponding to the Ru bulk plasmon feature. For Figure S19, spatial masks defining Au-rich and Ru-rich regions were derived from the corresponding energy-filtered maps by thresholding regions of enhanced spectral intensity relative to the local background, with vacuum and low-signal regions excluded based on ZLP-normalized low-loss intensity. These masks were applied consistently across energy windows to enable spatially correlated analysis of Au and Ru spectral features.

For linescan measurements, spectra were spatially binned by a factor of two along the scan direction (1.24 nm per pixel prior to binning) and binned in energy by a factor of three (0.01 eV per channel prior to binning). At each position along the linescan, spectra were averaged over the corresponding spatial bin prior to analysis. For visualization, intensity profiles were smoothed using a Savitzky–Golay filter (window length = 13 points, second-order polynomial). Linescan plots were generated by plotting the averaged, background-subtracted spectral intensity as a function of distance along the scan, with distances calculated from the calibrated probe step size.

*4D-STEM*

Pre-annealing 4D-STEM datasets were processed using the py4DSTEM library. [35] Detector calibration was performed using a polycrystalline Au film on carbon as a reference standard to determine detector pixel geometry and reciprocal-space calibration (pixel size = 0.0384 $Å^{-1}$). The raw 4D-STEM datacube (256 × 256 probe positions, 192 × 192 detector pixels) was cleaned of hot pixels and spatially binned to 128 × 128 probe positions. Mean and maximum diffraction patterns were used to define virtual detector geometries, and circular bright-field and annular dark-field images were generated to guide selection of a vacuum region for probe reconstruction. The probe semi-angle and diffraction-center coordinates extracted from the vacuum probe were used to generate a matched kernel for Bragg disk detection. Bragg disks were identified across all probe positions using template matching with subpixel refinement.[36] Automated crystal orientation mapping (ACOM)[36] was performed by matching experimentally detected Bragg disk configurations to simulated diffraction templates to generate spatially resolved orientation maps.

Post-annealing 4D-STEM datasets were analyzed to distinguish Au and Ru domains. Experimentally measured Bragg vector magnitudes were compared to theoretically expected reciprocal-lattice spacings for FCC Au and HCP Ru, calculated from known lattice parameters, enabling assignment of detected Bragg disks to Au- or Ru-like scattering families. Phase-resolved ACOM was performed by matching local Bragg disk configurations to simulated diffraction templates for each phase, generating spatially resolved maps encoding crystallographic orientation and phase identity. Quantitative phase fractions were obtained by tallying probe positions assigned to each phase. Phase-selective reciprocal-space masks and annular integrations were applied to Bragg vector maps to independently isolate Au- and Ru-associated scattering ranges.

*X-EDS*

X-EDS spectrum images were processed using Python-based workflows with the HyperSpy library to generate background-subtracted, spatially resolved elemental maps of AuRu nanoparticles. Following spatial and spectral rebinning, element-specific spectral models were constructed by defining the relevant Au and Ru X-ray emission lines (Au Lα, Au Mα, Ru Lα, Ru Kα), with Mg Kα and O Kα lines included for particles supported on MgO. For each pixel, integrated peak intensities were obtained after local background subtraction, performed by estimating the background from pre- and post-edge energy windows adjacent to each characteristic line and subtracting this contribution from the peak region. Background-subtracted elemental maps were generated for each line. To isolate the nanoparticle from the vacuum and ensure consistent spatial comparison, a binary particle mask was constructed by thresholding the combined background-subtracted Au and Ru signals and applied uniformly across all elemental channels. Masked elemental maps were combined to visualize and quantify the spatial segregation of Au-rich and Ru-rich regions.

*STEM and TEM image alignment and image cropping*

Image centering and cropping in all HAADF-STEM and HRTEM images were performed by phase correlation in Fourier space using Python-based workflows in JupyterLab with NumPy, Matplotlib, and HyperSpy. Images were rescaled and rebinned to a common pixel size, padded to minimize edge effects, and aligned using subpixel shifts extracted from the cross-correlation peak. Aligned images were cropped to a common region centered on the nanoparticle prior to analysis.

*Nanocrystal segmentation and size determination using radial intensity line profiles*

Projected nanocrystal areas and associated errors from Figure 4d-f were measured using Python workflows implemented in JupyterLab with the NumPy, Pandas, Matplotlib, scikit-image, and SciPy packages. For each HAADF-STEM image, a particle mask was generated using Otsu thresholding [69]. Using the centroid of this generated mask as the origin, radial HAADF intensity profiles, $I_i(r)$, were taken from the particle center past the border defined by the mask (Figure S20a, marked in orange). The particle edge was then determined from the radial intensity gradient, with the boundary assigned to the steepest point in the radial line profile (Figure S20b)[70]. The resulting particle boundaries for all particles from Figure 4a–c can be seen in Figure S21. The uncertainty in radial edge position was estimated from the width of the corresponding gradient peak. The full width at half maximum (FWHM) of the gradient peak was measured and converted to an associated error using $\sigma_r = FWHM/2.355$. The uncertainty in edge position was then propagated to the area measurement using $A \approx 1/2 \, \Sigma \, r_i^2 \, \Delta\theta$. Thus, the error bars in Figure 4d–f represent the propagated uncertainty in localizing the nanocrystal boundary from the radial gradient peaks.

*Tomography*

The tomographic reconstructions in this work used the open-source Python package *quantEM.* [45] Using implicit neural representations with a neural network as a model regularizer, we jointly optimize the volume and the relative orientations of each tilt image, allowing for the most accurate reconstruction possible. This framework also compensates for the large missing wedge, minimizing streaking along the missing wedge direction. We also perform background subtraction using Bernstein polynomial fitting and a coarse cross-correlation alignment prior to reconstruction. The reconstruction ran for 100 iterations, where the volume and alignment parameters were fully converged.

*Software and computational tools*

The manuscript was proofread for clarity and grammar using a large language model (ChatGPT, OpenAI). The authors reviewed and edited the output where appropriate; all scientific content, interpretations, and conclusions are the authors' own.

## Supporting Information

Additional experimental methods and gas-cell operating procedures; structural and compositional characterization of pristine and annealed AuRu nanocrystals using STEM, HRTEM, X-EDS, 4D-STEM, EELS, and electron tomography; in situ vacuum- and gas-environment annealing experiments across vacuum, sub-atmospheric, and atmospheric-pressure conditions; controls evaluating electron-beam exposure, support effects, gas composition, hydrogen pressure, pure Au nanocrystals, and post-reaction air exposure; and quantitative analyses of nanocrystal size, morphology, boundary evolution, nanovoid formation, and post-annealing composition (PDF).

## Acknowledgements

The authors acknowledge Pinaki Mukherjee for TEM assistance and all members of the Dionne group for their scientific feedback and support. **Funding:** J.A.D., A.S.M.-G., and P.M. acknowledge salary support from the Office of Basic Energy Sciences, US Department of Energy, Division of Materials Science and Engineering, DE-AC02-76SF00515. We also acknowledge the financial support from the U.S. Department of Energy Office of Science National Quantum Information Science Research Centers as part of the Q-NEXT center, which provided support for the multimodal TEM instrumentation. A.S.M.-G. acknowledges the support of the National Science Foundation Graduate

Research fellowship (NSF-GRFP) under grant no. DGE-2146755 and the John Stauffer Graduate Fellowship. This research used resources of the National Energy Research Scientific Computing Center, a DOE Office of Science User Facility supported by the Office of Science of the U.S. Department of Energy under Contract No. DE-AC02-05CH11231 using AI4Sci@NERSC NERSC award NERSC DDR-ERCAP0038157 and BES-ERCAP0024127. Z.Z. is supported by the Stanford Energy Fellowship from the Precourt Institute for Energy, Doerr School of Sustainability. F.A-P. acknowledges support from the U.S. Department of Energy, Office of Science, Office of Basic Energy Sciences, Chemical Sciences, Geosciences, and Biosciences Division, Catalysis Science Program to the SUNCAT Center for Interface Science and Catalysis. Part of this work was performed at nano@stanford RRID:SCR_026695.

## TOC (graphic)

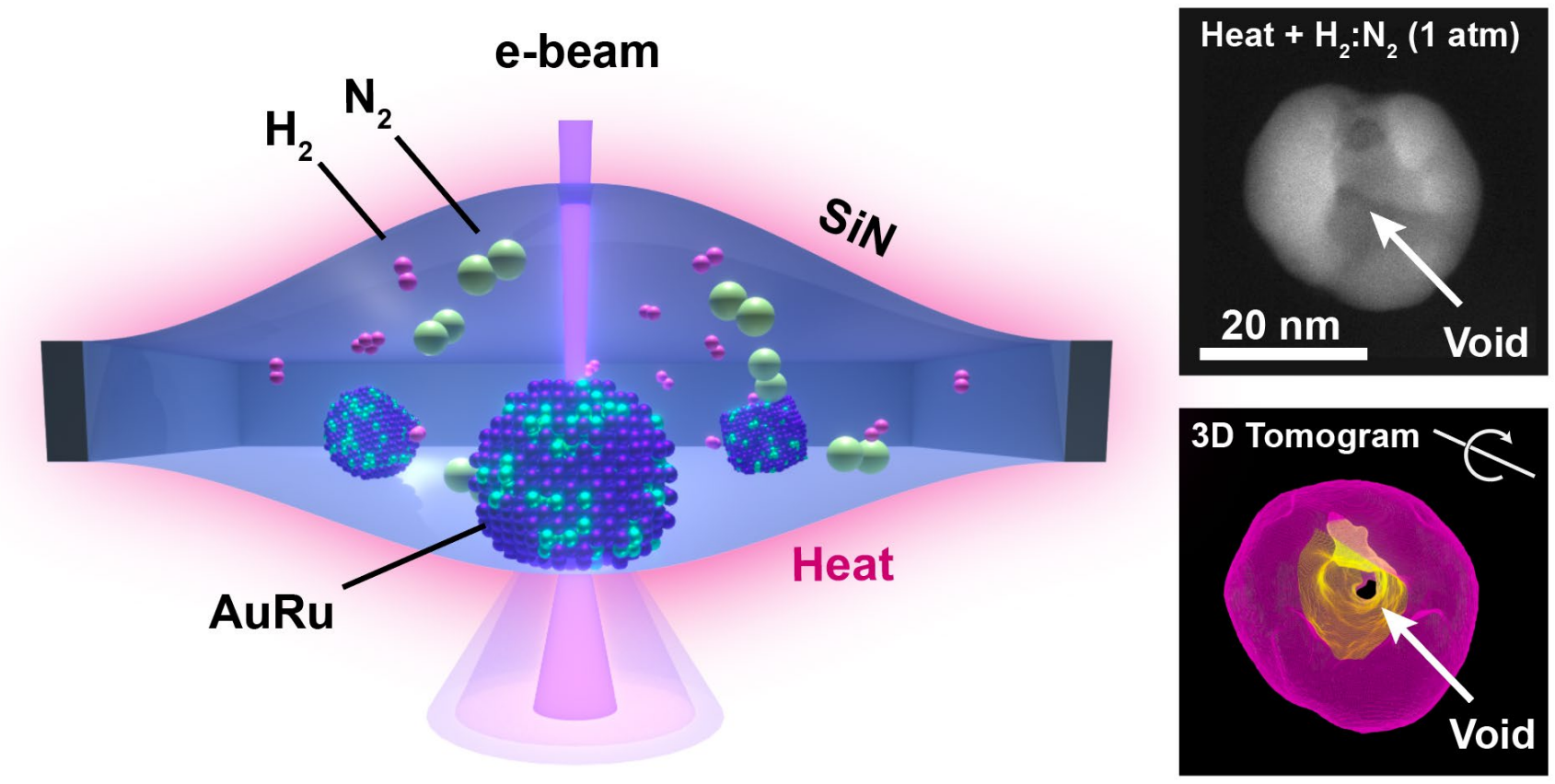

# Supporting Information

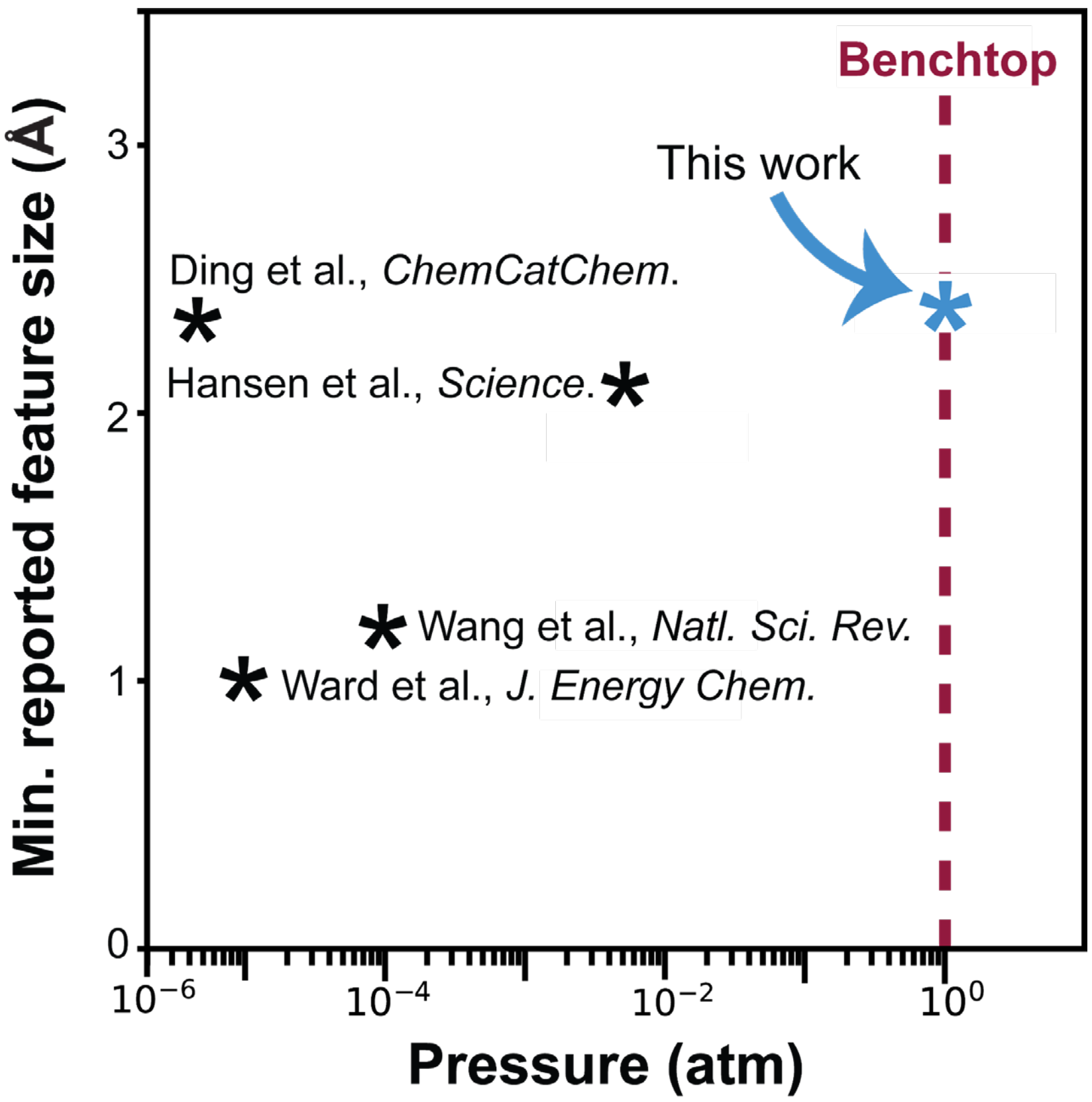

Figure S1. Comparison of reported spatial resolution and gas pressure in *in situ* TEM studies of ammonia synthesis catalysts. Minimum reported feature size is plotted as a function of the $H_2$:$N_2$ gas pressure used in prior *in situ* TEM studies, highlighting the lack of near-atmospheric-pressure measurements in the literature [1–4]. The blue marker indicates the conditions used in this work, and the dashed line denotes 1 atm, representative of benchtop reactor pressure[5].

Figure S1 places the present gas-cell STEM measurements within the pressure and spatial-resolution regimes reported for prior *in situ* TEM studies of ammonia-synthesis catalysts. Conventional differentially pumped ETEM has enabled direct observation of catalyst restructuring in reactive gases with high spatial resolution, but these measurements have generally been limited to pressures of approximately $10^{-5}$–$10^{-3}$ atm[1–4]. This range remains substantially below the atmospheric pressures used in emerging benchtop $NH_3$-synthesis systems and the still higher pressures employed industrially. Windowed gas-cell platforms extend *in situ* electron microscopy to atmospheric pressure, although electron scattering by the gas and SiN membranes reduces spatial resolution relative to conventional ETEM. The present study occupies this complementary high-pressure regime, providing nanoscale imaging of individual AuRu nanocrystals at pressures up to approximately 1 atm. Rather than assuming that restructuring observed under reduced pressure persists unchanged at higher pressure, this expanded experimental range enables direct comparison of catalyst evolution across vacuum, sub-atmospheric, and atmospheric-pressure environments, where differences in adsorbate coverage, adsorption kinetics, and surface diffusion can select distinct structural pathways.

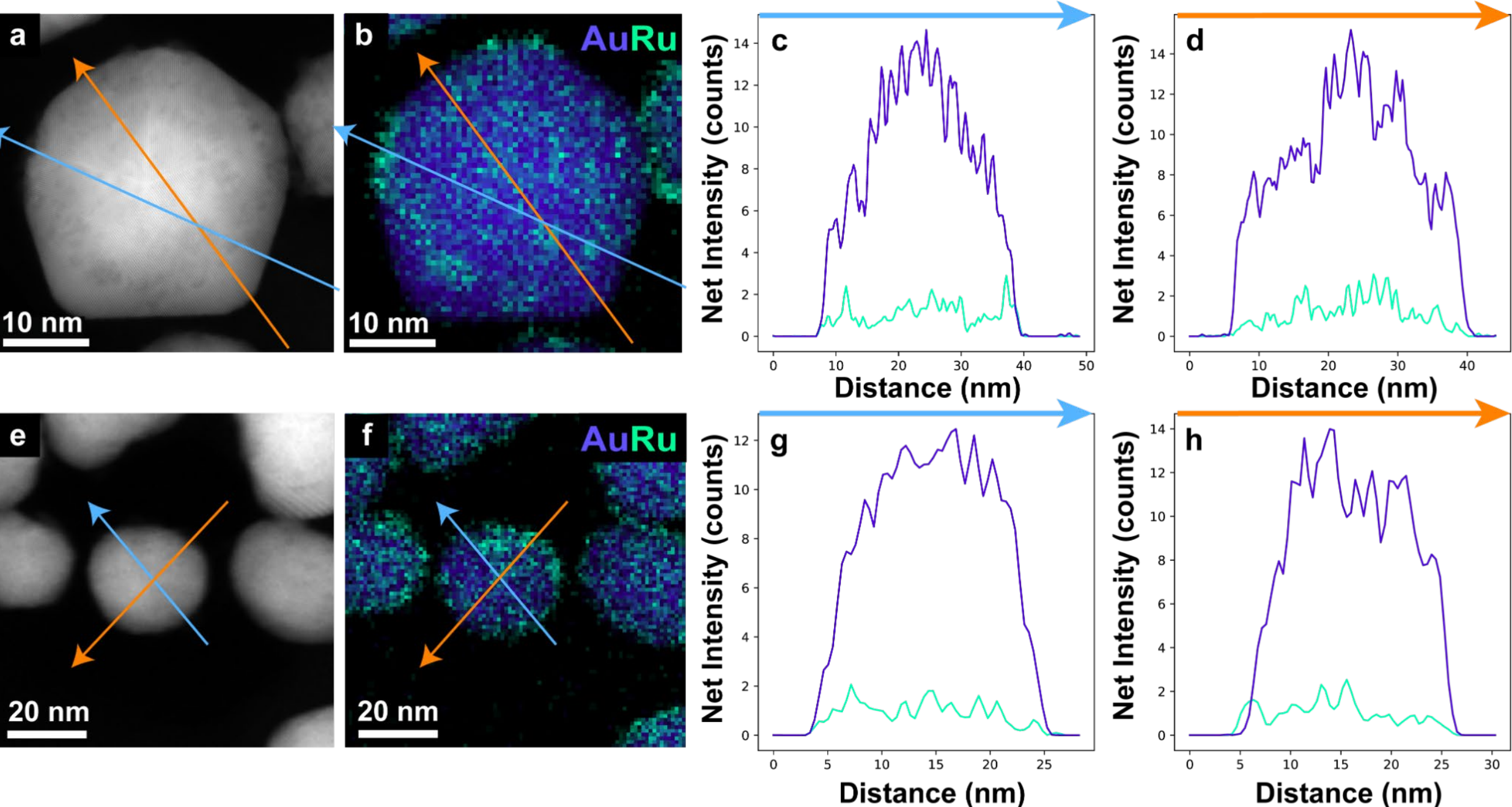

Figure S2. X-EDS analysis of Au–Ru intermixing pre-heating. (a) HAADF-STEM image of an AuRu nanocrystal from Figure 1, with line-scan paths indicated (blue and orange). (b) Corresponding X-EDS elemental map (reproduced from the main text) with line-scan locations marked. (c, d) X-EDS line profiles along the blue and orange paths, showing net intensities of the Au Lα and Ru Kα lines, respectively. (e–h) Additional HAADF-STEM images, X-EDS maps, and corresponding line scans from different particles, confirming Au–Ru intermixing across the sample. For the composite X-EDS map, each X-ray line (Au (Lα/Mα) and Ru (Kα/Lα)) in the composite X-EDS map was independently scaled to its 99.9th percentile intensity.

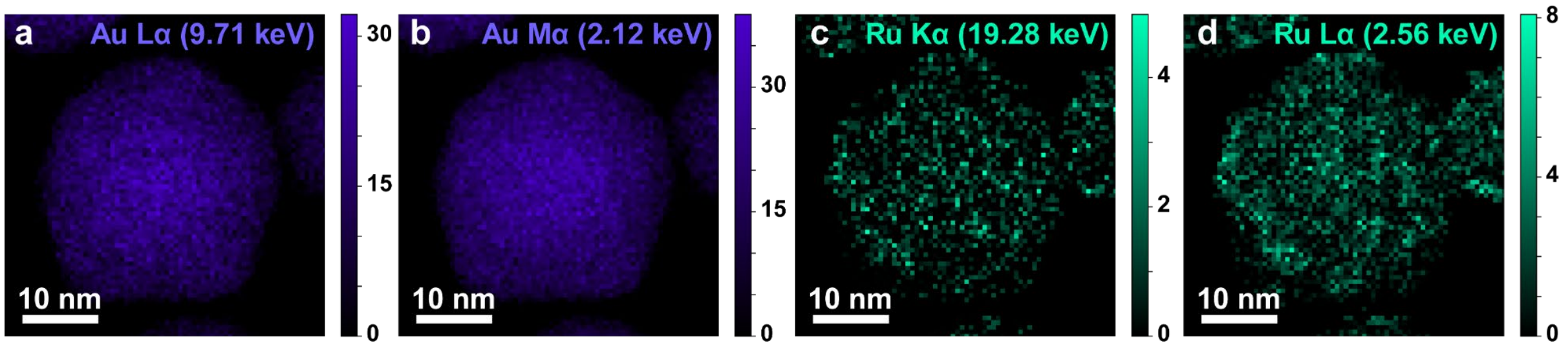

Figure S3. Individual Au and Ru X-EDS maps of the AuRu nanocrystal from Figure 1d. X-EDS maps showing the net intensities of the (a) Au Lα, (b) Au Mα, (c) Ru Kα, and (d) Ru Lα emission lines used to construct the composite Au(Lα/Mα)/Ru(Kα/Lα) elemental map in Figure 1d. Each panel is displayed with its own net-intensity color scale, with the upper limit set to the 99.9th percentile intensity for that emission line. Although localized regions of higher Ru intensity are observed, the Au and Ru signals remain spatially intermixed across the nanocrystal.

**Note on X-EDS map normalization.** X-EDS map intensities are affected by element- and line-specific factors, including ionization cross sections, fluorescence yields, detector response, and absorption/matrix effects. As a result, displaying Au and Ru maps on a common raw-count scale can either obscure the lower-intensity Ru signal or saturate the higher-intensity Au signal. For visualization of composite Au/Ru maps, each emission-line map was independently scaled, with the upper intensity limit set to the 99.9th percentile intensity for that x-ray line using built-in HyperSpy analysis workflows. This normalization was applied only for display and was not used for quantitative composition analysis. Quantification was performed separately from X-EDS spectra in Thermo Fisher Velox software using the Brown–Powell ionization cross-section model.

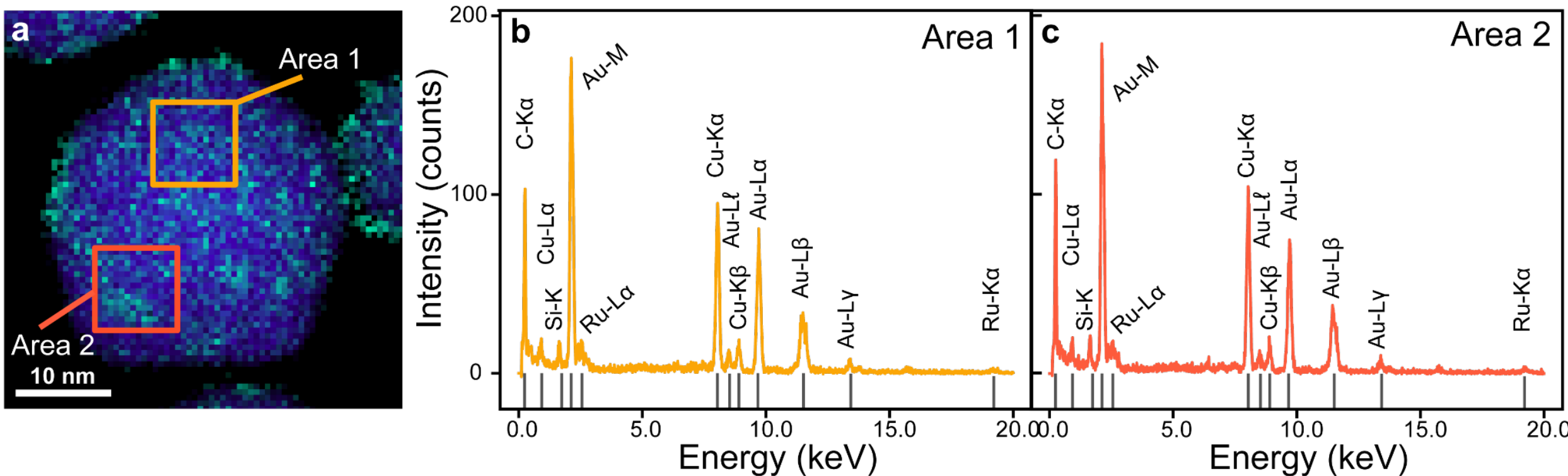

Figure S4. Spatial distribution of Au and Ru in pristine-state AuRu nanocrystals. (a) Composite Au(Lα/Mα)/Ru(Kα/Lα) X-EDS elemental map of the AuRu nanocrystal from Figure 1d, with regions selected for local spectral analysis indicated by the boxes labeled Area 1 and Area 2. (b,c) X-EDS spectra from Area 1 and Area 2, respectively, showing the presence of Au (Lα/Mα) and Ru (Kα/Lα) peaks of comparable intensity across both areas. Quantification using the Au Lα and Ru Lα lines is provided in Table S1. Strong Cu Kα and Kβ signals arise from the carbon-coated copper grid. For the composite X-EDS map, each X-ray line (Au (Lα/Mα) and Ru (Kα/Lα)) was independently scaled to its 99.9th percentile intensity.

Table S1. X-EDS quantification of AuRu nanocrystal from Areas 1 & 2 in Figure S4 as well as the entire nanocrystal from Figure 1d. Atomic percentages determined from X-EDS spectra using the Brown–Powell ionization cross-section model. Au and Ru contents were quantified using the Au Lα and Ru Lα lines, respectively.

| **Location** | **Element** | **Pre-annealing atomic %** |
|---|---|---|
| Area 1 | Au | 89 ± 1 |
| | Ru | 11 ± 1 |
| Area 2 | Au | 87 ± 1 |
| | Ru | 13 ± 1 |
| Total Particle | Au | 87 ± 1 |
| | Ru | 13 ± 1 |

**Assessment of Au–Ru intermixing in pristine nanocrystals.** Au and Ru are immiscible in bulk and are not expected to form a thermodynamic solid solution[6]. However, the X-EDS elemental maps, line profiles, local spectra, and quantitative analysis indicate that the pristine AuRu nanocrystals exhibit nanoscale Au–Ru intermixing rather than complete phase separation. In the X-EDS line profiles in Figure S2c–d and S2g–h, a baseline Ru signal is present across the width of each nanocrystal. However, the Ru signal varies about this baseline, indicating local regions of higher and lower Ru content. This behavior is particularly evident in Figure S2c, where a sharp peak in Ru intensity between 35 and 40 nm correlates with the localized Ru-rich region observed in the corresponding X-EDS map from Figure S2b.

The individual Au and Ru X-EDS maps in Figure S3 further support this interpretation. The Au Lα and Au Mα maps show Au evenly distributed across the nanocrystal. While the Ru Kα and Ru Lα maps show consistent Ru signal throughout the particle, there are localized regions of enhanced Ru intensity. These Ru-rich hotspots indicate modest local Ru enrichment, but the persistence of Ru signal across the nanocrystal is inconsistent with complete phase separation in the pristine state.

Local spectral analysis in Figure S4 provides a similar picture: the summed spectra from Area 1 and Area 2 are qualitatively comparable while being from different regions of the particle (Figure S4b–c). The X-EDS quantification in Table S1 shows that Area 2 contains ~2 at.% more Ru than Area 1, but both local compositions remain close to the composition measured for the full nanocrystal. Together, these results support substantial nanoscale Au–Ru intermixing in the pristine nanocrystals, with modest local variation in Ru content consistent with the expected limited miscibility of Au and Ru.

These features are distinct from the pronounced, spatially separated Au-rich and Ru-rich domains that emerge after annealing in both vacuum and gas environments.

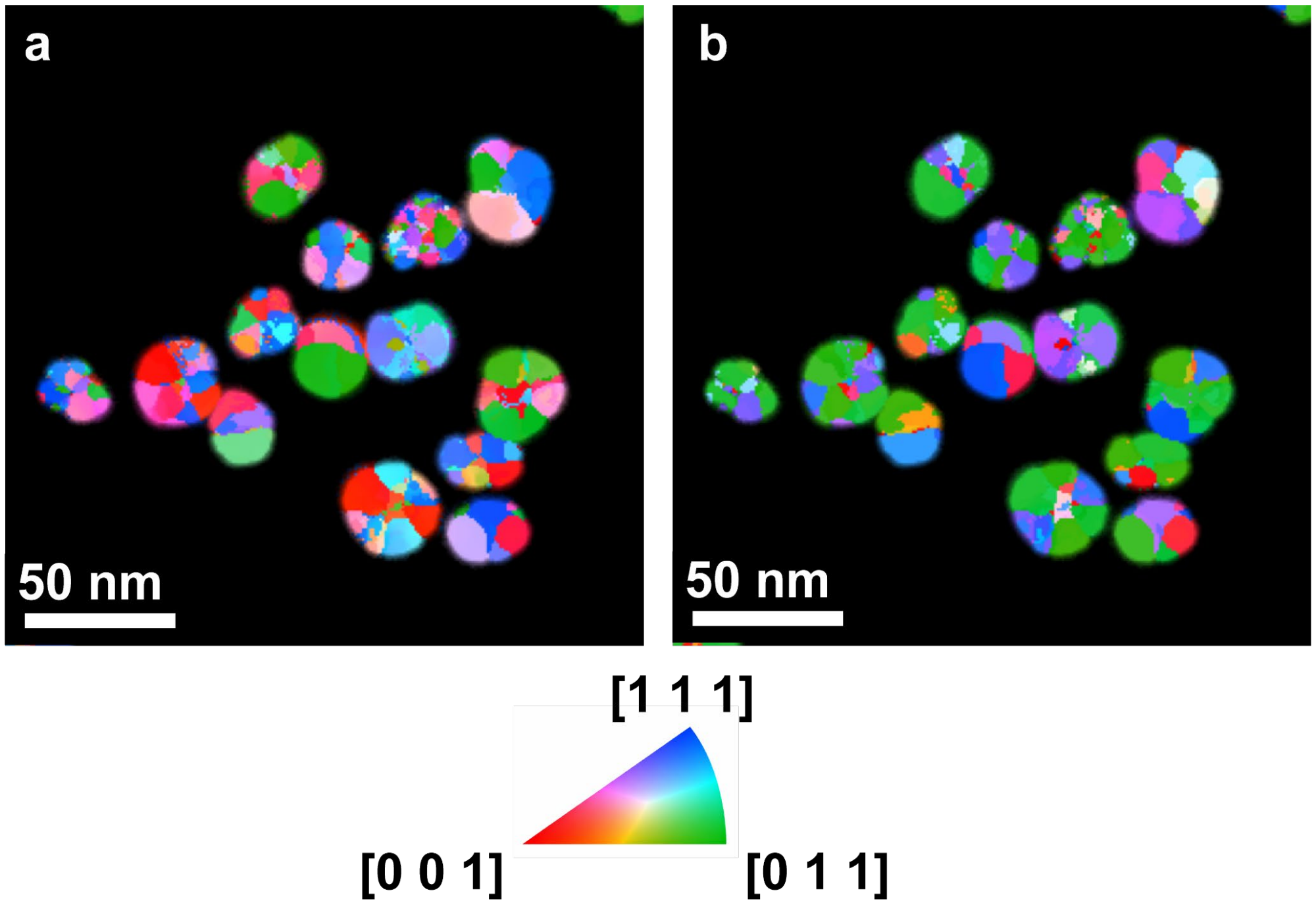

Figure S5. ACOM orientation maps corresponding to the main-text 4D-STEM dataset acquired with a patterned bullseye aperture. (a) In-plane and (b) out-of-plane orientation maps of pristine AuRu nanocrystals, showing the prevalence of pentatwinned particles and internal twin-boundary structure. Colors are assigned according to the principal crystallographic zone axes shown in the inverse pole figure color key.

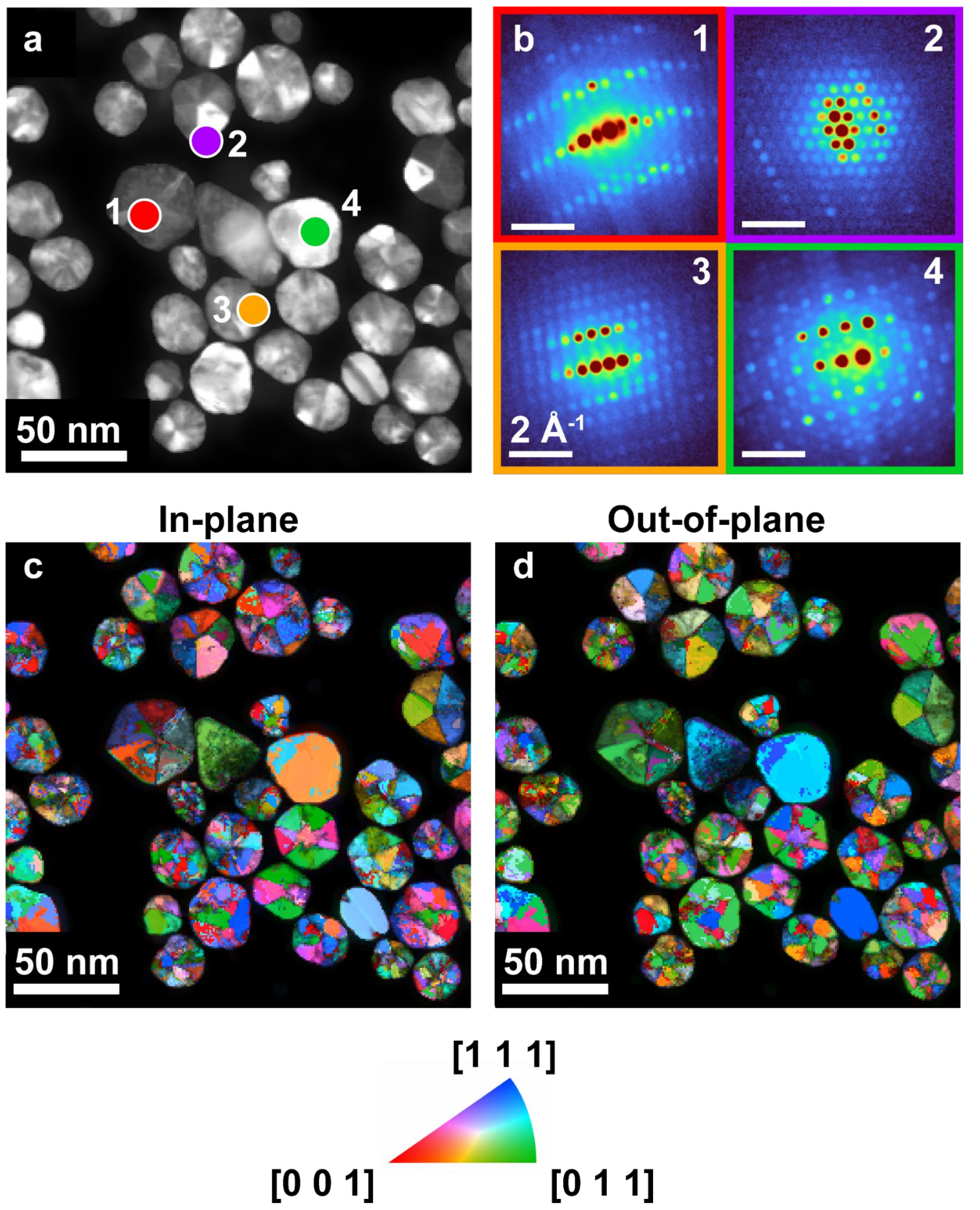

Figure S6. Higher-spatial-resolution 4D-STEM and ACOM analysis of pristine AuRu nanocrystals acquired at a larger convergence angle. (a) 4D-STEM dark-field image of pristine AuRu nanocrystals showing structural heterogeneity across the particle ensemble, with colored markers indicating selected probe positions. (b) Single-point diffraction patterns extracted from the labeled probe positions in (a). (c,d) Corresponding ACOM-derived (c) in-plane and (d) out-of-plane orientation maps, showing the prevalence of pentatwinned particles and internal twin-boundary structure, with colors assigned according to the principal crystallographic zone axes shown in the color key.

**4D-STEM orientation maps show heterogeneity.** Figure S6 shows a dark-field image, representative diffraction patterns, and in-plane and out-of-plane orientation maps of AuRu nanocrystals in their pristine state taken using a larger convergence angle of 2.5 mrad and shorter 73 mm camera length. These parameters enable the higher real-space resolution present in the dark-field image (Figure S6a) and greater orientation sensitivity due to the presence of higher-order diffraction spots (Figure S6b). This allows us to observe the slight variation and strain within individual grains of the AuRu nanocrystals, displayed as color variation in the in-plane and out-of-plane orientation maps in Figure S6c–d.

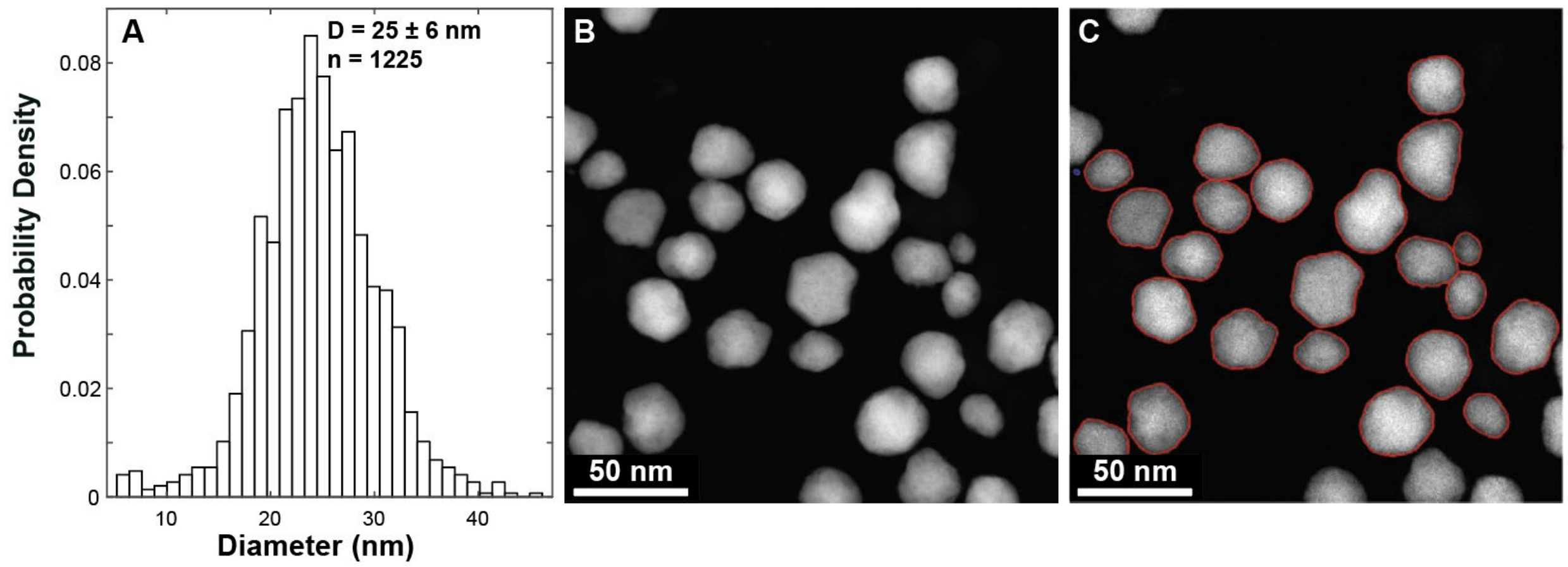

Figure S7. Native size distribution of AuRu nanocrystals. (a) Diameter distribution of 1225 AuRu nanocrystals measured from HAADF-STEM images, with a mean $\mu_0$ = 25 nm and a standard deviation $\sigma_0$ = 6 nm. (b) Representative HAADF-STEM image used for particle sizing. (c) Corresponding thresholded image showing the particle boundary used to extract diameters.

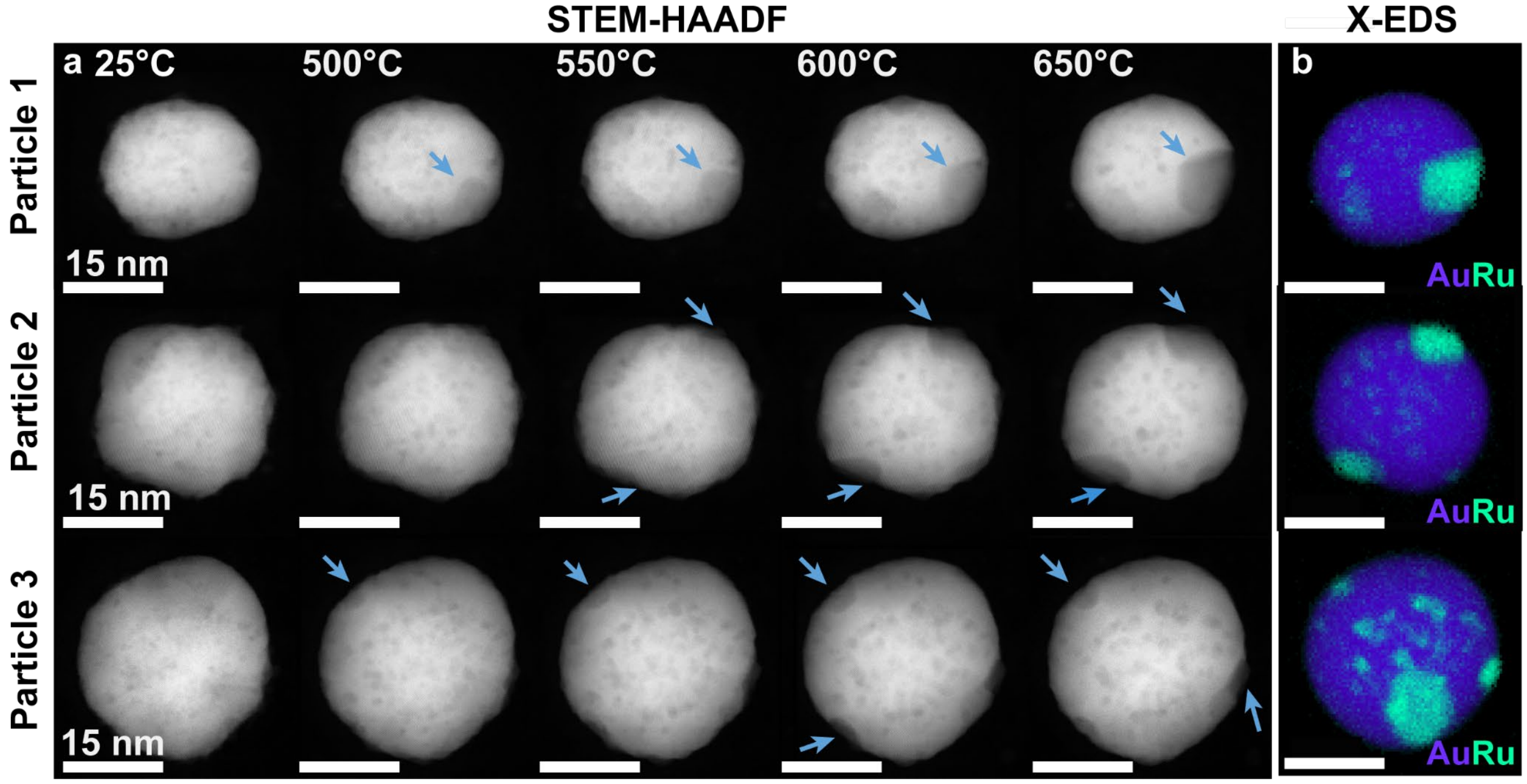

Figure S8. Vacuum-annealing-induced Au–Ru phase segregation in AuRu nanocrystals. (a) HAADF-STEM image series of three representative AuRu nanocrystals during stepwise heating from 25 to 650 °C at a ramp rate of 20 °C/min. Images were acquired at 50 °C intervals, with selected images at 25, 500, 550, 600, and 650 °C shown. Progressive Au–Ru phase segregation is visible as regions of distinct Z-contrast, with lower-Z, dark-contrast Ru-rich domains indicated by blue arrows. (b) Post-annealing X-EDS elemental maps of the same nanocrystals, confirming spatial segregation into Au-rich and Ru-rich regions based on the characteristic Au Lα/Mα and Ru Kα/Lα x-ray lines. Multiple Ru-rich islands are observed within each nanocrystal, consistent with an intermediate segregation state after heating was terminated at 650 °C. For the composite X-EDS maps, each X-ray line, Au Lα/Mα and Ru Kα/Lα, was independently scaled to its 99.9th percentile intensity.

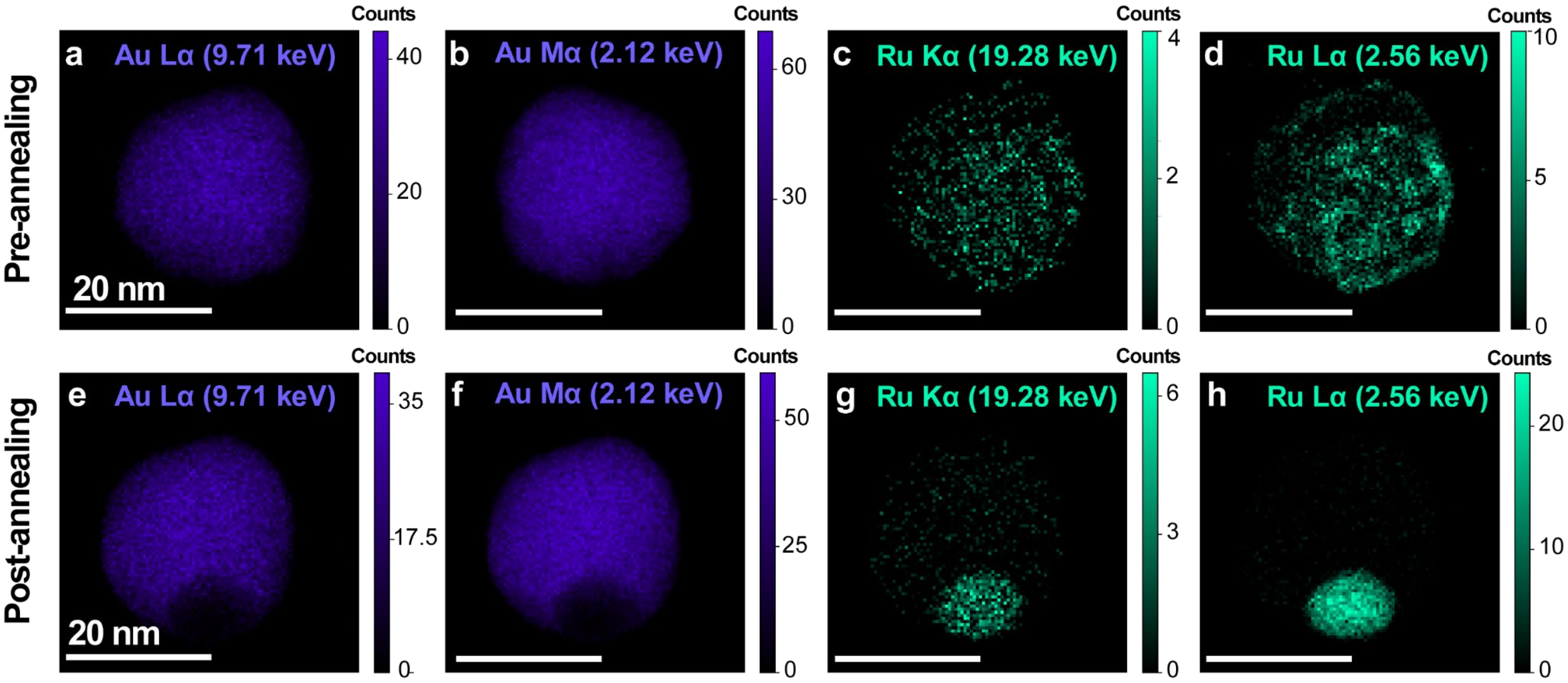

Figure S9. Individual Au and Ru X-EDS maps of the same AuRu nanocrystal pre- and post-annealing. X-EDS maps showing the net intensities of the (a,e) Au Lα, (b,f) Au Mα, (c,g) Ru Kα, and (d,h) Ru Lα emission lines used to construct the composite Au(Lα/Mα)/Ru(Kα/Lα) elemental maps in Figure 2. Panels a–d show the nanocrystal pre-annealing, and panels e–h show the same nanocrystal post-annealing. Each panel is displayed with its own net-intensity color scale, with the upper limit set to the 99.9th percentile intensity for that emission line. Pre-annealing, the Au and Ru signals are spatially intermixed across the nanocrystal, although local regions of higher Ru intensity are present, consistent with the limited bulk miscibility of Au and Ru. Post-annealing, the Ru Kα and Ru Lα maps show localized Ru enrichment near the lower portion of the nanocrystal, accompanied by the absence of Au Lα and Au Mα intensity in the same region.

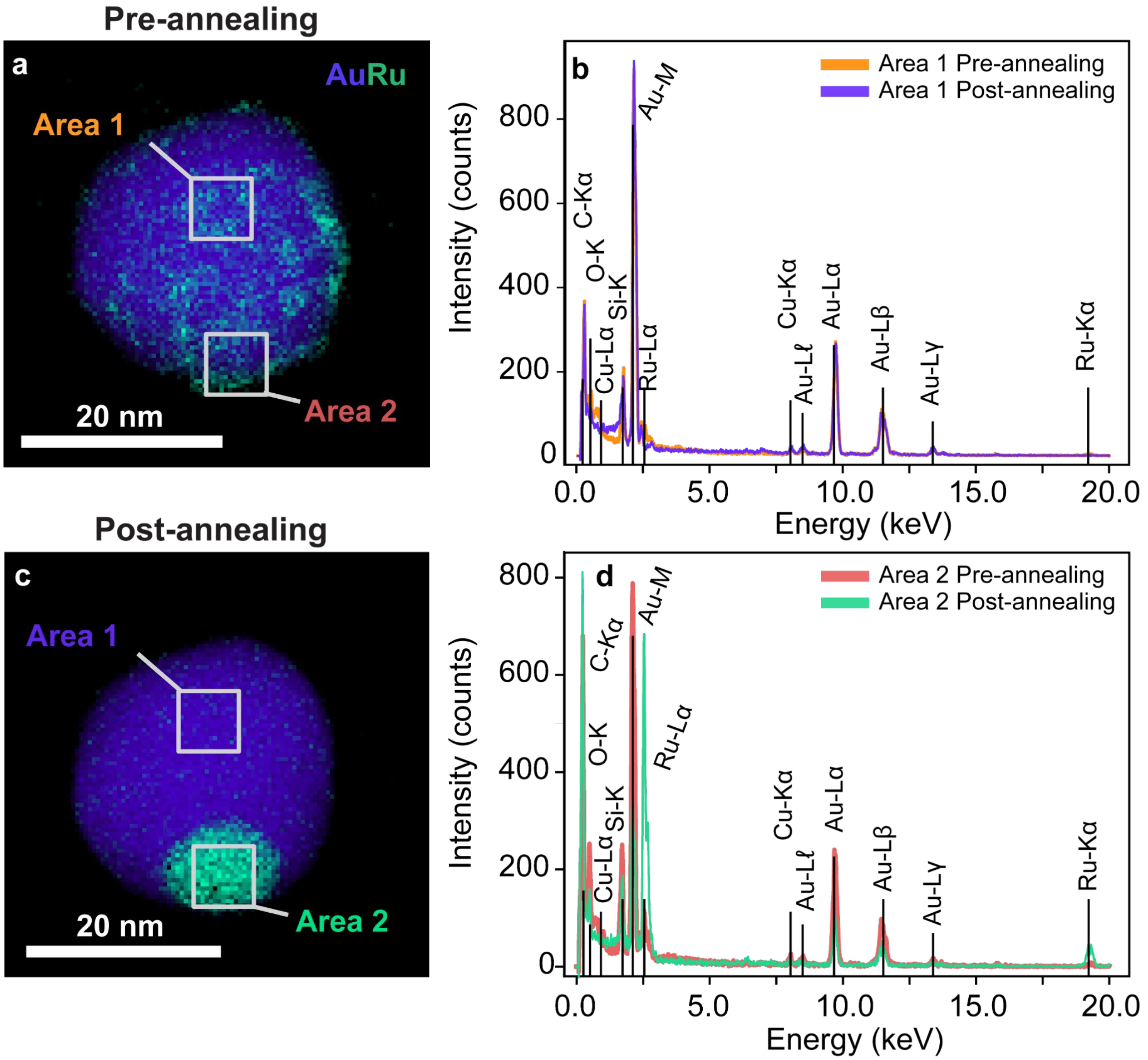

Figure S10. Local X-EDS analysis of AuRu nanocrystal composition pre- and post-annealing. (a,c) Composite Au(Lα/Mα)/Ru(Kα/Lα) X-EDS elemental maps of the same AuRu nanocrystal pre- (a) and post-annealing (c), with regions selected for local spectral analysis indicated by the white boxes labeled Area 1 and Area 2. Each X-EDS map is displayed with an independently scaled intensity range, with the maximum set to the 99.9th percentile intensity for that X-ray emission line. (b) X-EDS spectra from Area 1 pre- and post-annealing show that this region remains Au-rich, with a reduced Ru contribution post-annealing, evident from slight decreases in the Ru Lα and Ru Kα signals and confirmed by quantification in Table S2. (d) X-EDS spectra from Area 2 pre- and post-annealing show a pronounced increase in the Ru Lα signal post-annealing, consistent with local Ru enrichment. C, O, Si, and Cu signals are attributed to background and grid contributions. Quantification using the Au Lα and Ru Lα lines is provided in Table S2.

Table S2. Local X-EDS quantification of compositional changes within the AuRu nanocrystal shown in Figure S10, pre- and post-annealing. Atomic percentages for Area 1 and Area 2 were determined from the local X-EDS spectra acquired from the boxed regions in Figure S10 using the Brown–Powell ionization cross-section model. Au and Ru contents were quantified using the Au Lα and Ru Lα lines, respectively.

| **Location** | **Element** | **Pre-annealing atomic %** | **Post-annealing atomic %** |
|---|---|---|---|
| Area 1 | Au | 91 ± 1 | 100.0 ± 0.1 |
| | Ru | 9 ± 1 | 0.0 ± 0.1 |
| Area 2 | Au | 81 ± 2 | 18 ± 2 |
| | Ru | 19 ± 2 | 82 ± 2 |

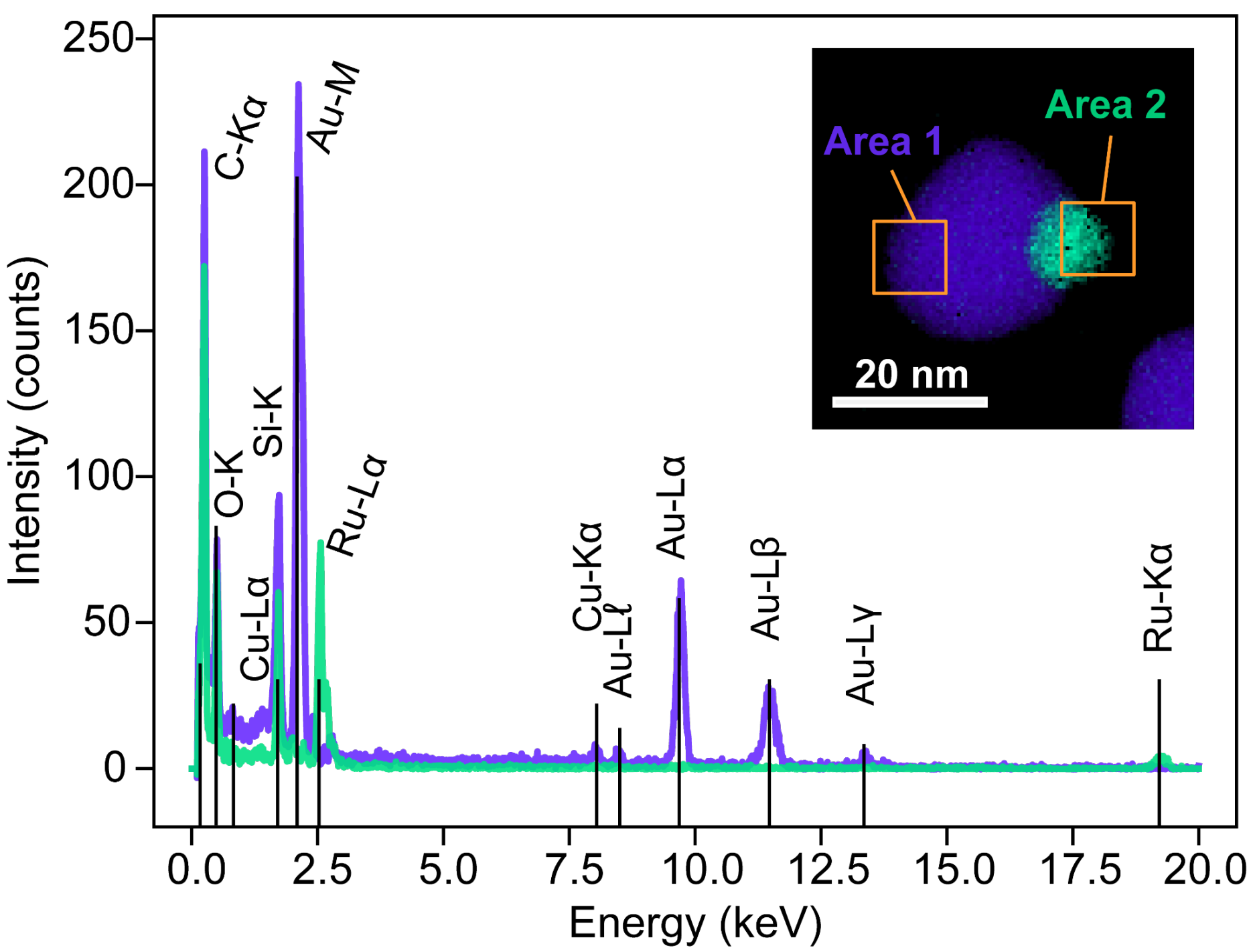

Figure S11. Local X-EDS analysis of near-complete Au/Ru segregation in a post-annealed AuRu nanocrystal. Inset composite Au(Lα/Mα)/Ru(Kα/Lα) X-EDS elemental map identifies regions selected for local spectral analysis. The selected areas are spatially separated in the plane of the TEM grid, such that the Ru-rich region is not strongly overlapped by Au-rich material along the electron-beam direction, enabling clearer assessment of the local extent of elemental separation. X-EDS spectra acquired from these regions show distinct local compositions within the same post-annealed AuRu nanocrystal. Area 1 is Au-rich, with strong Au Lα/Mα signal and no appreciable Ru Kα/Lα peaks. In contrast, Area 2 is Ru-enriched, with strong Ru Kα/Lα signal and no appreciable Au Lα/Mα intensity. C, O, Si, and Cu signals are attributed to background, support, or grid contributions. Quantification using the Au Lα and Ru Lα lines is provided in Table S3. For the composite X-EDS map, each X-ray line, Au Lα/Mα and Ru Kα/Lα, was independently scaled to its 99.9th percentile intensity.

Table S3. Local X-EDS quantification of Areas 1 and 2 in the post-annealed AuRu nanocrystal shown in Figure S11. Atomic percentages were determined from the local X-EDS spectra acquired from the boxed regions in Figure S11 using the Brown–Powell

ionization cross-section model. Au and Ru contents were quantified using the Au Lα and Ru Lα lines, respectively.

| Region | Element | Atomic % |
|---|---|---|
| Area 1 | Ru | 0.0 ± 0.1 |
| | Au | 100 ± 0.1 |
| Area 2 | Ru | 98 ± 2 |
| | Au | 2 ± 2 |

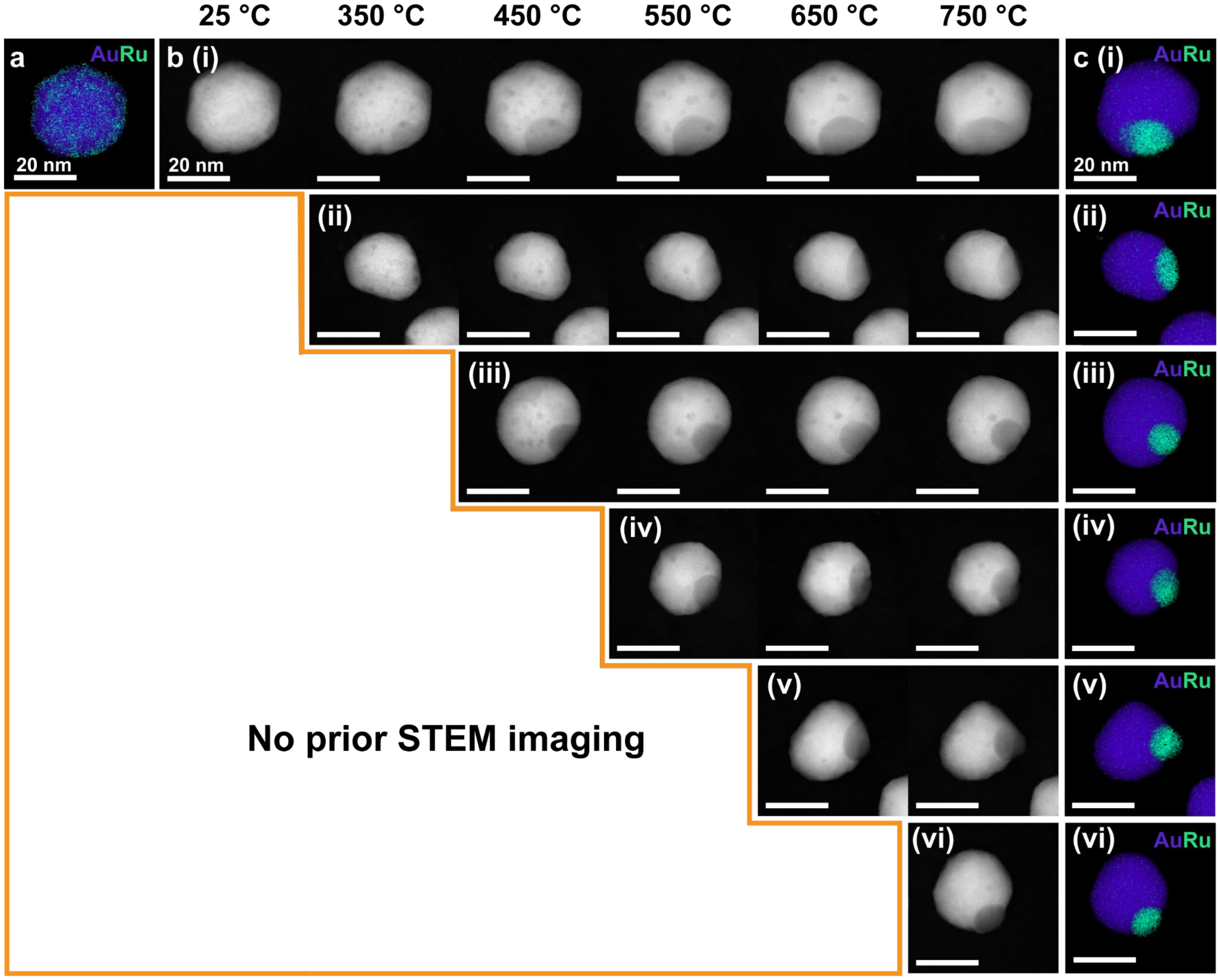

Figure S12. Vacuum annealing series assessing the effect of repeated STEM probe exposure on AuRu nanocrystal phase segregation. (a) Composite Au(Lα/Mα)/Ru(Kα/Lα) X-EDS elemental map of a target AuRu nanocrystal pre-annealing. Au and Ru are intermixed but there are distinct Ru clusters that are distributed around the nanocrystal. (b)(i–vi) HAADF-STEM image series of representative AuRu nanocrystals during stepwise vacuum annealing from 25 to 750 °C. To assess repeated STEM-probe exposure, previously unimaged nanocrystals were introduced at progressively higher temperatures and tracked during subsequent heating. Nanocrystals imaged throughout the full series and those first imaged at higher temperatures show qualitatively similar final morphologies, including particle shape and Ru-rich segregate formation. (c)(i–vi) Composite Au(Lα/Mα)/Ru(Kα/Lα) X-EDS elemental maps of the same nanocrystals post-annealing, showing Au/Ru spatial segregation through the characteristic Au Lα/Mα and Ru Kα/Lα X-ray emission lines. For the composite X-EDS maps, each X-ray line, Au Lα/Mα and Ru Kα/Lα, was independently scaled to its 99.9th percentile intensity.

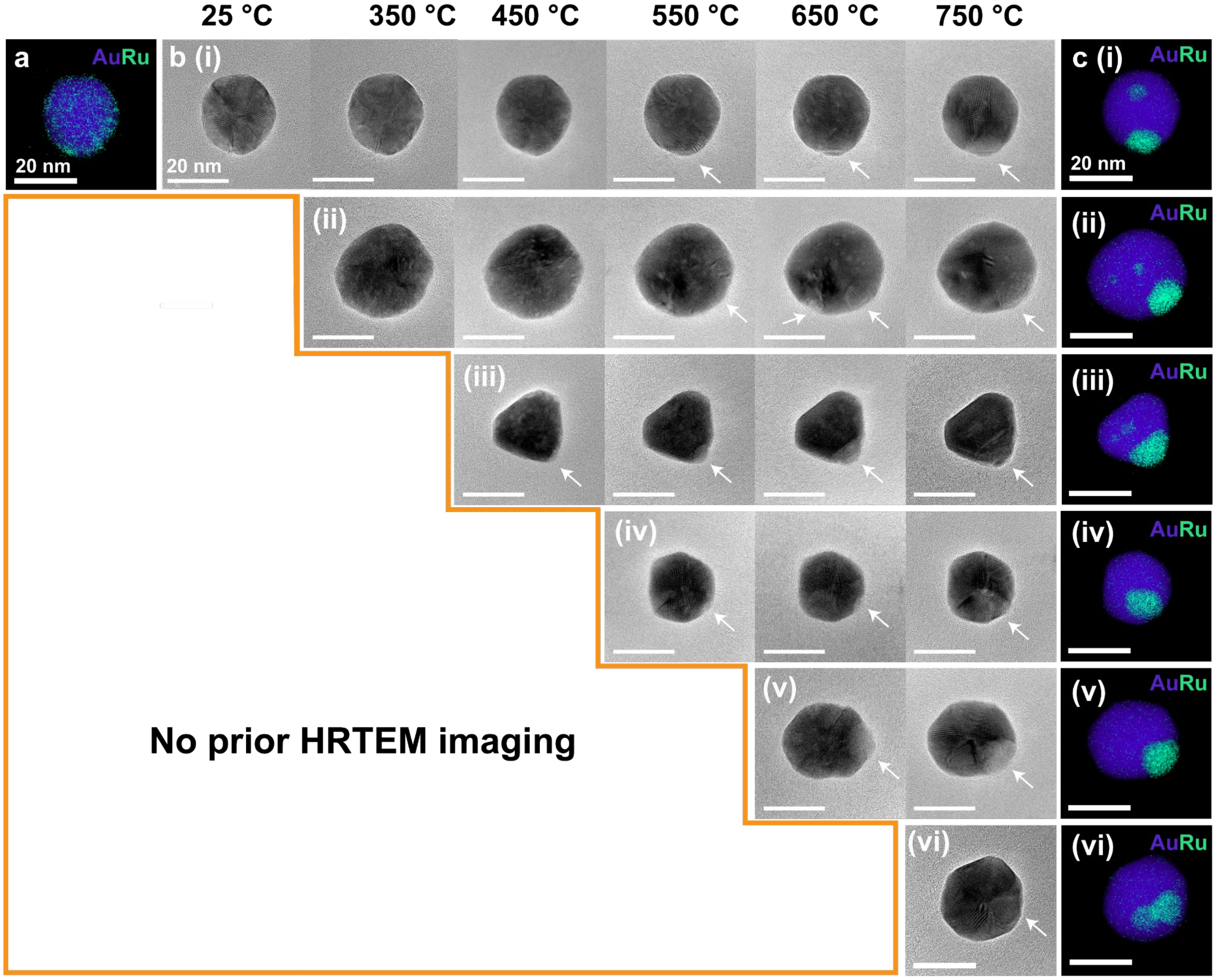

Figure S13. Vacuum annealing series assessing the effect of repeated HRTEM electron beam exposure on AuRu nanocrystal phase segregation. (a) Composite Au(Lα/Mα)/Ru(Kα/Lα) X-EDS elemental map of a target AuRu nanocrystal pre-annealing. Au and Ru are intermixed but there are distinct Ru clusters that are distributed around the nanocrystal. (b)(i–vi) HRTEM image series of representative AuRu nanocrystals during stepwise vacuum annealing from 25 to 750 °C. To assess repeated HRTEM beam exposure, previously unimaged nanocrystals were introduced at progressively higher temperatures and tracked during subsequent heating. Nanocrystals imaged throughout the full series and those first imaged at higher temperatures show qualitatively similar final morphologies, including particle shape and Ru-rich segregate formation. (c)(i–vi) Composite Au(Lα/Mα)/Ru(Kα/Lα) X-EDS elemental maps of the same nanocrystals post-annealing, showing Au/Ru spatial segregation through the characteristic Au Lα/Mα and Ru Kα/Lα X-ray emission lines. For the composite X-EDS maps, each X-ray line, Au Lα/Mα and Ru Kα/Lα, was independently scaled to its 99.9th percentile intensity.

**Discussion of beam-history controls.** In the HRTEM beam-history control series, several nanocrystals developed smaller secondary Ru-rich islands within the particle after heating (Figure S13c(i–iii, vi)). These local features were not observed in the nanocrystals shown in Figure S13c(iv, v), nor were they observed in the STEM beam-history control series (Figure S12). The difference in secondary-island formation between the HRTEM and STEM beam-history control series could suggest a possible imaging-mode-dependent beam effect. However, similar Ru-rich islands were also observed in AuRu nanocrystals heated to 650 °C under STEM imaging conditions, demonstrating that secondary Ru-rich islands are not unique to continuously imaged HRTEM particles (Figure S8). In this case, the Ru-rich islands are attributed to an intermediate segregation state after heating was terminated at 650 °C. We therefore ascribe the presence or absence of smaller secondary Ru-rich islands to particle-to-particle variation and local differences in the annealing environment across the grid, rather than to a systematic beam-induced effect. Across the full control series, AuRu nanocrystals with and without prior beam exposure show the same primary restructuring behavior: temperature-dependent Au–Ru phase segregation into distinct Au-rich and Ru-rich regions.

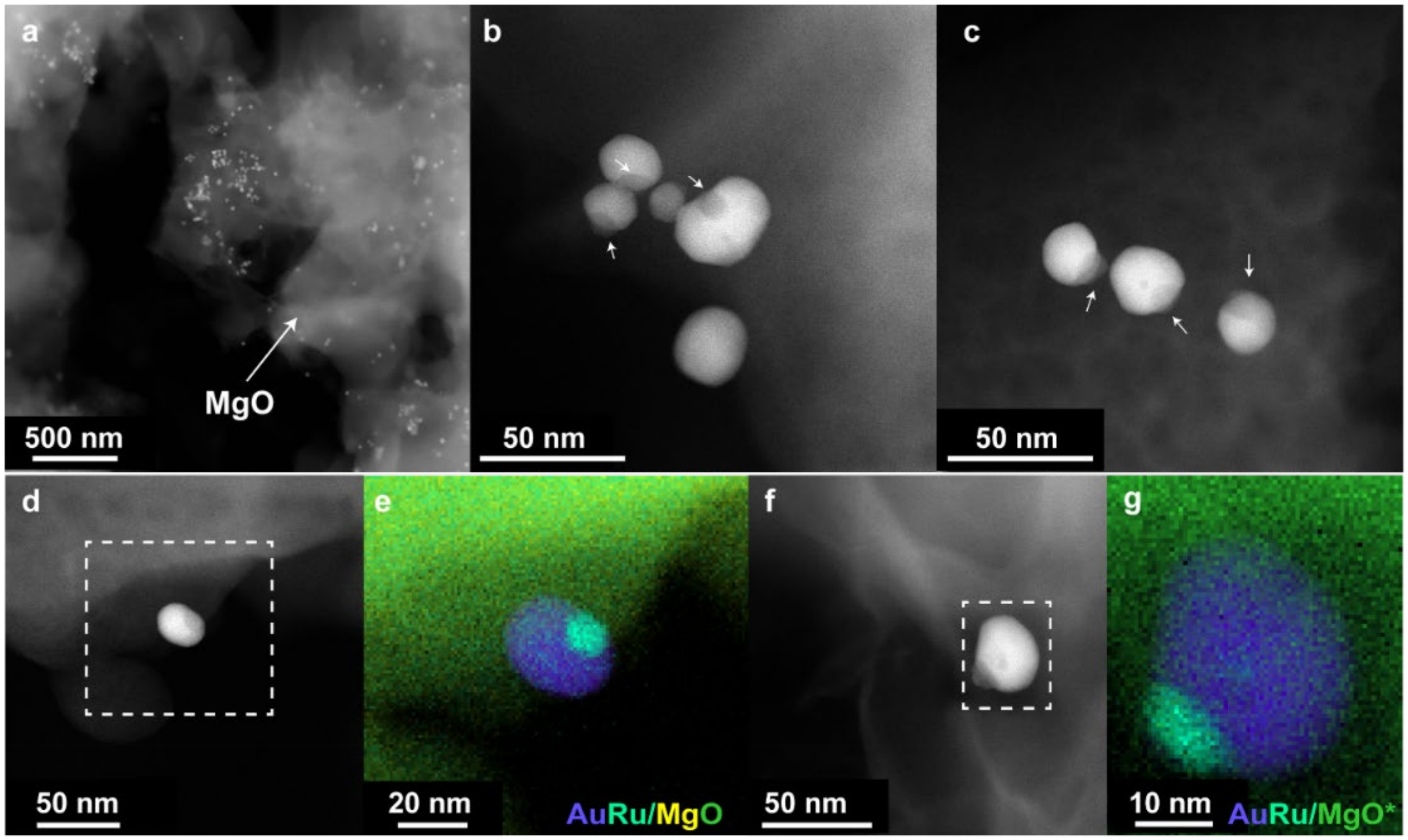

Figure S14. Vacuum annealing induces phase segregation of AuRu nanocrystals on MgO support. (a) HAADF-STEM image of AuRu nanocrystals supported on MgO. (b, c) HAADF-STEM images of multiple supported particles after vacuum annealing, showing Au–Ru phase segregation identified by Z-contrast; white arrows mark Ru-rich regions. (d) HAADF-STEM image of a representative segregated AuRu nanocrystal on MgO. (e) Corresponding X-EDS elemental maps showing Au (Lα/Mα), Ru (Kα/Lα), and Mg and O (Kα) raw signals. (f) HAADF-STEM image of an additional segregated AuRu nanocrystal on MgO. (g) X-EDS component map derived using non-negative matrix factorization (NMF), separating Au, Ru, and MgO* regions. *NMF was necessary due to the thickness of MgO in this region. The sample was not exposed to the electron beam during heating.

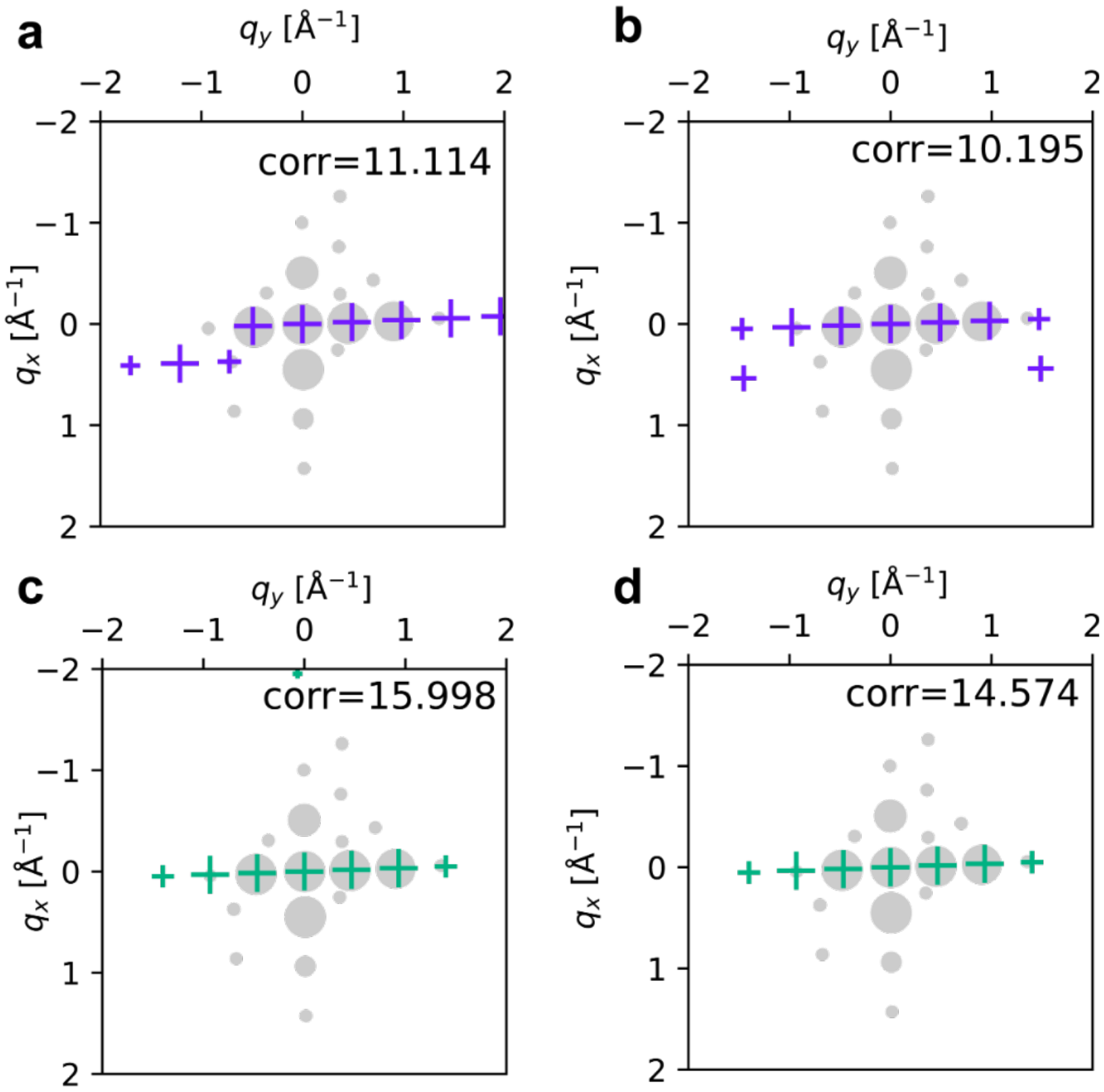

Figure S15. Phase assignment for the single-point diffraction pattern shown in Figure 3e. (a, b) Best and second-best orientation matches assuming an Au FCC structure, with inset values indicating the corresponding correlation strengths. (c, d) Best and second-best orientation matches assuming a Ru HCP structure, with inset correlation strengths. The higher correlation strengths obtained for the Ru HCP assignments indicate that the diffraction pattern is best indexed as Ru.

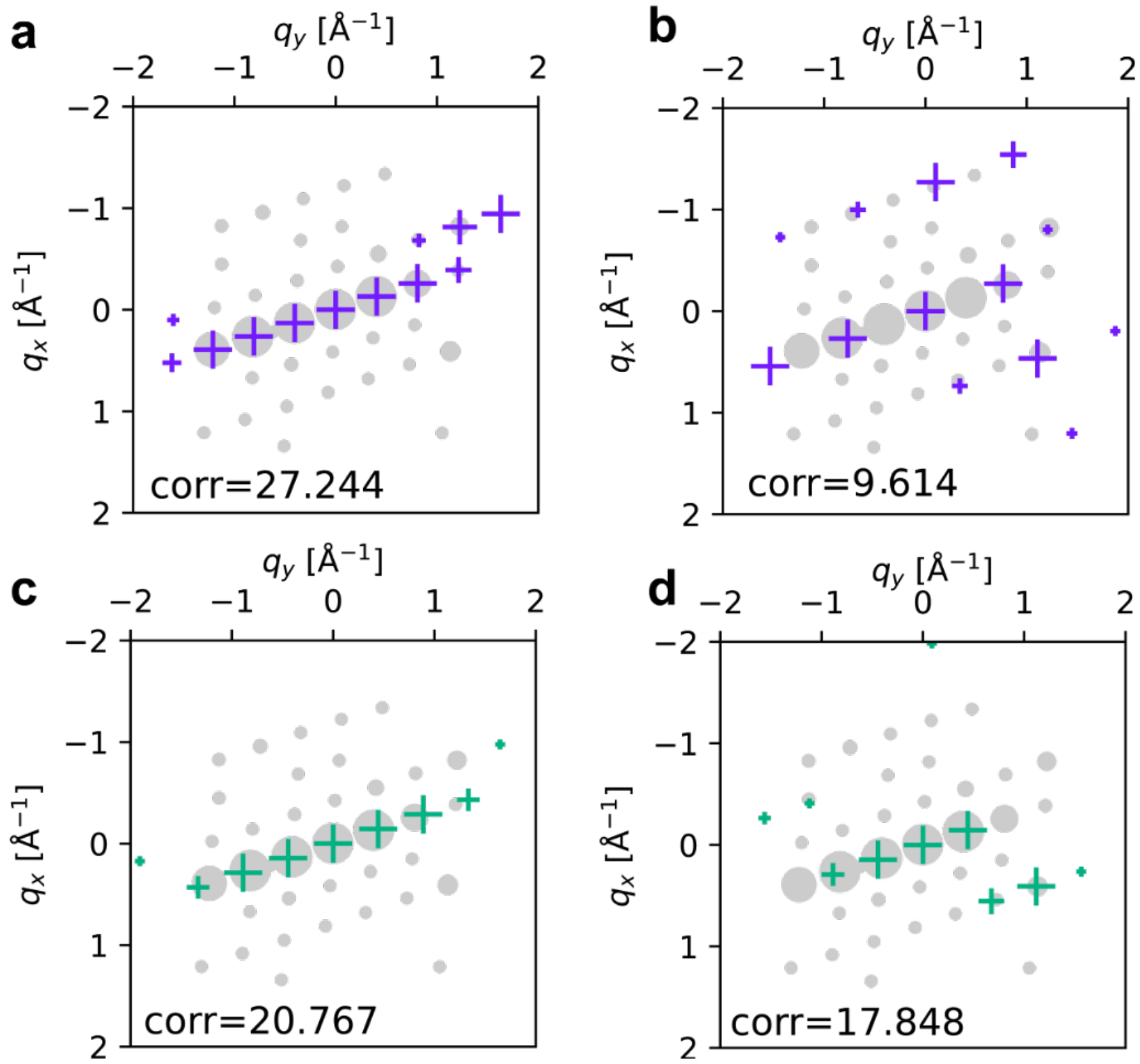

Figure S16. Phase assignment for the single-point diffraction pattern shown in Figure 3f. (a, b) Best and second-best orientation matches assuming an Au FCC structure, with inset values indicating the corresponding correlation strengths. (c, d) Best and second-best orientation matches assuming a Ru HCP structure, with inset correlation strengths. The higher correlation strengths obtained for the Au FCC assignments indicate that the diffraction pattern is best indexed as Au.

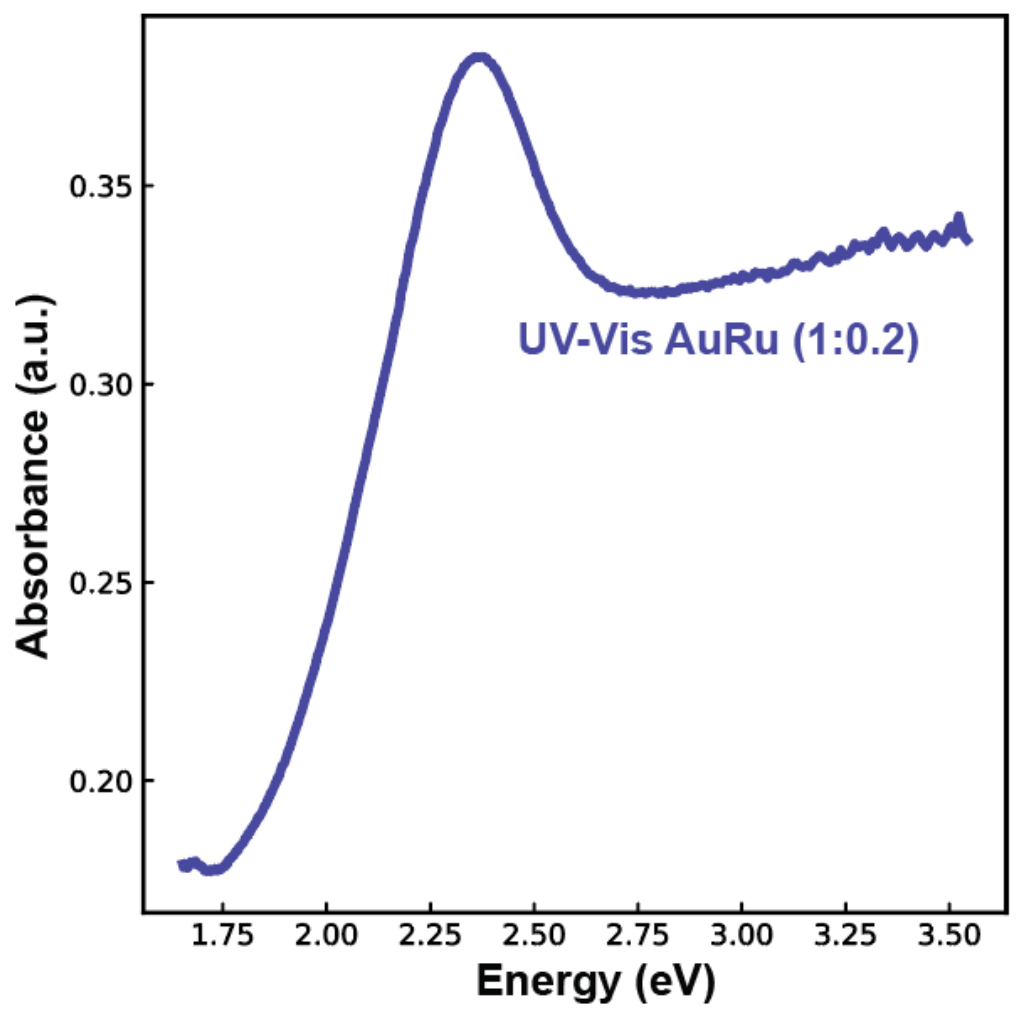

Figure S17. UV–Vis absorbance spectra of AuRu (1:0.2) nanocrystals showing a surface plasmon resonance centered at ~2.4 eV, slightly red-shifted relative to the vacuum monochromated EELS resonance due to the aqueous environment.

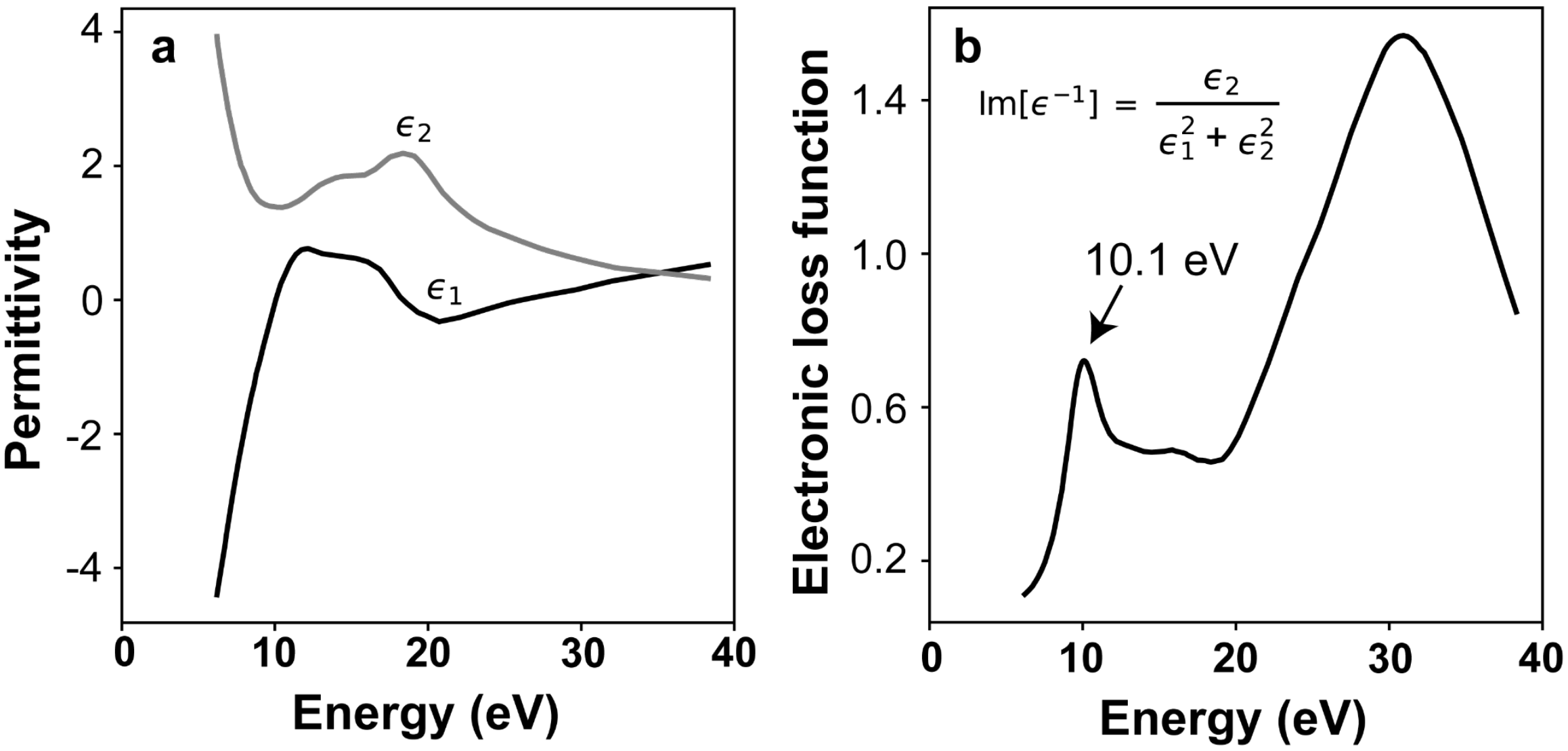

Figure S18. Electronic loss function of Ru showing a bulk plasmon feature. (a) Bulk refractive index data digitized from the published measurements of Cox *et al.* and replotted[7]. (b) Corresponding electronic loss function calculated from the refractive index data in (a), revealing a Ru bulk plasmon at 10.1 eV.

The weak loss feature at 10–11 eV is observed in low-loss EELS after phase segregation and is spatially confined to the Ru-rich domains (Figure 3i, k, l). This spatial localization is inconsistent with an Au-origin loss mechanism: the dominant collective valence excitation of metallic Au is the bulk plasmon at 9.02 eV[8], which would be expected to be strongest in Au-rich regions and to scale with local Au thickness rather than anticorrelate with the Au plasmon map.

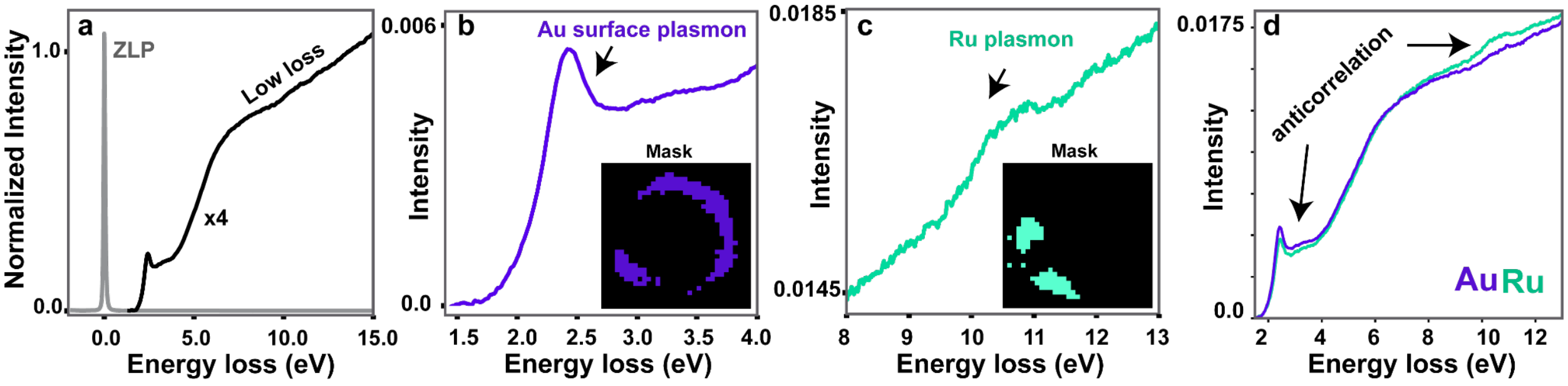

Figure S19. Low-loss EELS analysis of Au and Ru plasmon modes. (a) Overlaid low-loss EELS spectra showing the zero-loss peak (ZLP) and plasmon features associated with Au and Ru. (b) Averaged Au surface plasmon spectrum near 2.5 eV, extracted from regions of high Au signal indicated by the mask in the inset. (c) Averaged Ru plasmon spectrum near 10 eV, extracted from regions of high Ru signal indicated by the mask in the inset. (d) Overlaid full low-loss spectra from the two masked regions, highlighting the anticorrelation between Au and Ru plasmon features. The sample was not exposed to the electron beam during heating.

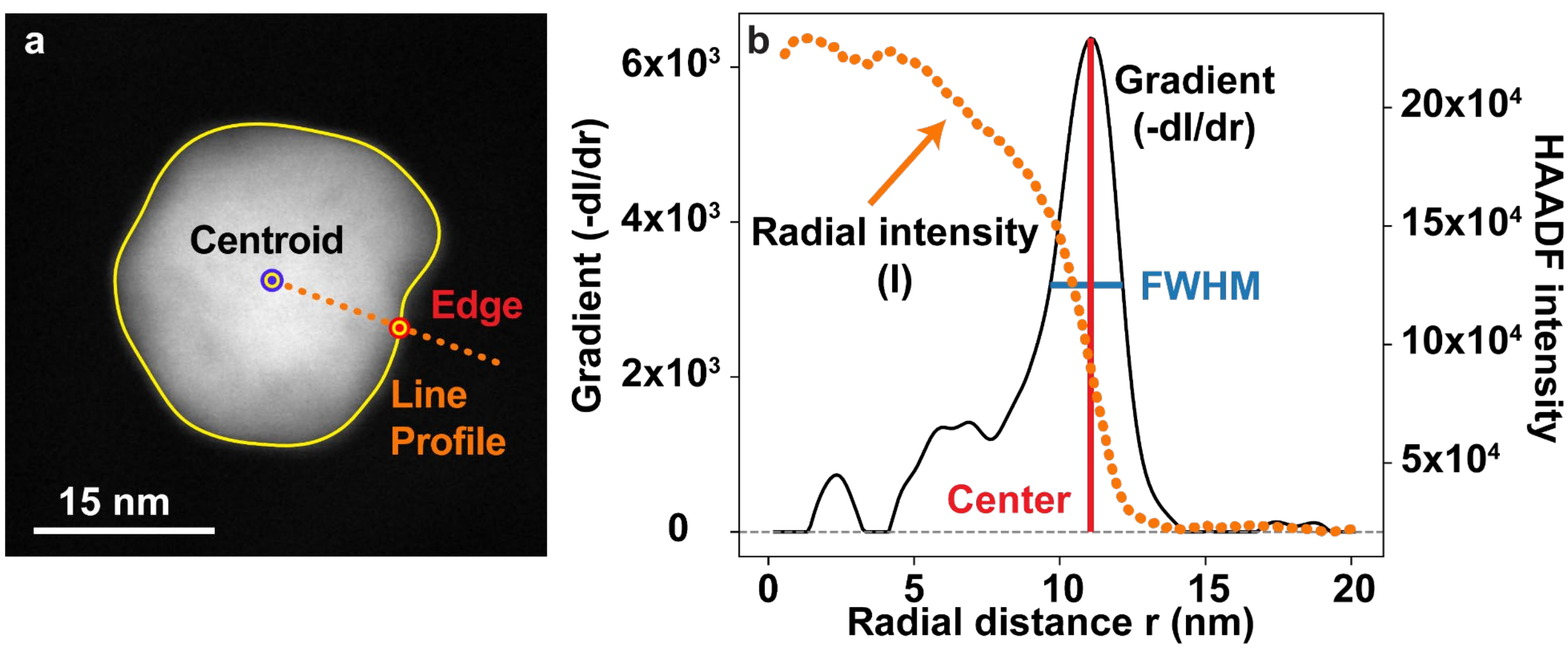

Figure S20. Radial gradient edge detection to determine particle area. (a) HAADF-STEM image of the nanocrystal from Figure 4a with annotated centroid (blue), radial line profile (orange), and edge location (red). (b) Plot of the radial line intensity from the particle center (orange) and the gradient of the radial intensity (black), with the determined peak center (red) and FWHM (blue) used to estimate error.

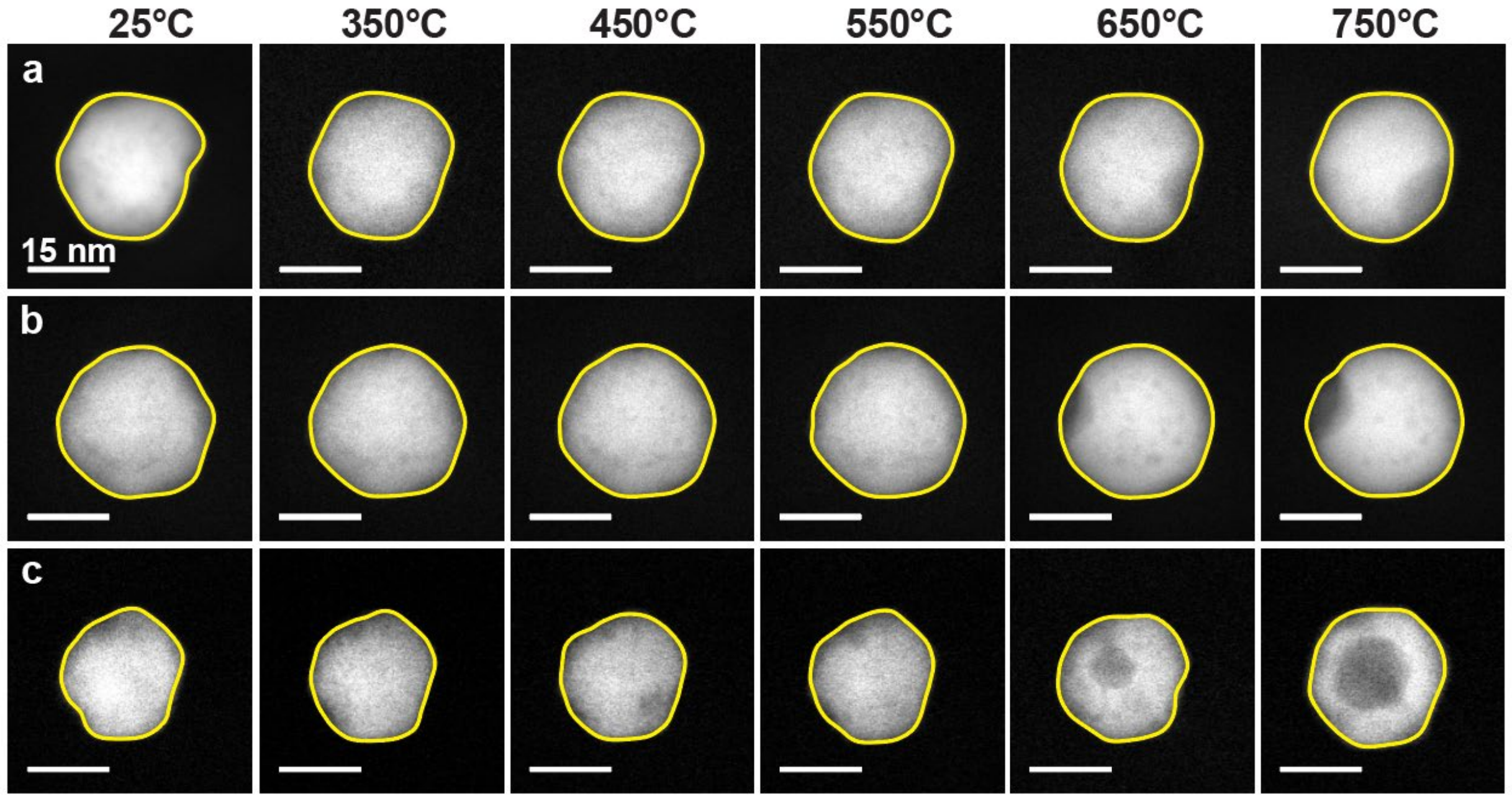

Figure S21. HAADF-STEM image series showing the particle boundary determined using radial gradient edge detection. Yellow contours indicate the boundaries of particles from Figure 4, which were heated at a $H_2$:$N_2$ (3:1) pressure of (a) 50 Torr, (b) 350 Torr, and (c) 780 Torr respectively.

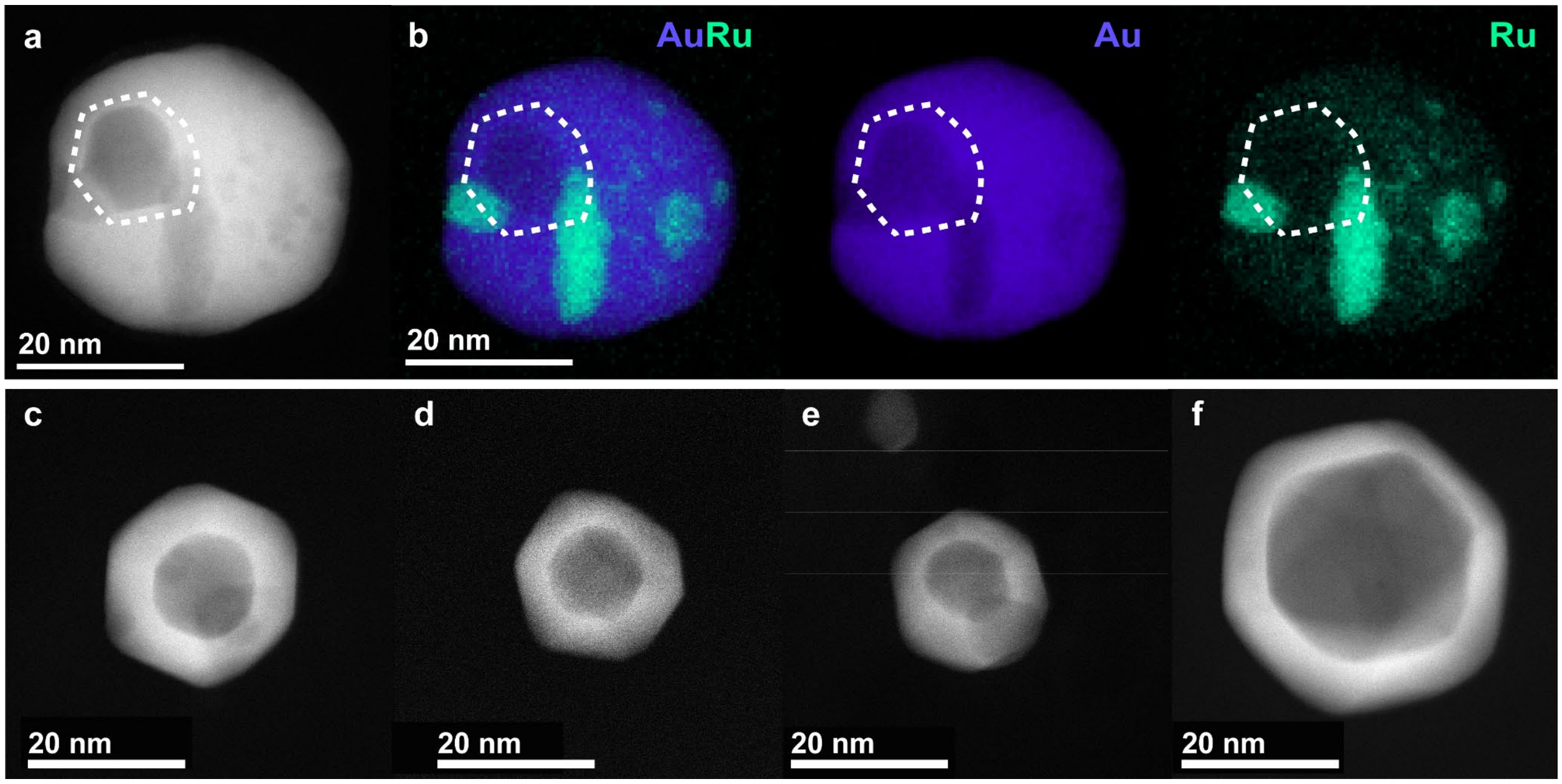

Figure S22. Additional examples of nanocrystals showing void formation in the presence of a $H_2$:$N_2$ (3:1) gas mixture. (a) HAADF-STEM image of an AuRu nanocrystal heated in a $H_2$:$N_2$ (3:1) gas environment, showing a faceted internal void (dashed outline). (b) Corresponding X-EDS elemental maps (Au Mα/Lα and Ru Kα/Lα), overlaid and separated by element, confirming the absence of signal within the void region. (c-f) HAADF-STEM images of AuRu nanocrystals that were not exposed to the electron beam during heating in $H_2$:$N_2$ (3:1) gas.

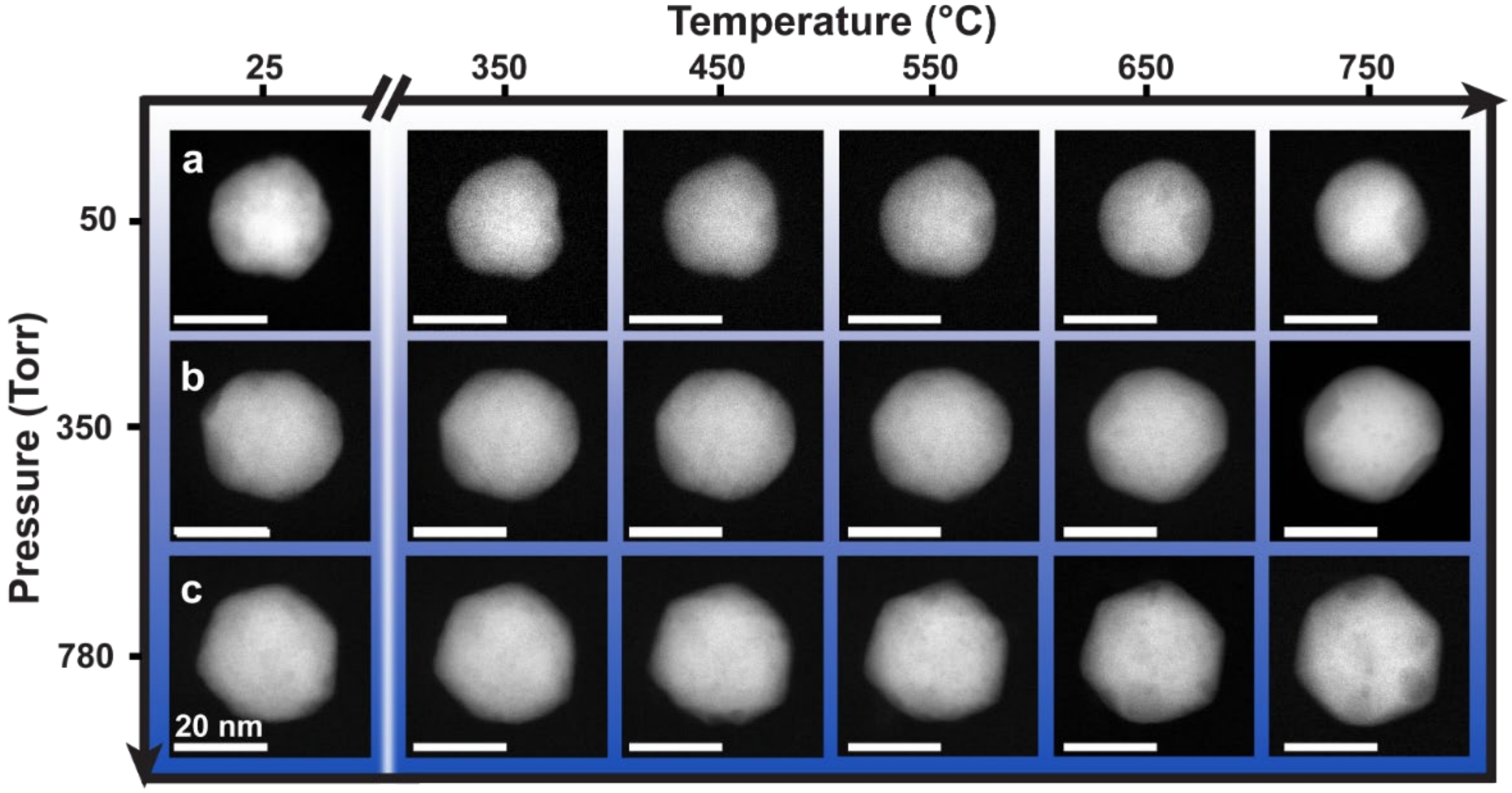

Figure S23. Additional examples of pressure-dependent restructuring of AuRu nanocrystals in a 3:1 $H_2$:$N_2$ atmosphere. (a-c) *In situ* HAADF-STEM image sequences of representative AuRu nanocrystals during stepwise heating to 750 °C (100 °C increments) under increasing $H_2$:$N_2$ (3:1) pressure. (a) At 50 Torr (0.07 atm), Au–Ru phase segregation resembles vacuum annealing (Figure 2b). (b) At 350 Torr (0.46 atm), segregation persists. (c) At 780 Torr (~1 atm), the particle maintains its hexagonal-type faceting, and multiple Ru regions can be observed.

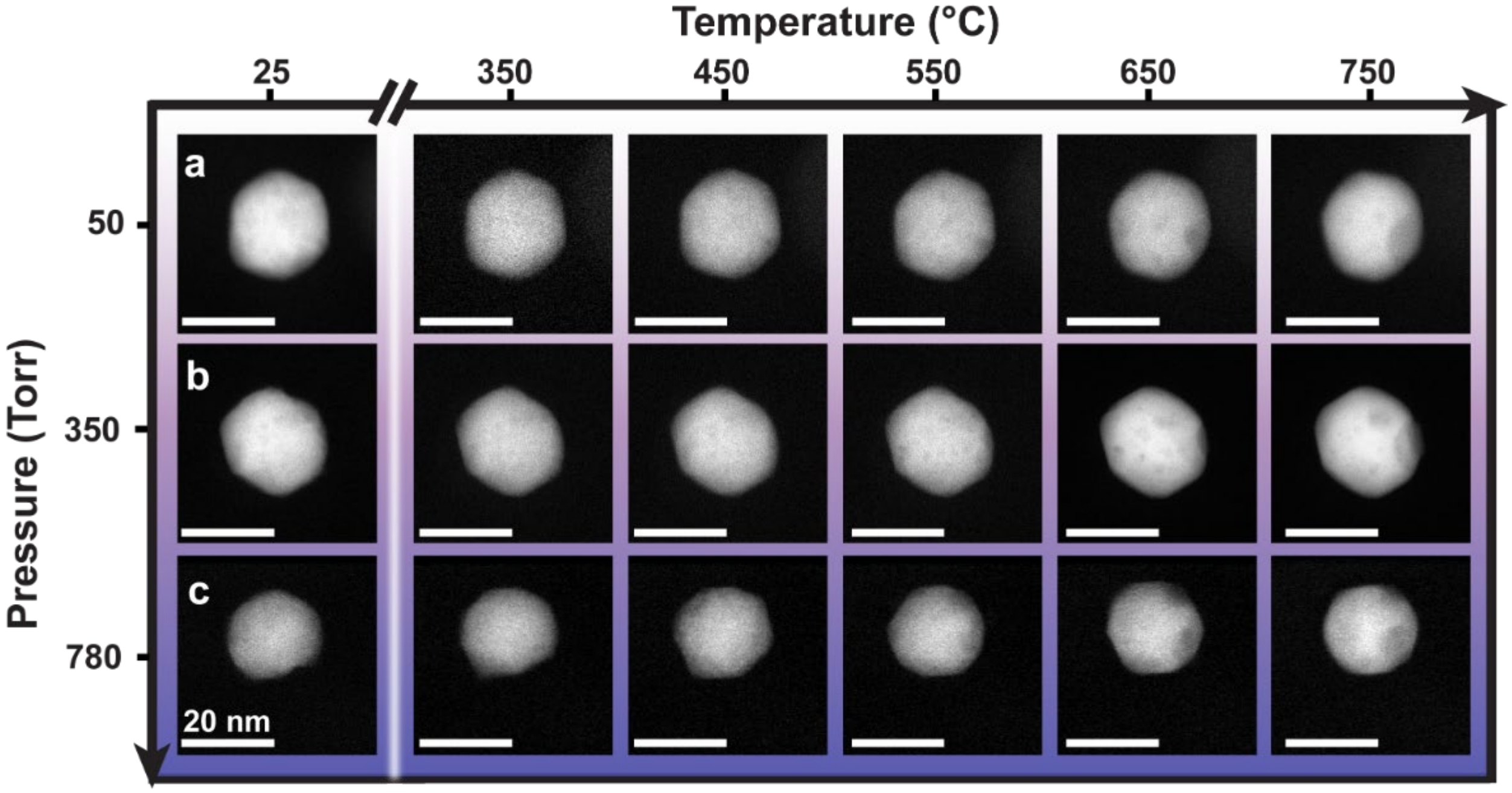

Figure S24. Additional examples of pressure-dependent restructuring of AuRu nanocrystals in a 3:1 $H_2$:$N_2$ atmosphere. (a-c) *In situ* HAADF-STEM image sequences of representative AuRu nanocrystals during stepwise heating to 750 °C (100 °C increments) under increasing $H_2$:$N_2$ (3:1) pressure. (a) At 50 Torr (0.07 atm), Au–Ru phase segregation resembles vacuum annealing (Figure 2b). (b) At 350 Torr (0.46 atm), segregation persists. (c) At 780 Torr (~1 atm), the particle moves towards a more hexagonal-type faceting around 450 °C, which is maintained towards higher temperatures.

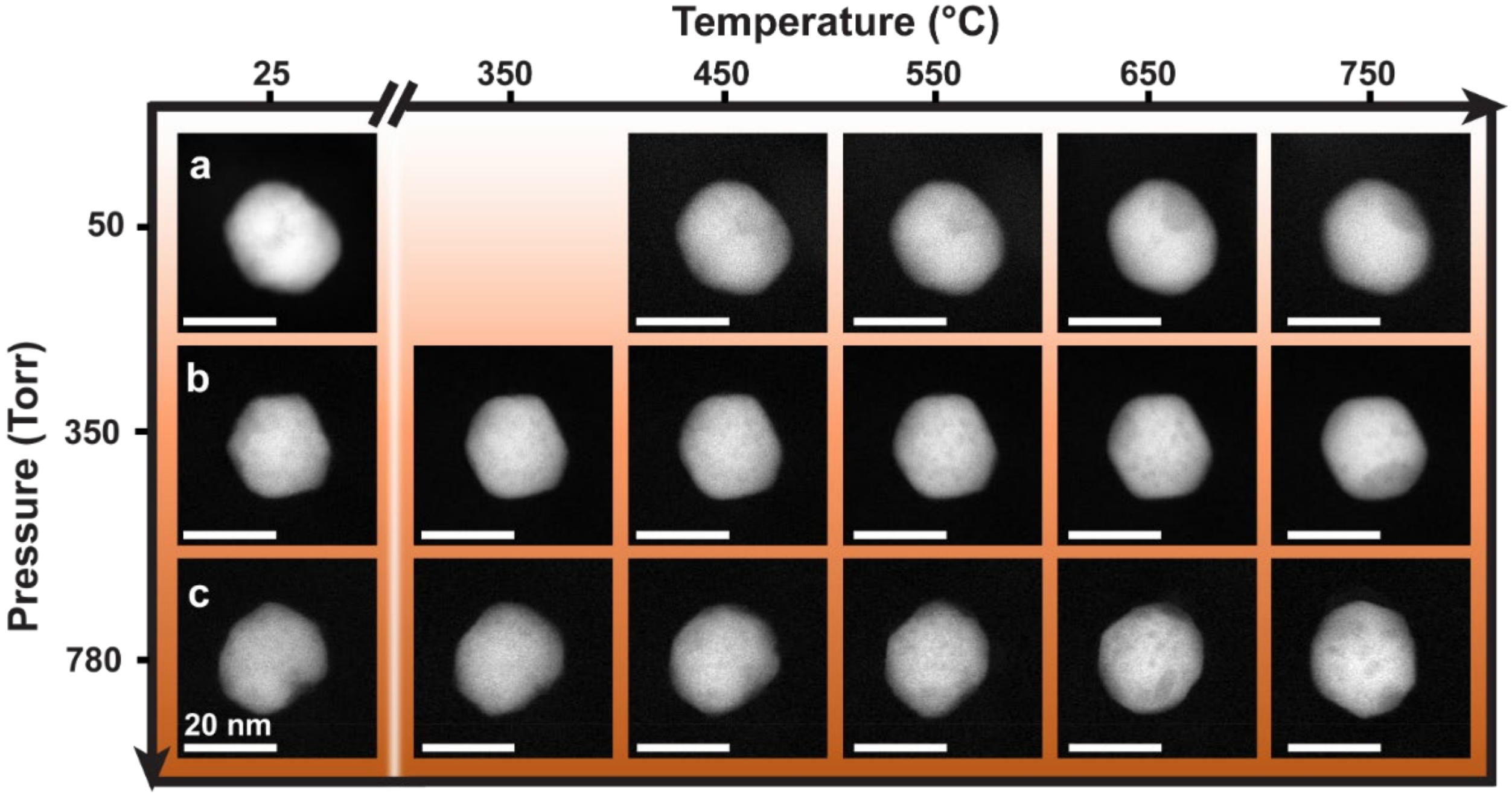

Figure S25. Additional examples of pressure-dependent restructuring of AuRu nanocrystals in a 3:1 $H_2$:$N_2$ atmosphere. (a-c) *In situ* HAADF-STEM image sequences of representative AuRu nanocrystals during stepwise heating to 750 °C (100 °C increments) under increasing $H_2$:$N_2$ (3:1) pressure. (a) At 50 Torr (0.07 atm), Au–Ru phase segregation resembles vacuum annealing (Figure 2b). *No image was collected at 350 °C. (b) At 350 Torr (0.46 atm), segregation persists. (c) At 780 Torr (~1 atm), the particle morphology begins to evolve around 550 °C towards the more faceted final structure seen at 750 °C.

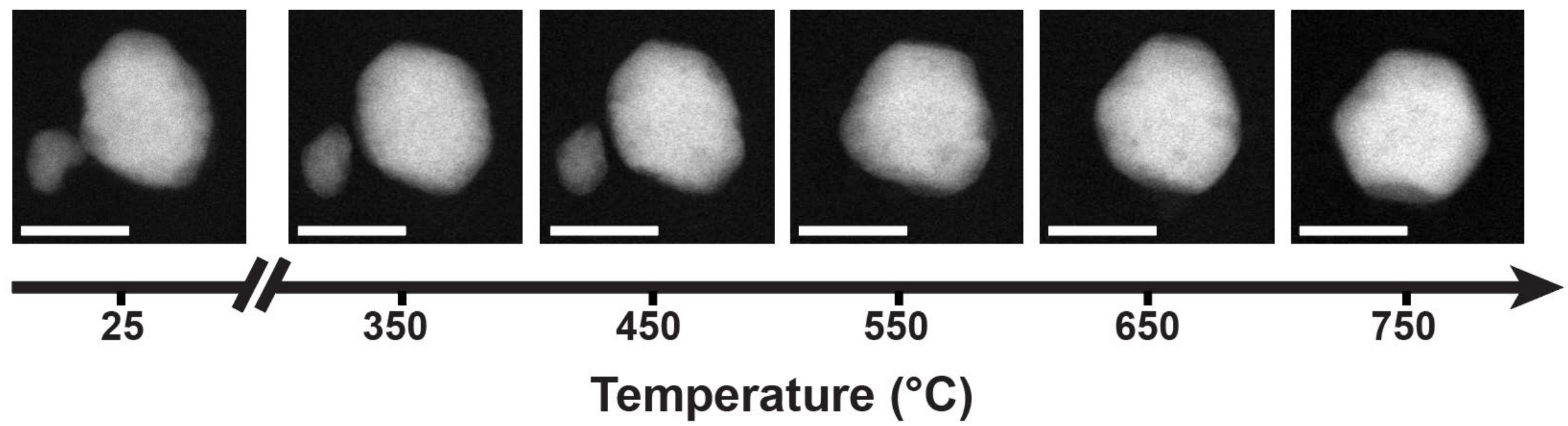

Figure S26. Facet evolution and particle coalescence in a 3:1 $H_2$:$N_2$ atmosphere at 780 Torr. HAADF-STEM image series showing the coalescence of neighboring AuRu nanocrystals at 550 °C, followed by morphological evolution into a hexagonally faceted particle with a segregated Ru-rich region.

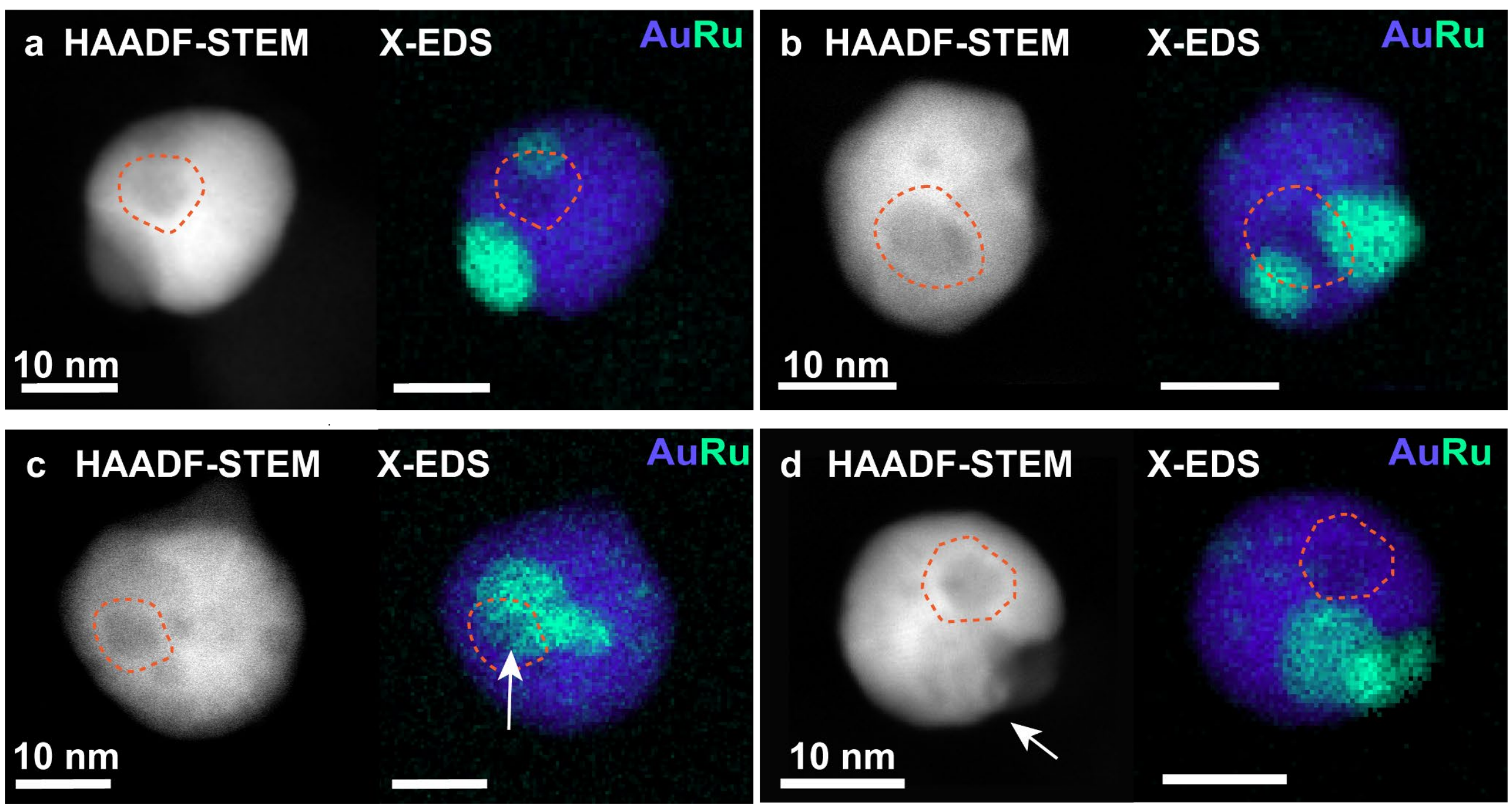

Figure S27. Morphological diversity of voided AuRu nanocrystals after atmospheric-pressure $H_2$ heating not previously exposed to the electron beam. HAADF-STEM images and corresponding composite Au(Lα/Mα)/Ru(Kα/Lα) X-EDS elemental maps of representative AuRu nanocrystals after annealing in pure $H_2$ at ~1 atm. Each nanocrystal contains an internal nanovoid, outlined in orange. The final voided morphologies vary across the ensemble, with some nanocrystals retaining more defined faceting and others adopting less regular final geometries. All analysis was performed post-reaction after opening the gas cell and examining the sample-containing top chip post-reaction. For the composite X-EDS maps, each X-ray line, Au Lα/Mα and Ru Kα/Lα, was independently scaled to its 99.9th percentile intensity. These nanocrystals were not exposed to the beam during heating in $H_2$.

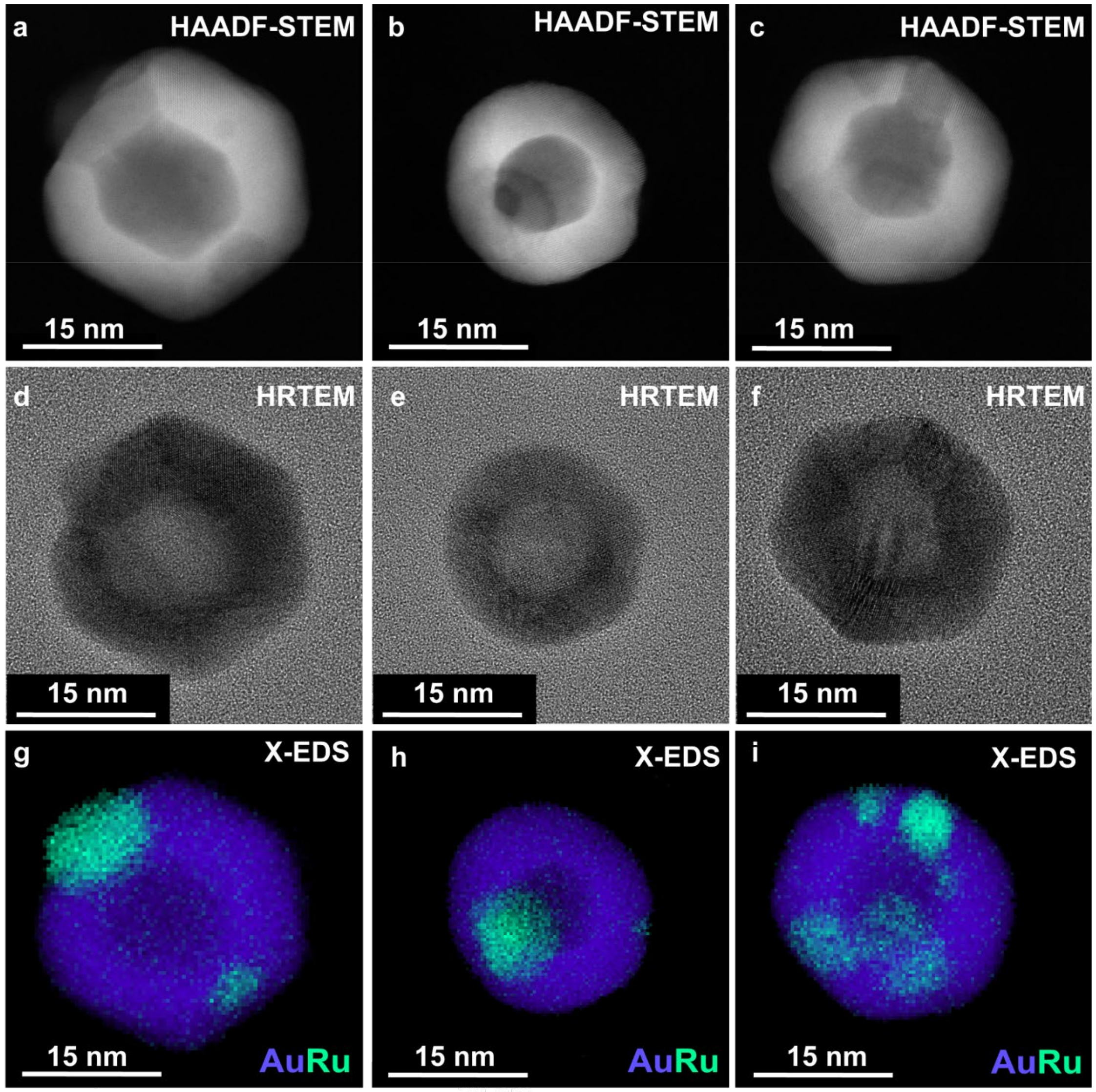

Figure S28. Faceted voided AuRu nanocrystals after atmospheric-pressure $H_2$ heating without electron-beam exposure during annealing. (a–c) HAADF-STEM images of representative voided AuRu nanocrystals with faceted, near-hexagonal projected morphologies after $H_2$ heating. (d–f) Corresponding HRTEM images showing internal void morphology, faceted particle outlines, and lattice-fringe contrast consistent with internal planar defects. (g–i) Corresponding composite Au(Lα/Mα)/Ru(Kα/Lα) X-EDS elemental maps showing Ru-rich regions intersecting or adjacent to the internal voids. These examples show that void formation can occur in faceted, near-hexagonal AuRu nanocrystals and is spatially associated with Ru-rich regions. The particles were not exposed to the electron beam during heating. All analysis was performed post-reaction after opening the gas cell and examining the sample-containing top chip post-reaction. For the composite X-EDS maps, each X-ray line, Au Lα/Mα and Ru Kα/Lα, was independently scaled to its 99.9th percentile intensity.

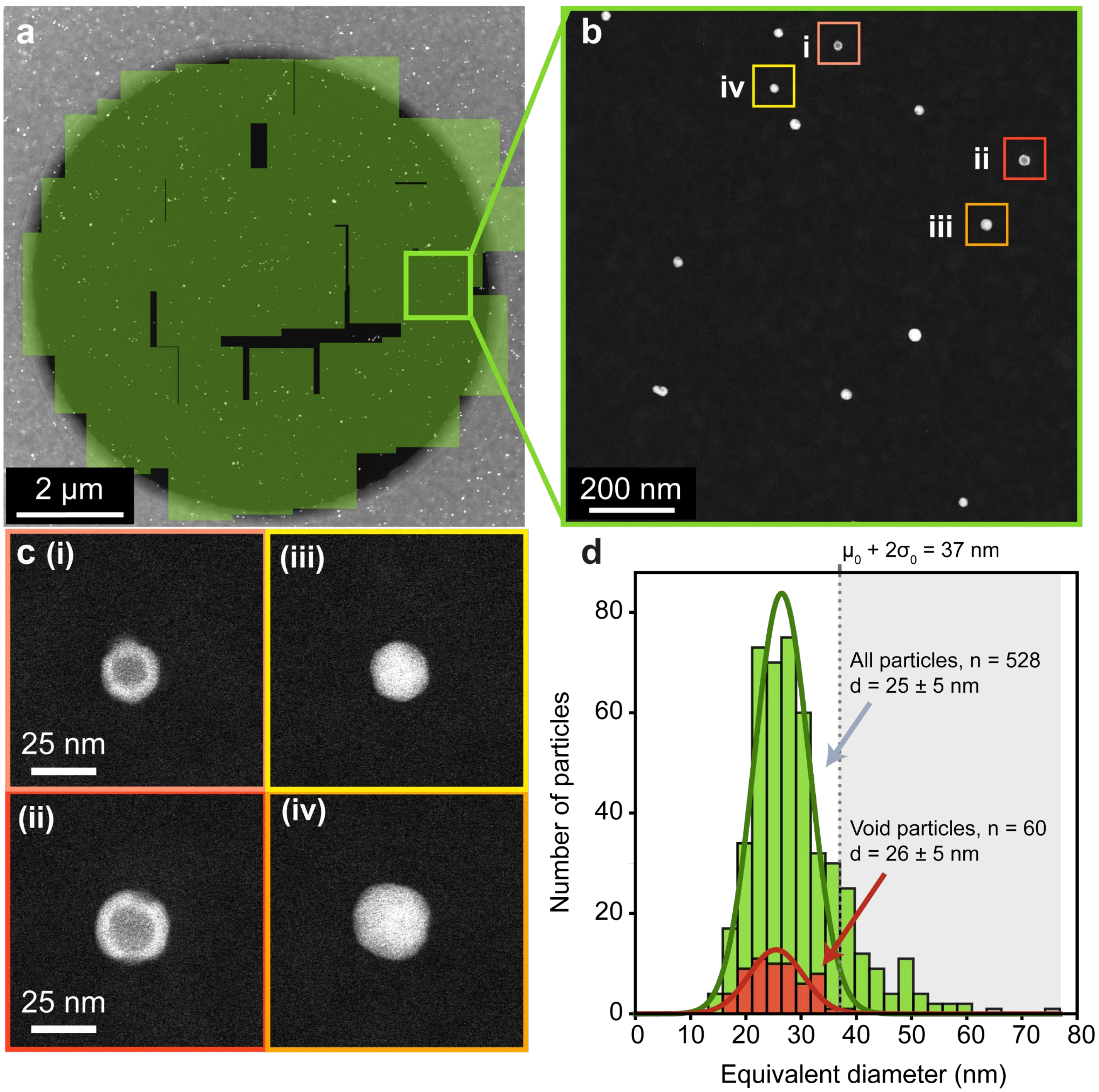

Figure S29. Identification of void-containing AuRu nanocrystals after annealing in 1 atm $H_2$ without prior electron beam exposure. (a) Low-magnification HAADF-STEM image of one of the six SiN windows on a Protochips Atmosphere top chip post-heating in 1 atm $H_2$. The green overlay marks the set of higher-magnification images acquired to tile the window for post-heating analysis, and the outlined box indicates one representative image shown in (b). (b) Higher-magnification HAADF-STEM image of the boxed region in (a), with void-containing AuRu nanocrystals highlighted by two-toned orange boxes and non-void-containing AuRu nanocrystals highlighted by two-toned yellow boxes. (c) HAADF-STEM images of the four regions of interest from (b) showing that it is possible to clearly distinguish AuRu nanocrystals with voids and those without. These tiled images were used to identify void-containing nanocrystals for the population statistics

shown in (d). (d) Diameter distribution of AuRu nanocrystals measured from HAADF-STEM images of the single SiN window shown in (a). The green distribution includes all measured AuRu nanocrystals (n = 528), and the orange distribution shows the subset of nanocrystals that contain voids (n = 60). The dotted vertical line marks $\mu_0 + 2\sigma_0 = 37$ nm, where $\mu_0$ and $\sigma_0$ are the mean and standard deviation of the native size distribution shown in Figure S7. Post-heating nanocrystals larger than $\mu_0 + 2\sigma_0$ were excluded to guard against including large nanocrystals that formed by coalescence of neighboring nanocrystals during heating. Figure S26 shows an example of this type of coalescence event.

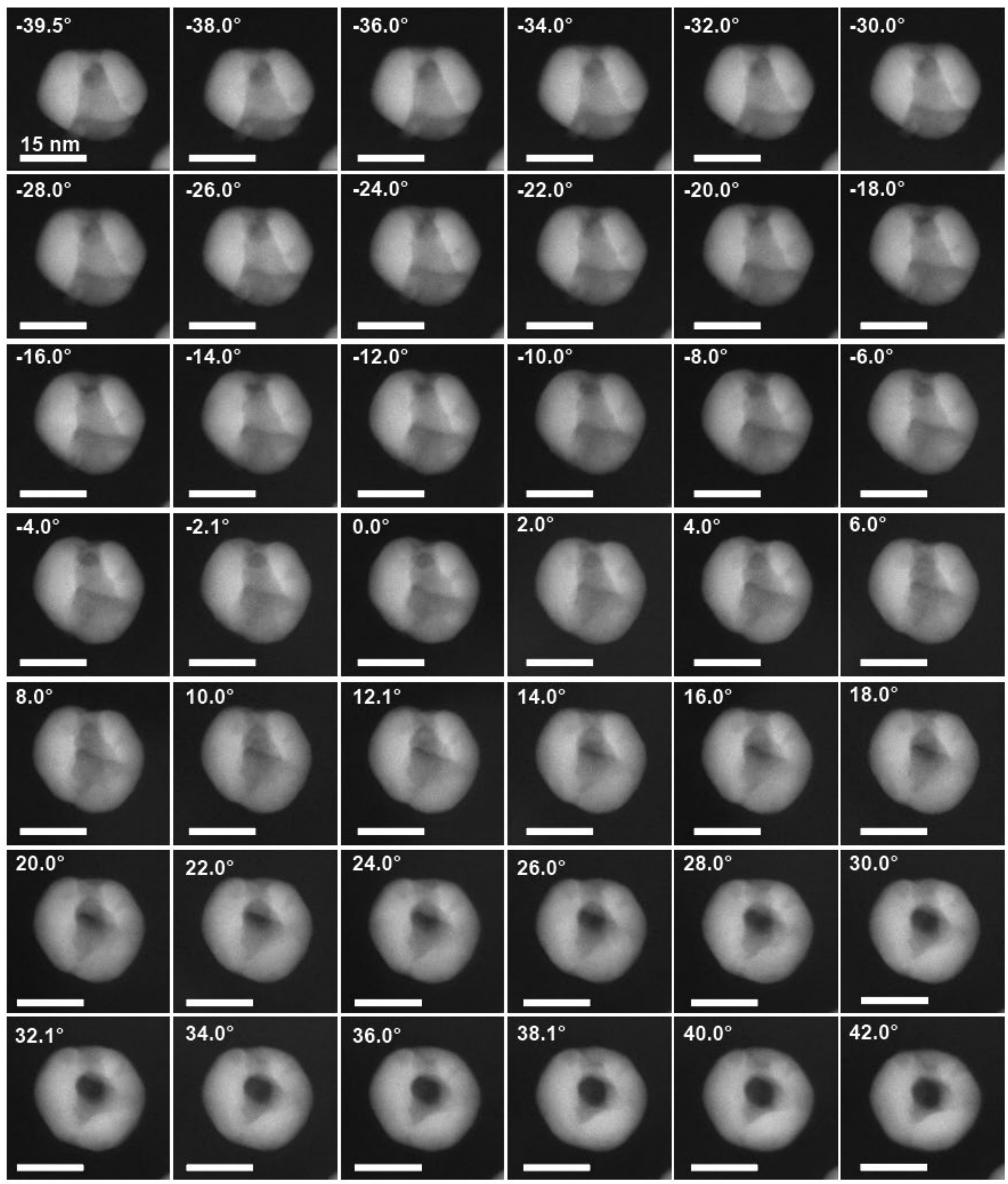

Figure S30. Experimental tomographic tilt series of the AuRu nanocrystal post-exposure to $H_2$:$N_2$ (3:1) gas at 780 Torr. 42 HAADF-STEM images of the nanocrystal with a tilt range from -39.5° to +42.0°. Scale bar, 15 nm.

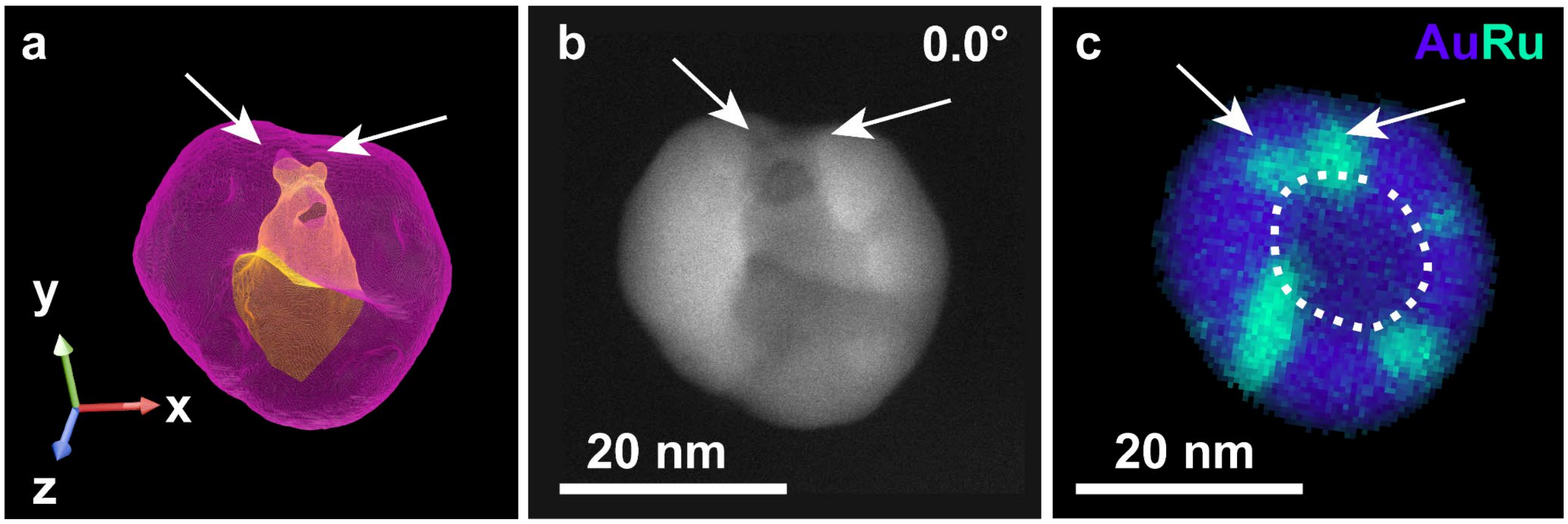

Figure S31. Cross-correlation of tomography, HAADF-STEM, and X-EDS showing the presence of a Ru channel. (a) Machine-learning–assisted tomographic reconstruction showing internal (yellow) and external (magenta) particle surfaces; this view correlates to the 0.0° tilt HAADF particle orientation. (b) HAADF-STEM image from the tilt series at 0.0°. (c) Corresponding X-EDS elemental map reproduced from Figure 4g, with arrows indicating the mid-particle Ru-rich channel intersecting the void.

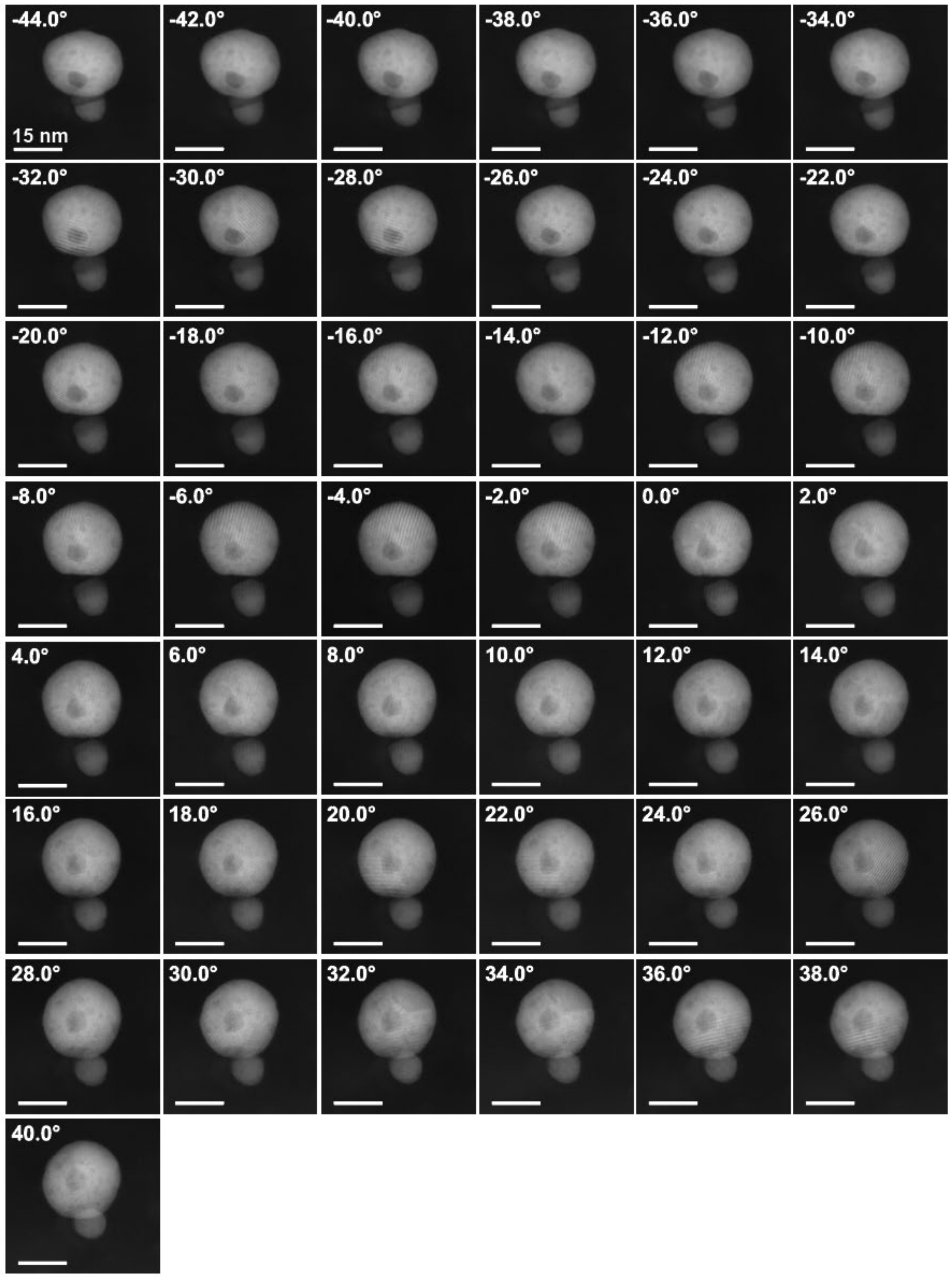

Figure S32. Experimental tomographic tilt series of the AuRu nanocrystal post-exposure to pure $H_2$ gas at 760 Torr. 43 HAADF-STEM images of the nanocrystal with a tilt range from -44.0° to +40.0°. Scale bar, 15 nm.

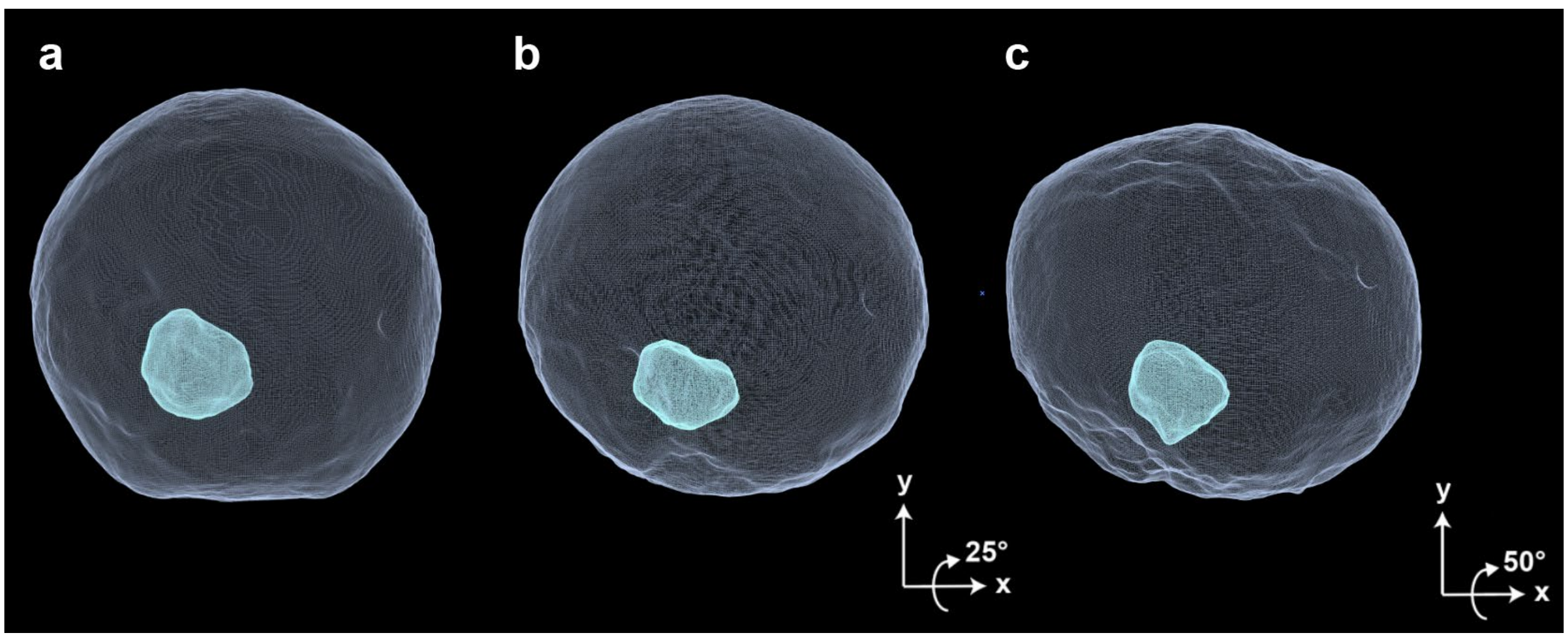

Figure S33. Machine-learning-assisted tomographic reconstruction from multiple viewing directions revealing the 3D geometry of the internal nanovoid. Isosurface at 1.05 converted to wire mesh, with particle exterior (white) and internal void (light-cyan). (a) Face-one view matching the 0.0° tilt from Figure S32. (b) View with 25° tilt along the x-axis. (c) View with a 50° tilt along the x-axis.

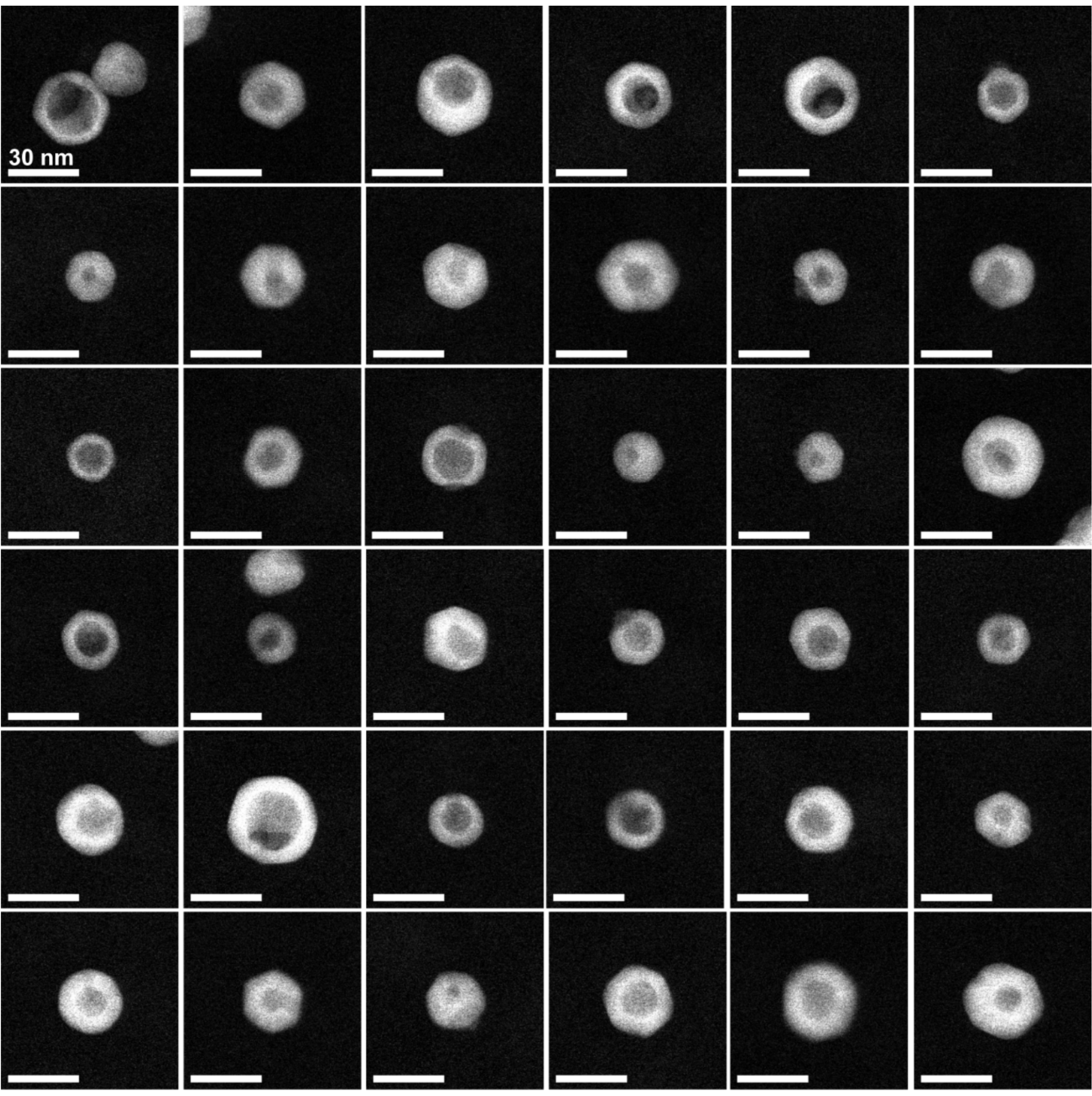

Figure S34. HAADF-STEM survey of void-containing AuRu nanocrystals after annealing in 1 atm $H_2$ without prior electron beam exposure. Representative HAADF-STEM images of void-containing nanocrystals from statistics presented in Figure S29.

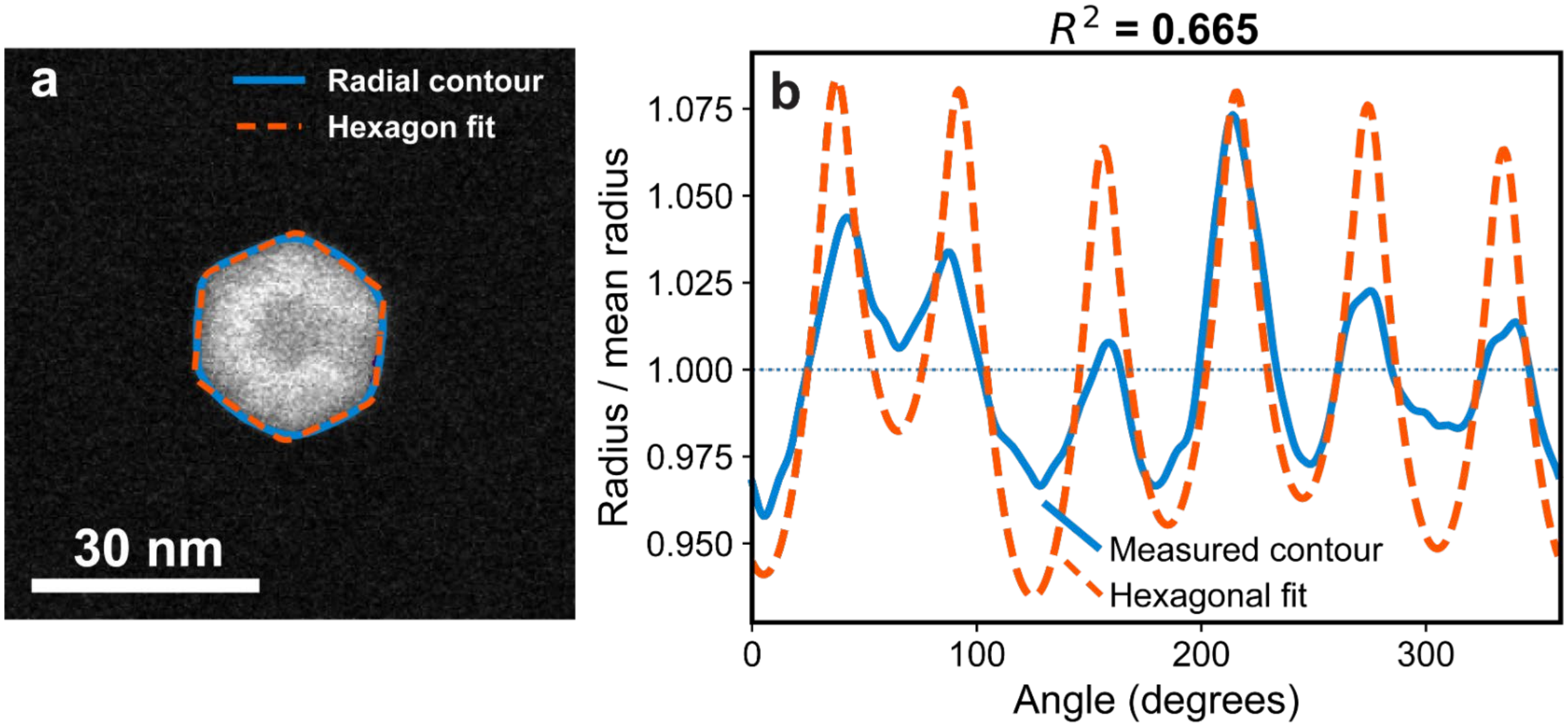

Figure S35. Hexagonal fitting of nanocrystals using a radial contour. (a) HAADF-STEM image with radial contour overlay (blue) and hexagonal fit (orange). (b) Normalized radial profile as a function of angle comparing nanocrystal radial contour (blue) to hexagonal fit (orange). The reported $R^2$ is the coefficient of determination between the measured mean-normalized radial contour and the best-fitting regular hexagon profile that allowed for rounded vertices.

Nanocrystal shape was quantified by extracting the particle boundary from each HAADF-STEM image and expressing the contour as the radial distance from the particle center as a function of angle. The measured radial profile was normalized by its mean radius and fitted to a hexagonal model while optimizing its orientation and scale. The fitting procedure allowed for rounding at the vertices to account for deviations from an ideal regular hexagon. Goodness of fit was quantified using the coefficient of determination, $R^2$, which represents the fraction of angular variation in the measured radial contour explained by the fitted hexagonal model. Higher $R^2$ values indicate greater agreement with a hexagonal shape. See Supporting Methods for details.

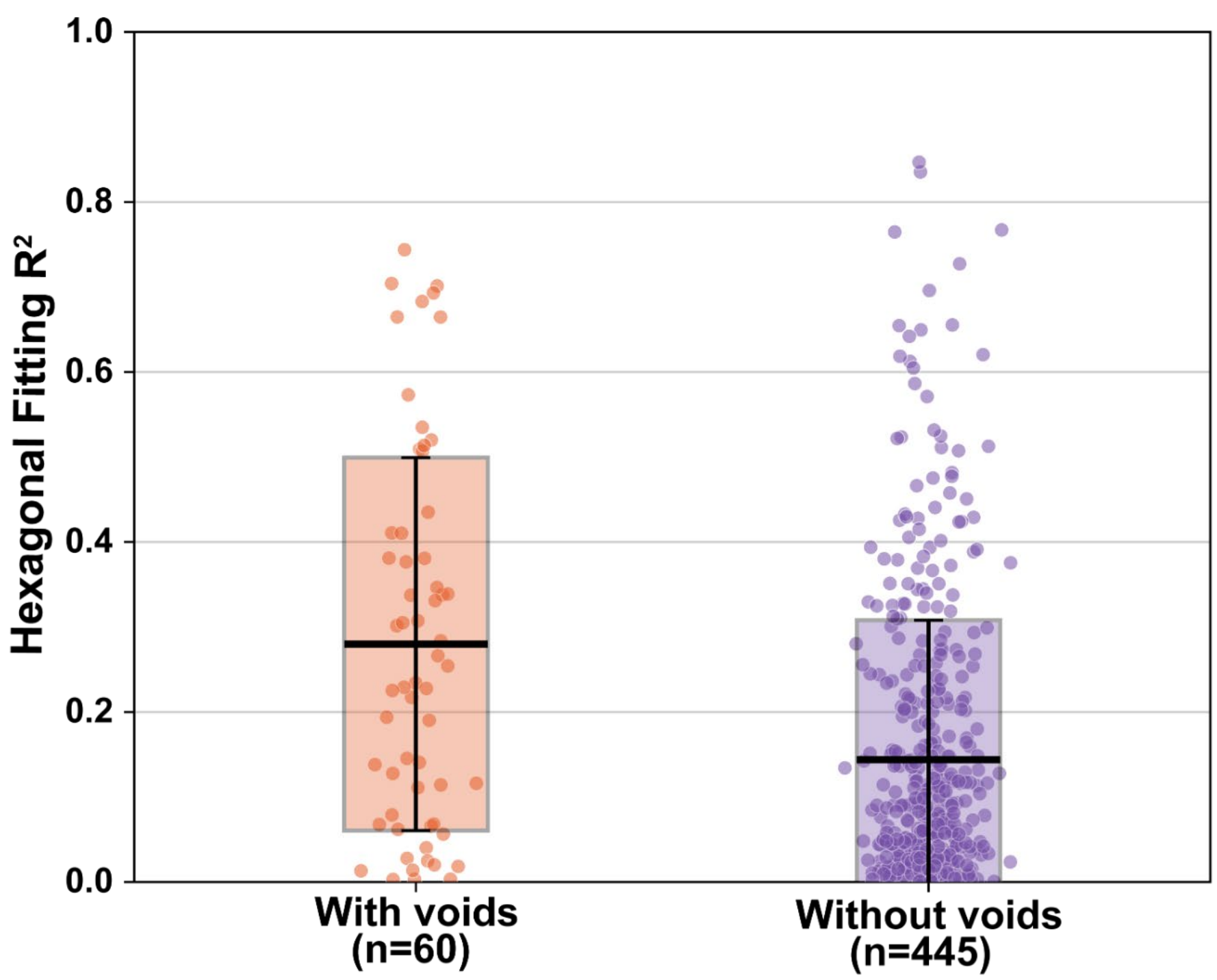

Figure S36. Comparison of hexagonal-fit $R^2$ values for void-containing and non-void-containing AuRu nanocrystals after atmospheric-pressure $H^2$ heating. Individual points represent measured nanocrystals, and the shaded distributions indicate the relative density of the data. Boxes and whiskers span one standard deviation above and below the mean, and horizontal lines indicate the mean. Void-containing nanocrystals exhibited a higher mean $R^2$ (0.3 ± 0.2, n = 60) than nanocrystals without voids (0.1 ± 0.2, n = 445). Of the 468 non-void-containing nanocrystals, 445 passed the contour-extraction and fit-quality criteria and were included in the shape analysis. The remaining 23 were excluded because they either intersected the edge of the image or had neighboring particles that were too close to allow for their shape to be isolated. Values are reported as mean ± standard deviation.

Nanocrystals containing voids exhibited greater average agreement with the fitted hexagonal model than particles without voids. The mean hexagonal-fit $R^2$ was 0.3 ± 0.2 for particles with voids (n = 60), compared with 0.1 ± 0.2 for particles without voids (n = 445). However, the broad, overlapping distributions indicate substantial variation within both populations.

Table S4. X-EDS quantification of Ru:Au ratios across pristine, post-vacuum, and voided gas-heated AuRu nanocrystals. Average Ru:Au atomic ratios were determined from X-EDS quantification of AuRu nanocrystals in the pristine state, after vacuum annealing, and after $H_2$ or $H_2$:$N_2$ heating with void formation. Atomic compositions were quantified using the Brown–Powell ionization cross-section model, with Au and Ru contents determined from the Au Lα and Ru Lα lines, respectively. Values are reported as mean ± standard deviation, with the number of nanocrystals analyzed listed for each category.

| **Category** | **Average Ru:Au** | **Number of particles** |
| --- | --- | --- |
| Pristine | 0.14 ± 0.04 | 17 |
| Post-vacuum annealing | 0.12 ± 0.02 | 17 |
| Post $H_2$ heating (voids present) | 0.15 ± 0.02 | 8 |
| Post $H_2$:$N_2$ heating (voids present) | 0.13 ± 0.03 | 2 |

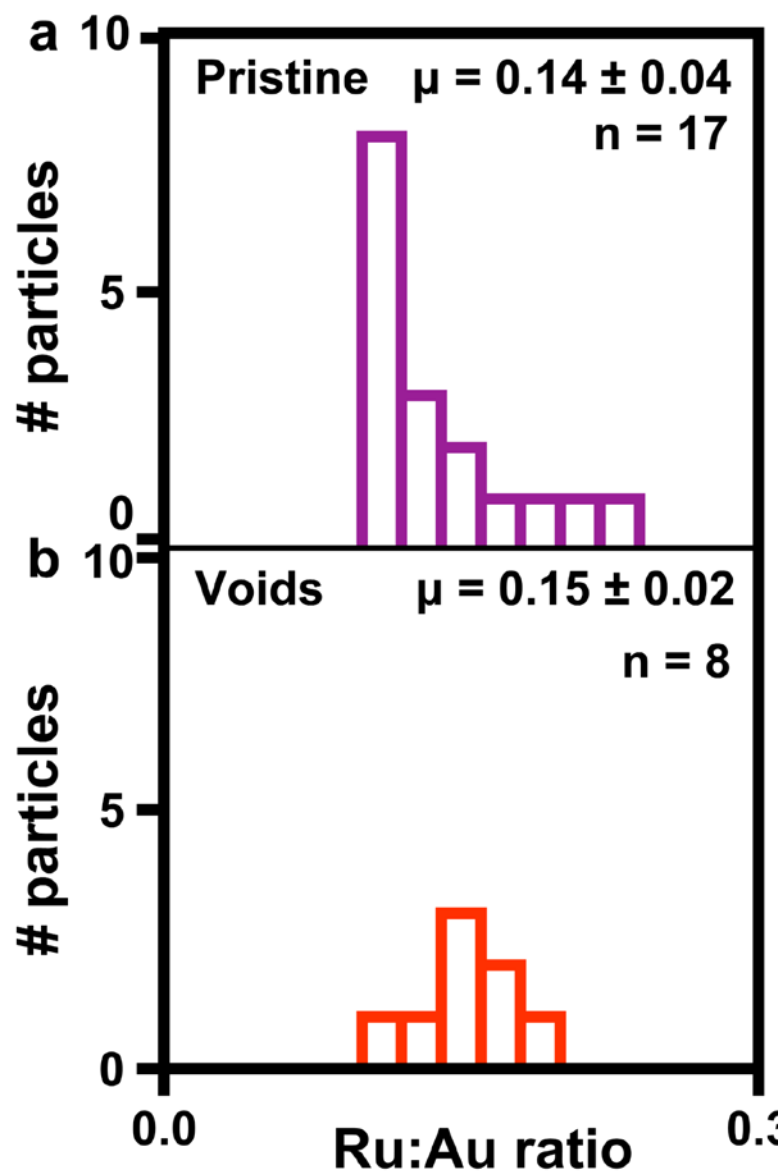

Figure S37. Ru:Au ratio distributions for pristine AuRu nanocrystals and AuRu nanocrystals with voids post-heating in $H_2$. Histograms show Ru:Au atomic ratios determined from X-EDS quantification of individual AuRu nanocrystals in the pristine state (a) and void-containing AuRu nanocrystals following heating in $H_2$ (b). The mean Ru:Au ratios were 0.14 ± 0.04 for pristine nanocrystals (n = 17) and 0.15 ± 0.02 for void-containing nanocrystals (n = 8), where values are reported as mean ± standard deviation.

Post-reaction correlative analysis shows that voids form across a range of final nanocrystal morphologies after atmospheric-pressure $H_2$ heating (Figure S27), including faceted, near-hexagonal AuRu nanocrystals containing internal planar-defect contrast (Figure S28). In these representative particles, voids are spatially associated with Ru-rich regions, suggesting that Ru-rich restructuring pathways contribute to nanovoid formation (Figure S28). X-EDS quantification further shows that voided $H_2$-heated nanocrystals retain Ru:Au ratios comparable to the pristine ensemble (Figure S37; Table S4). Together, these results indicate that void formation is associated with Ru-rich restructuring pathways but is not driven by measurable Ru loss from the nanocrystals.

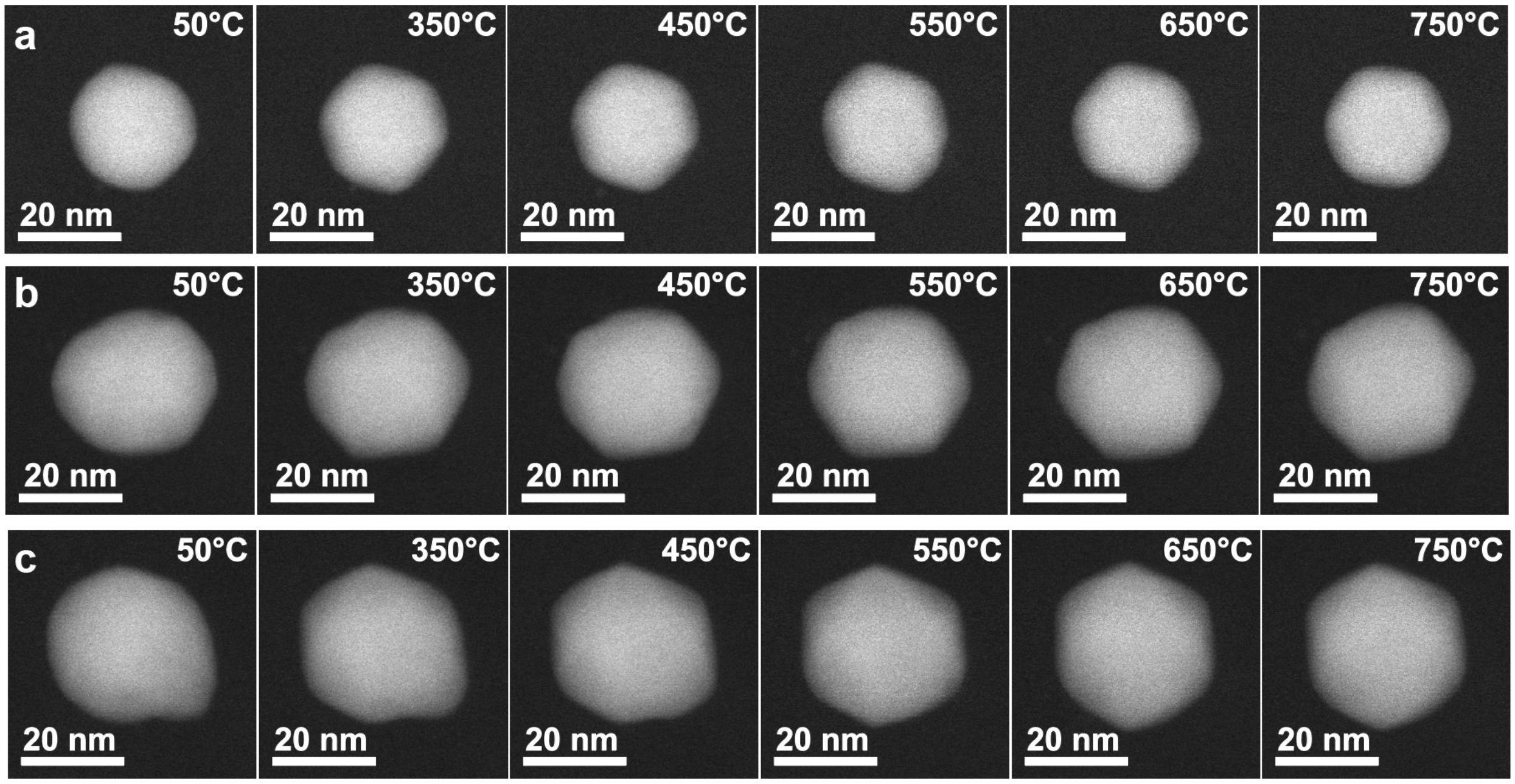

Figure S38. Pure Au nanocrystals heated in 774 Torr $H_2$ show no nanovoid formation. Sequential HAADF-STEM images of three representative pure Au nanocrystals (a–c) with different initial sizes and morphologies during annealing in atmospheric-pressure $H_2$ from 50 to 750 °C. Heating induces hexagonal faceting in all particles, but no internal nanovoids are observed, indicating that hydrogen alone is insufficient to drive void formation in pure Au.

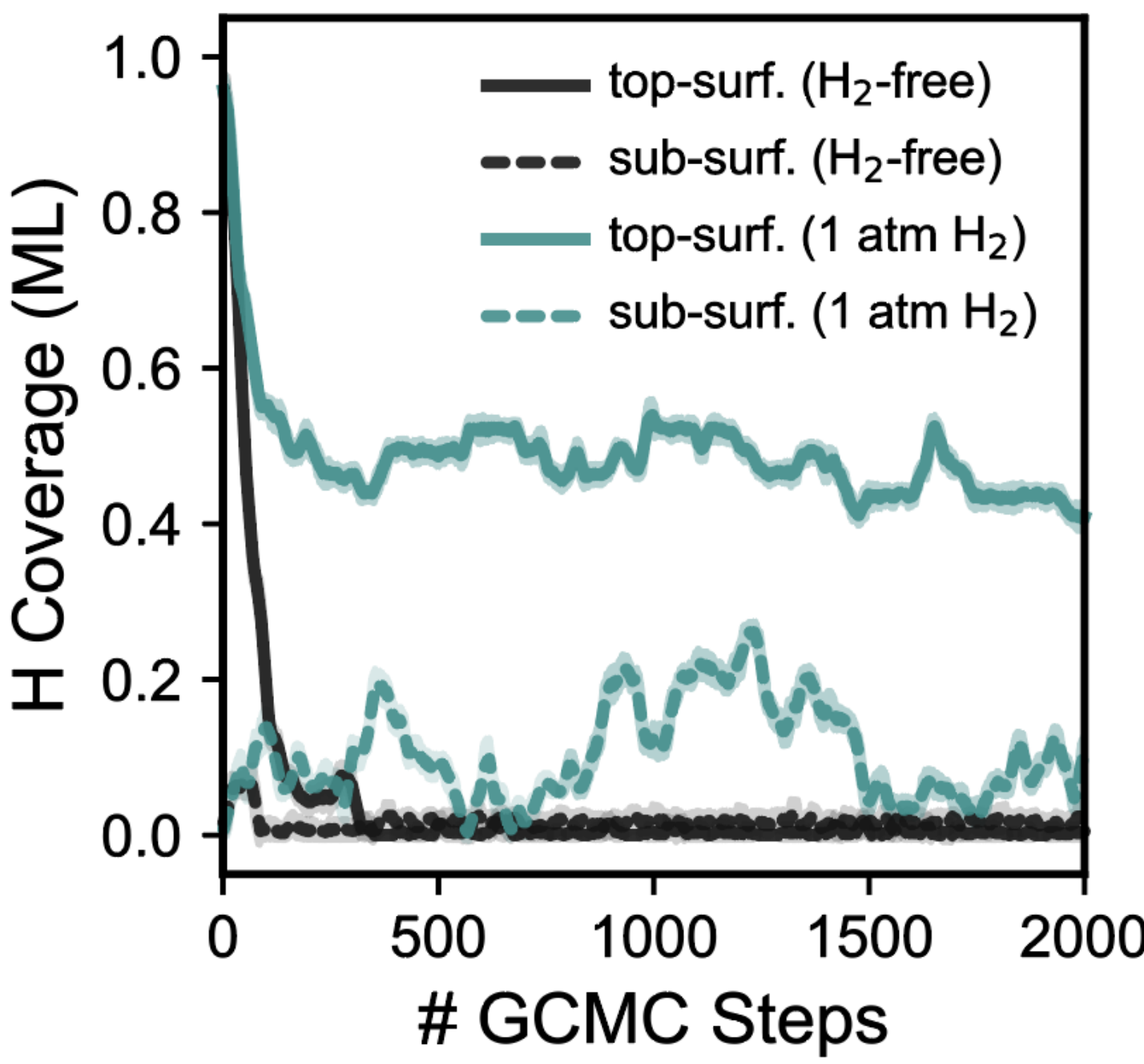

Figure S39. H-coverage in $H_2$-free and 1 atm $H_2$ conditions. In $H_2$-free conditions (black), H coverage converges to 0. In 1 atm $H_2$ conditions (green), we see the formation of ~0.5 ML of H atoms on the Ru surface and the presence of some subsurface H.

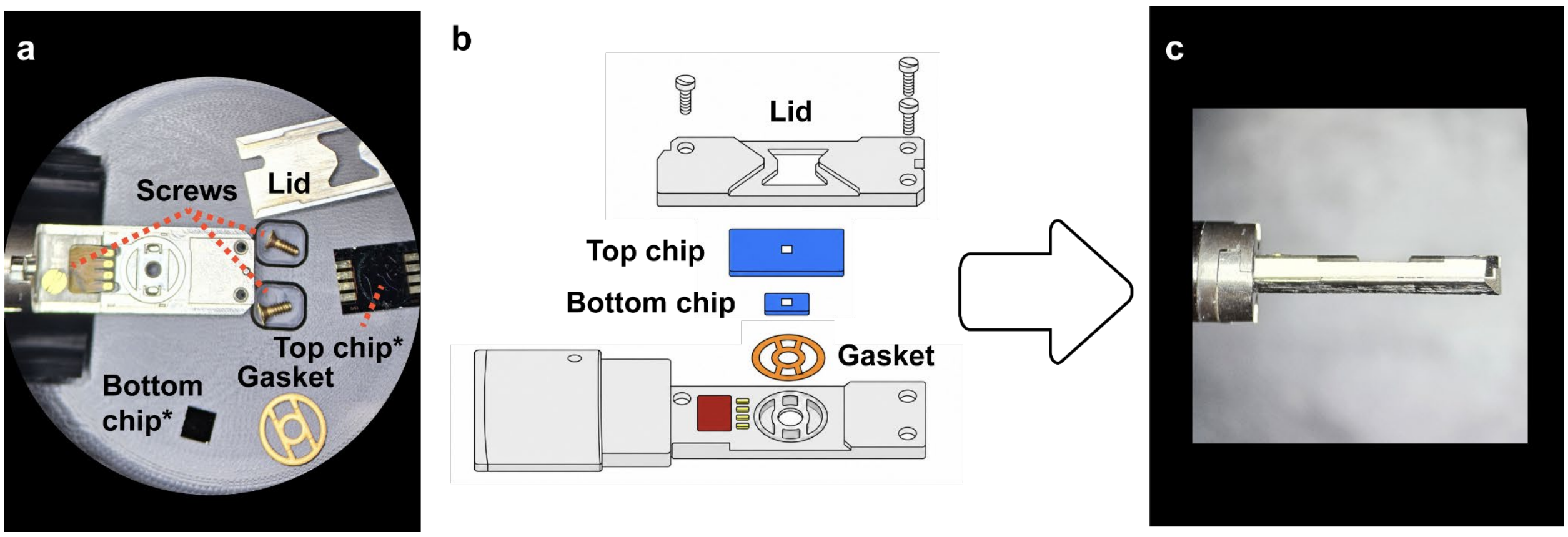

Figure S40. Protochips Atmosphere gas-cell holder assembly. (a) Stereoscope image of the Protochips Atmosphere holder components prior to assembly, including the lid, screws, gasket, top chip, and bottom chip. (b) Schematic showing the assembly order of the bottom chip, gasket, top chip, and lid in the Protochips Atmosphere gas-cell holder. (c) Side-view stereoscope image of the Protochips Atmosphere holder after gas-cell assembly.

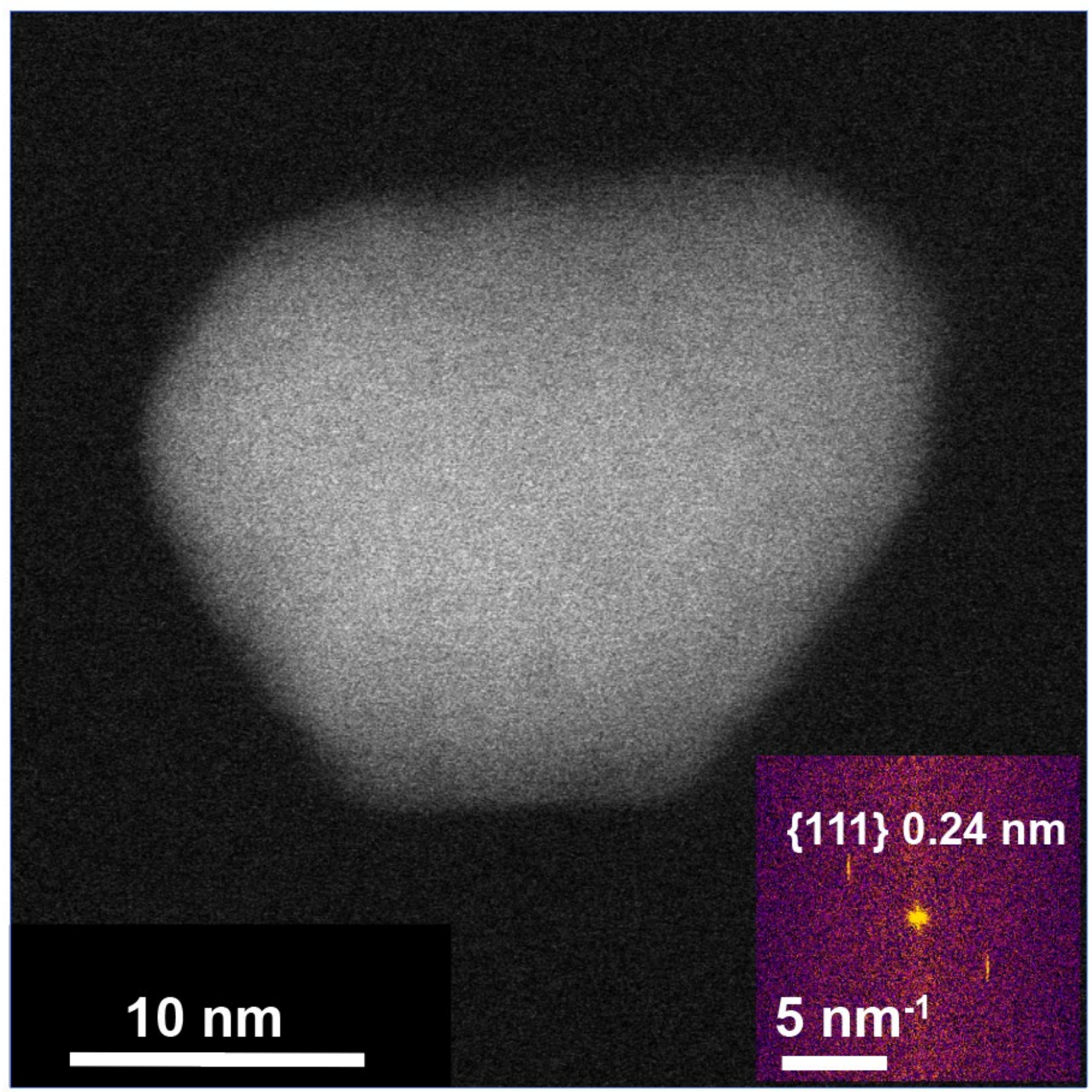

Figure S41. AuRu {111} lattice fringes resolved under 1 atm $H_2$. HAADF-STEM image of an AuRu nanocrystal before phase segregation, demonstrating the best lattice resolution achieved under atmospheric-pressure gas-cell conditions. Inset, FFT showing the {111} reflection at 0.24 nm.

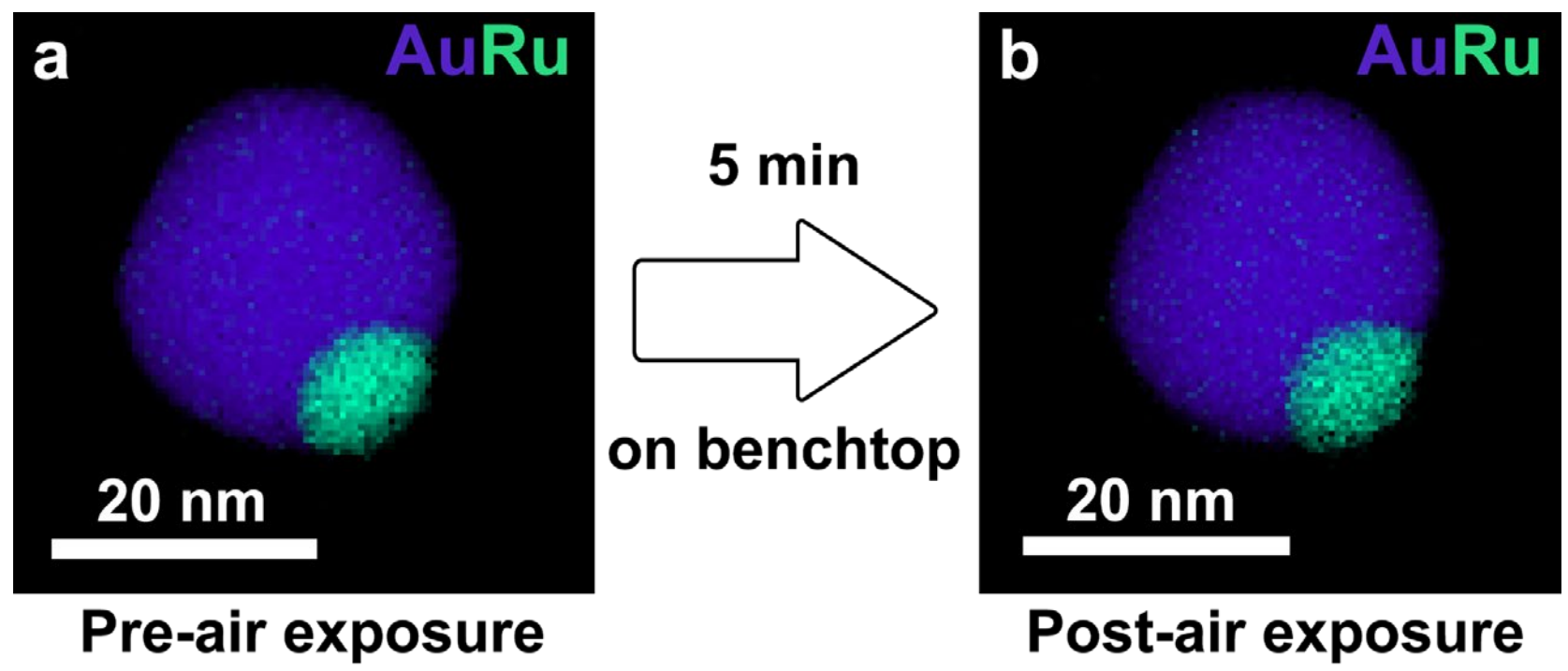

Figure S42. Understanding impact of air exposure on AuRu X-EDS elemental mapping. (a) Composite Au Lα/Mα and Ru Kα/Lα X-EDS elemental map of an AuRu nanocrystal acquired immediately after vacuum annealing without air exposure. (b) Composite Au Lα/Mα and Ru Kα/Lα X-EDS elemental map of the same nanocrystal after 5 min of air exposure on the benchtop, simulating gas-cell disassembly. No qualitative change in Au/Ru spatial distribution is observed.

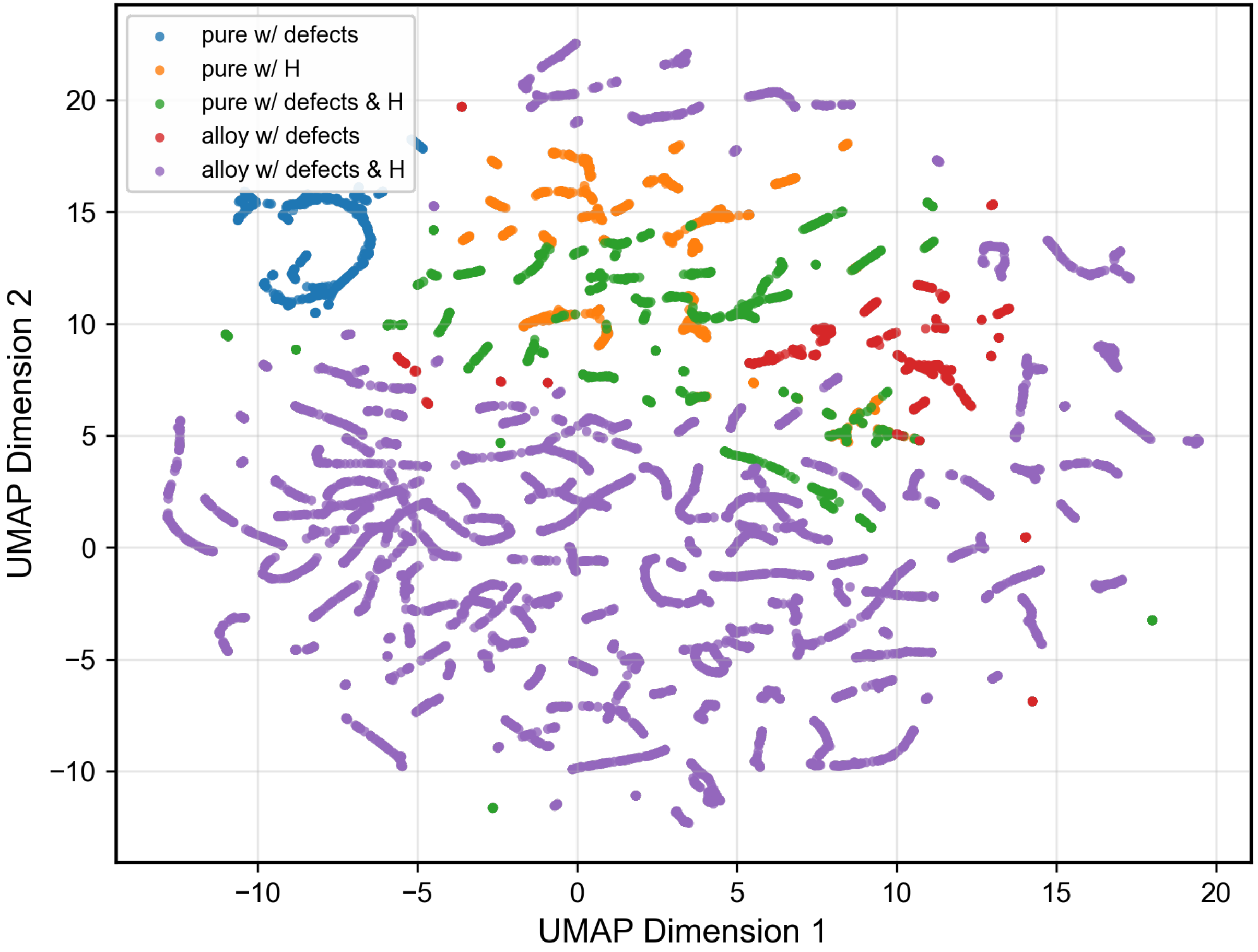

Figure S43. Visualization of the training dataset distribution. UMAP projection of 15,695 structures from five sampling groups (pure/alloy systems with/without defects and/or H). Structures were characterized using SOAP descriptors, reduced via Incremental PCA (50 components), and projected to 2D space using UMAP. Each point represents a sampled configuration, colored by its corresponding group.

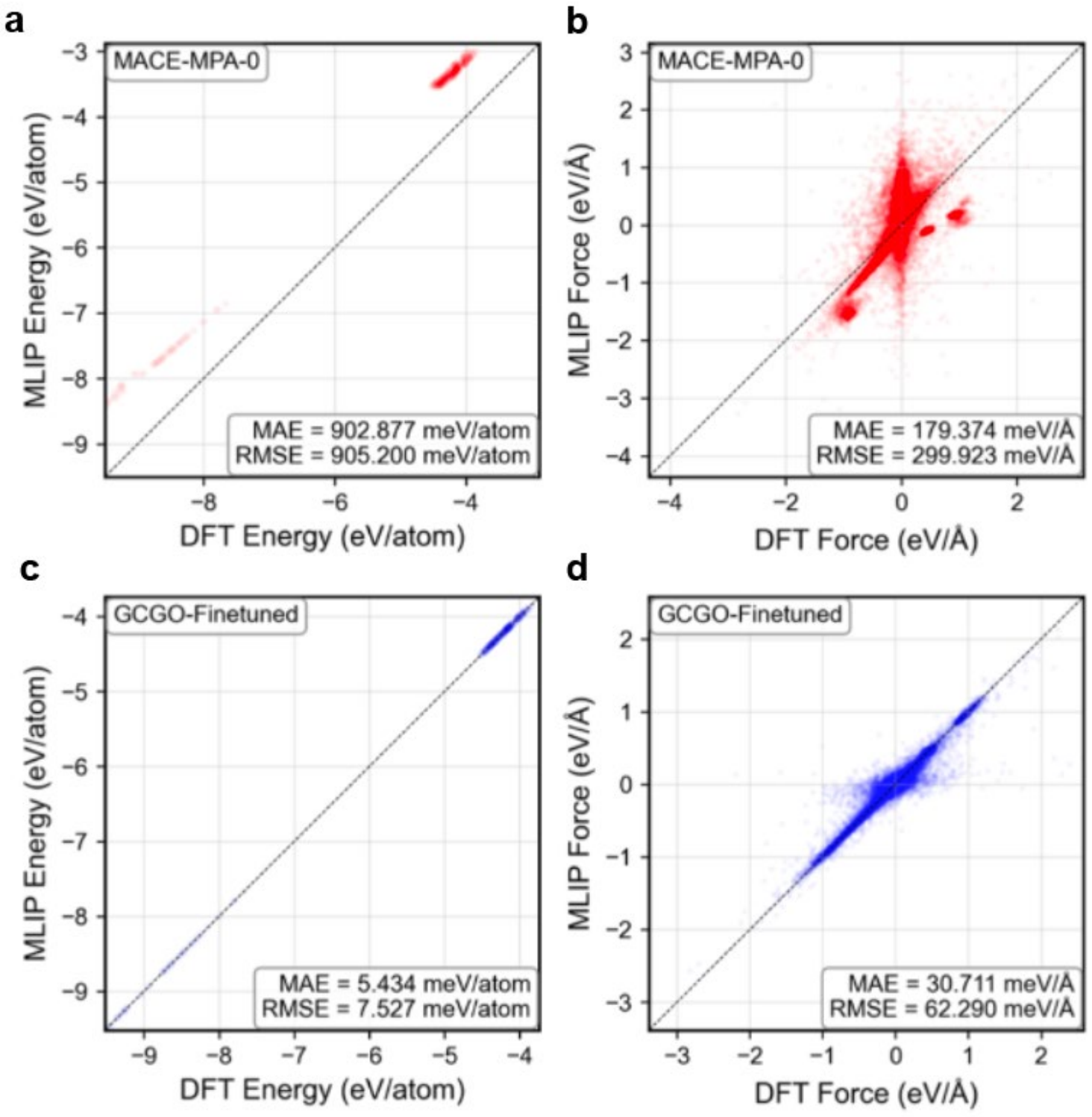

Figure S44. Performance of the MLIP on the test dataset. Parity plots for (a) energy and (b) force of the MACE-MPA-0 foundation model, and for (c) energy and (d) force of our fine-tuned version on GCGO samples. Each semi-transparent marker represents a data point.

## Supporting Methods:

### Image Registration and Particle Size Analysis

Image registration, field-of-view mapping, and particle size analysis were performed using Python workflows implemented in JupyterLab with the NumPy, Pandas, Matplotlib, scikit-image, and SciPy packages. High-resolution HAADF-STEM images (1.22 µm field of view) were registered to a large-field-of-view HAADF-STEM image (9.87 µm field of view) using normalized cross-correlation template matching (skimage.feature.match_template). Before registration, image intensities were normalized and a Gaussian filter ($\sigma = 1.0$ pixel) was applied. Registration consisted of an initial search using images downsampled by a factor of four, followed by full-resolution refinement within a local search region. To prevent double-counting nanocrystals in overlapping acquisitions, overlapping pixels were retained only in the image with the highest template-matching score and masked in all other images.

Particle sizes were measured from the resulting non-overlapping image regions. Images were Gaussian-filtered ($\sigma = 0.7$ pixel), and nanocrystals were segmented using a global intensity threshold scaled by a factor of 1.25. Binary masks were refined by removing objects and filling holes smaller than five pixels, and regions containing image annotations were excluded. Individual nanocrystals were identified using connected-component labeling and measured using a calibrated pixel size of 0.2986 nm $pixel^{-1}$.

### Hexagonal Shape Analysis

Nanocrystal shapes were analyzed using Python workflows implemented in JupyterLab with the NumPy, Pandas, Matplotlib, scikit-image, and SciPy packages. Nanocrystals were segmented from HAADF-STEM images using the same intensity-threshold masks generated for the particle-size analysis, and their external boundaries were represented by convex-hull contours. The convex hull bridges small indentations in the threshold mask, allowing the overall particle morphology to be analyzed without sensitivity to minor edge irregularities. Nanocrystals were excluded if they intersected an image boundary or if the threshold mask did not yield a complete, reliable contour, as could occur for overlapping, aggregated, or highly irregular particles.

Each accepted contour was fitted to a regular-hexagon model constrained to have six equal-length facets, 120° internal angles, and rounded vertices. Experimental contours were expressed as the radial distance from the particle center and sampled at 360 equally spaced angular increments. Model orientation was initialized through a 360° angular sweep and subsequently refined by least-squares optimization to minimize the residual sum of squares between the measured and modeled radial profiles.

The coefficient of determination ($R^2$) was calculated from the measured and fitted mean-normalized radial profiles. Higher $R^2$ values indicate closer correspondence to a regular hexagonal morphology but do not represent a general measure of faceting; strongly faceted particles with nonhexagonal or elongated shapes may therefore exhibit low $R^2$ values.